\RequirePackage{fix-cm}
\documentclass[smallextended]{svjour3}       % onecolumn (second format)
\smartqed  % flush right qed marks, e.g. at end of proof
\usepackage{graphicx}
\usepackage{ragged2e}
\usepackage[usenames]{color}
\usepackage[colorlinks=true,linkcolor=red,citecolor=blue]{hyperref}
\usepackage{amsmath,bm}      % use Times fonts if available on your TeX system
\usepackage{amssymb,multirow,multicol}
\usepackage[clockwise,figuresright]{rotating}
\usepackage{algorithm}
\usepackage{algorithmic}
\usepackage{caption}

\begin{document}

\title{Robust Estimation for Two-Dimensional Autoregressive Processes Based on Bounded Innovation Propagation Representations %On robust BMM estimation in two-dimensional autoregressive models%\thanks{Grants or other notes
}


\author{Grisel Maribel Britos         \and
        Silvia Mar\'{i}a Ojeda %etc.
}


\institute{Grisel Maribel Britos \at
              Facultad de Matem\'atica, Astronom\'ia, F\'isica y Computaci\'on, Universidad Nacional de C\'ordoba, C\'ordoba, Argentina\\
              \email{gbritos@famaf.unc.edu.ar}           %  \\
           \and
           Silvia Mar\'{i}a Ojeda \at
              Facultad de Matem\'atica, Astronom\'ia, F\'isica y Computaci\'on, Universidad Nacional de C\'ordoba, C\'ordoba, Argentina\\
              \email{ojeda@famaf.unc.edu.ar}
}

\date{Received: date / Accepted: date}

\maketitle

\begin{abstract}
Robust methods have been a successful approach to deal with contaminations and noises in image processing. In this paper, we introduce a new robust method for two-dimensional autoregressive models. Our method, called BMM-2D, relies on representing a two-dimensional autoregressive process with an auxiliary model to attenuate the effect of contamination (outliers). We compare the performance of our method with existing robust estimators and the least squares estimator via a comprehensive Monte Carlo simulation study which considers different levels of replacement contamination and window sizes. The results show that the new estimator is superior to the other estimators, both in accuracy and precision. An application to image filtering highlights the findings and illustrates how the estimator works in practical applications.

\end{abstract}

\section{Introduction}
\label{sec:intro}
Robust statistics methods arise in a wide range of applications; in particular, the words ``robust'' and ``robustness'' frequently appear in the context of image and signal processing. Filters based on robust statistics such as the median filter, provide basic mechanisms in image processing (\cite{Ays}; \cite{Hua}; \cite{Tzu}). In general, image processing based on robust strategies has shown remarkable ability in image restoration and segmentation (see, for example, \cite{Ben}; \cite{Ji}; \cite{Tar}).
High-level image analysis, or vision, also benefits from the use of robust estimation techniques, as it can be seen in \cite{Com}; \cite{Got}; \cite{Kim}; \cite{Pra} and \cite{Sin}.

There is a large number of examples such as signal analysis and processing on the plane, imaging remote sensing and design of experiments in agronomy, in which data is recorded in a grid or lattice in $\mathbb{Z}^2$. A class of 2D autoregressive processes has been proposed (\cite{Whi}) as a set of plausible models for the spatial correlation in such data (\cite{Tjo}). These models are natural extensions of the autoregressive processes used in time series analysis (\cite{Bas}).

Thus, most of the robust procedures suggested for time series parametric models have been generalized for spatial parametric models when the process has been contaminated with innovation or additive outliers (\cite{Kas}). Because a single extreme value is capable of introducing significant distortions in estimators, most initiatives are focused on providing estimators that are robust to the appearance of anomalous data.

In this context, there are at least three classes of robust estimators that have been studied. They are the estimators M, GM and RA. The M estimators were successfully used to develop and implement a computational algorithm for image restoration, based on 2D autoregressive models (\cite{Kas}). Later, \cite{All1} analyzed the development and implementation of Generalized M (GM) estimators for the same class of models. The robust estimators of Residual Autocovariance (RA) were presented by \cite{Bus3} in the context of the time series, where in the procedure of recursive estimation the residues are cleaned through the application of a robust function. The extension of the RA estimators for two-dimensional autoregressive models and their corresponding computational treatment were developed by \cite{Oje4}. Monte Carlo simulation studies show that the performance of the RA estimator is higher than the M estimator and slightly better than the GM estimator when the model has been contaminated with additive type outliers. In addition, \cite{Bus5} studied the asymptotic behavior of the RA estimator for unilateral two-dimensional autoregressive processes, generalizing the results for the asymptotic behavior of one-dimensional series previously established by \cite{Bus4}. We do not yet know the asymptotic properties associated with the M and GM estimators when the model is contaminated. 

One of the reasons why the spatial autoregressive model (AR-2D) has been extensively used in image analysis and processing is its remarkable ability to represent a variety of real scenarios without the need to use a large number of parameters. However, the robust estimators developed so far for the parameters of the AR-2D model have been constructed only under the assumption of innovative or additive type random noise; there are no proposals for parameter estimation when the model is contaminated from another more general pattern of noise. In this work we define a new class of robust estimators for the AR-2D contaminated models. This class of estimators is robust under replacement contamination that includes additive type contamination. To avoid the propagation of the effect of an outlier when calculating the innovative residuals in the AR-2D model we propose a new approach that consists of defining this residuals using an auxiliary model. For this reason, bounded innovation propagation AR-2D (BIP-AR-2D) models are proposed. With the help of these models we suggest an estimator for the AR-2D model that is robust when the process contains outliers.

Our proposal is the generalization of the two- dimensional case of the BMM estimators, developed by \cite{Mull} for ARMA time series models. The approach of generalizing one-dimensional proposals to the case of two or more dimensions is the strategy that is naturally used in many areas of science. The work is relevant because it offers a new way of robustly estimating the parameters in the two-dimensional autoregressive model. It is a contribution that can be beneficial in the field of signal processing, spatial statistics and, in general, in areas of applied mathematics that use this model intensively, for example, to represent texture images such as robotic vision, recognition of patterns and matching problems. Although estimators of the model parameters have been developed so far, the central objective of this work is to provide a new estimator that betters the estimation proposals already known.

The rest of the paper is organized as follows. In Section \ref{sec:preli}, the basic definitions are presented. We first presented some background material on bidimensional autoregressive processes (AR-2D) and parameter estimators of the model. We also define procedures for generating replacement contamination in such models. In Section \ref{sec:BMM}, the new model BIP-AR-2D for spatial processes and the new estimator of the AR-2D model parameters are presented. In Section \ref{sec:MC}, several Monte Carlo studies are carried out to evaluate the performance of the new estimator against different contamination schemes, compared to the LS, M, GM and RA estimators. Section \ref{sec:apli} presents two applications to real images that demonstrate the capabilities of the BMM-estimator to represent, segment and restore contaminated images. Conclusions and future works appear in Section \ref{sec:concl}. The results of the Monte Carlo studies (Section \ref{sec:MC}) are shown in the Appendix.

\section{Preliminaries}
\label{sec:preli}
\subsection{The spatial ARMA models}
\begin{figure}[h!]
\begin{center}
\begin{tabular}[c]{cccc}
\begin{minipage}{3.5cm}
\includegraphics[width=3.5cm, height=3.5cm]{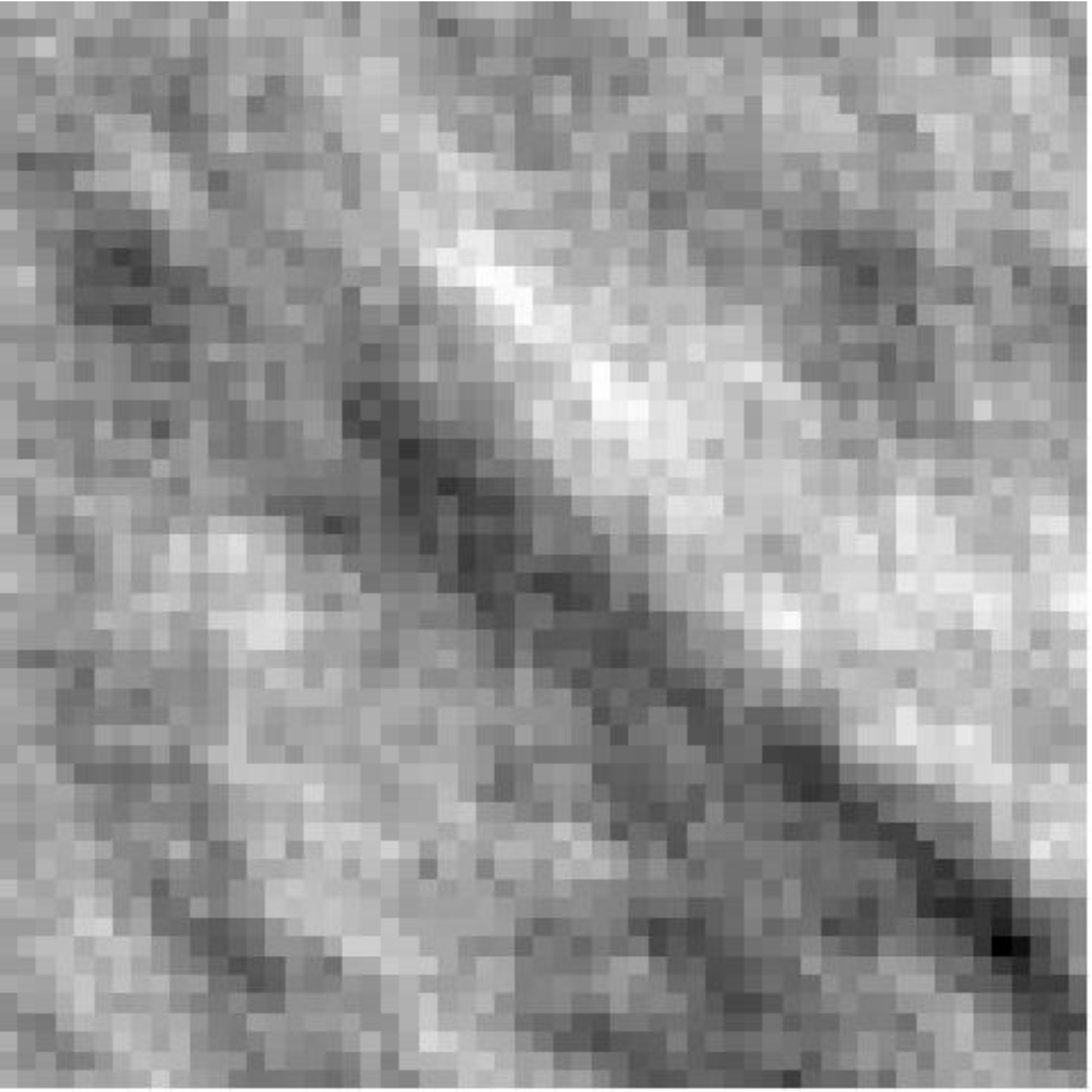}
\begin{center}
(a)
\end{center}
\end{minipage} & 
\begin{minipage}{3.5cm}
\includegraphics[width=3.5cm, height=3.5cm]{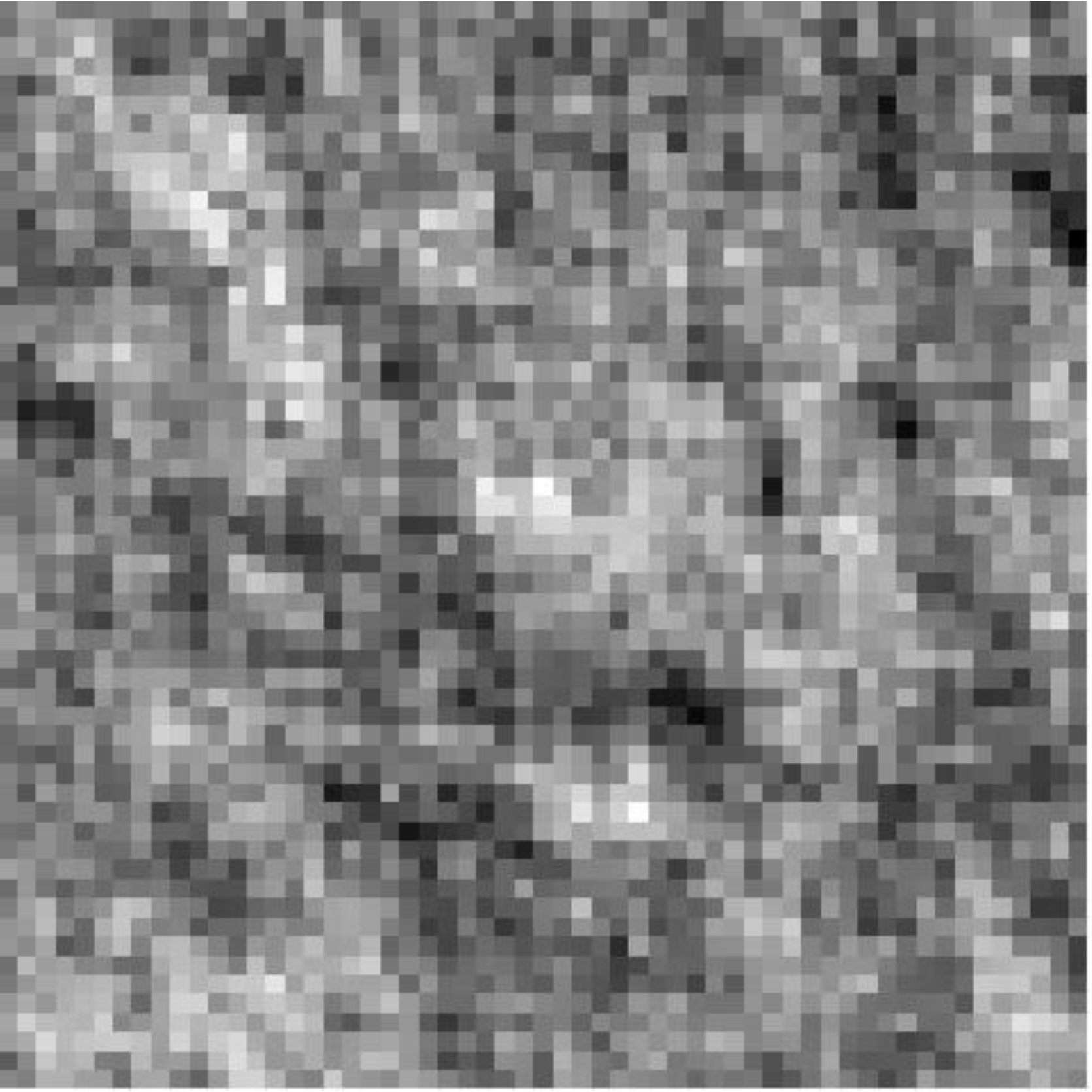}
\begin{center}
(b)
\end{center}
\end{minipage} &
\begin{minipage}{3.5cm}
\includegraphics[width=3.5cm, height=3.5cm]{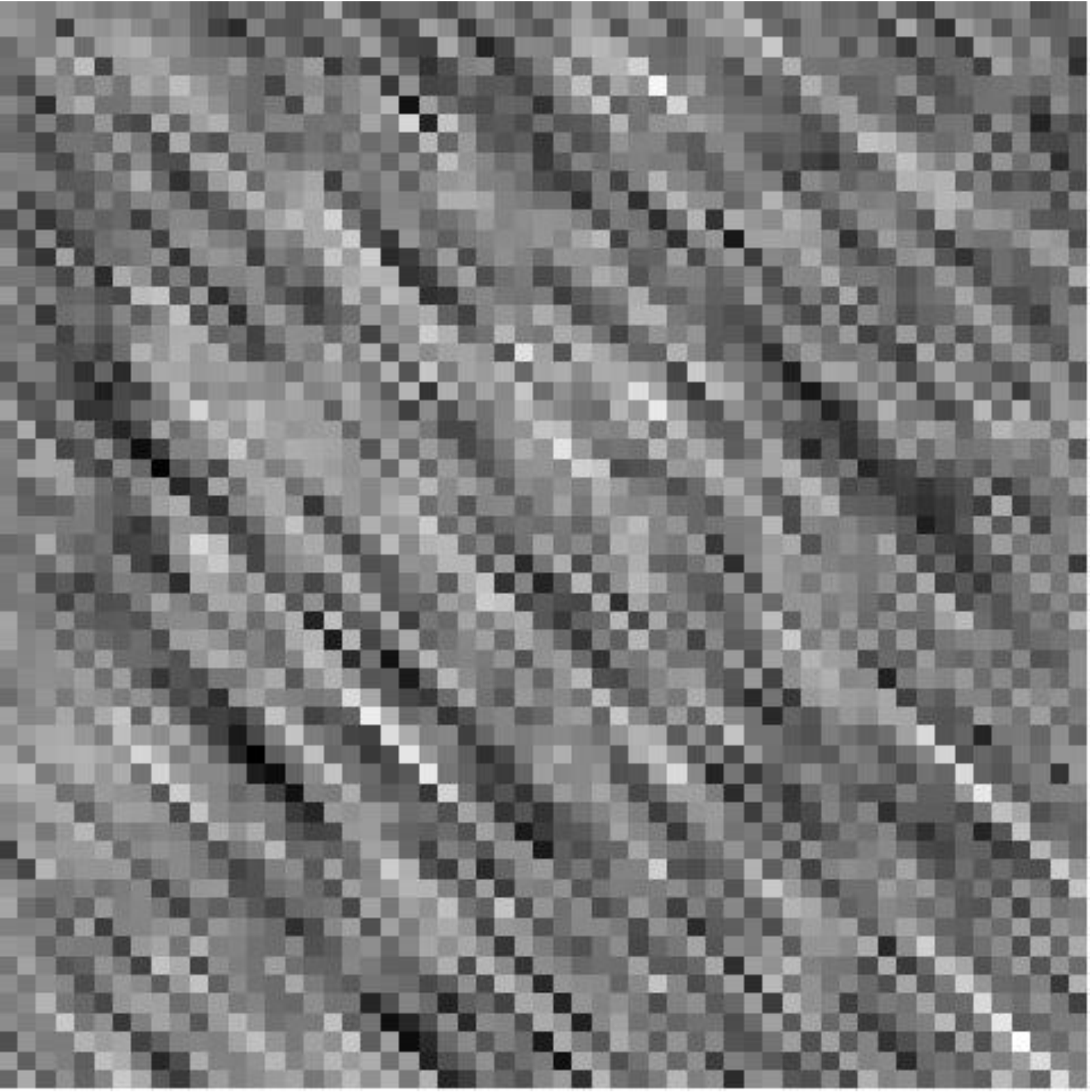}
\begin{center}
(c)
\end{center}
\end{minipage} &
\begin{minipage}{3.5cm}
\includegraphics[width=3.5cm, height=3.5cm]{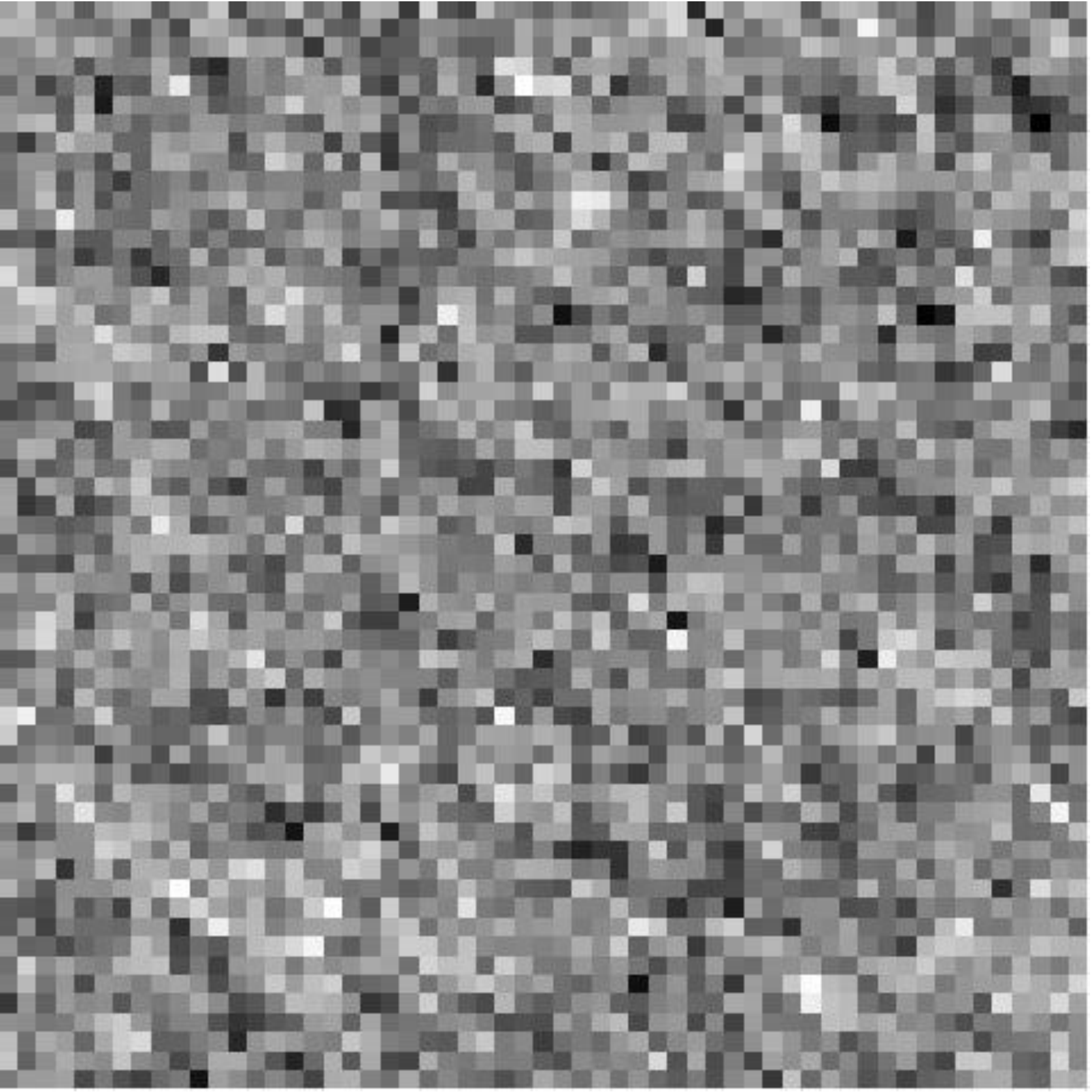}
\begin{center}
(d)
\end{center}
\end{minipage} \\

\end{tabular}

\caption{Autoregressive Processes. (a) $\phi_1=0.5$, $\phi_2=0.4999$, $\phi_3=0$; (b) $\phi_1=0.5$, $\phi_2=0.4$, $\phi_3=0$; (c) $\phi_1=0.01$, $\phi_2=0.01$, $\phi_3=0.8$; (d) $\phi_1=0.15$, $\phi_2=0.17$, $\phi_3=0.2$.}
\label{figTexturas}
\end{center}
\end{figure}

A great variety of texture can be generated through the two-dimensional autoregressive moving average models (ARMA-2D models). For example, Figure \ref{figTexturas} shows textures generated for the particular case of ARMA-2D processes, the   AR-2D process, with two and three parameters (Fig. \ref{figTexturas} (a)-(b) and (c)-(d) respectively). Besides, in the last years, several theoretical properties have been studied, and multiple applications have been developed for these processes (\cite{Bar,Vall3,Bus6,Qui,Zie,Yao,Sad}).

%\clearpage

Spatial autoregressive moving average (ARMA) processes can be defined over random fields indexed over $\mathbb{Z}^d, d \geq 2,$ where $\mathbb{Z}^d$ considers the usual partial order, i.e., for $s=(s_1,s_2,...,s_d), u=(u_1,u_2,..., u_d)$ in $\mathbb{Z}^d,$ $s\leq u$ if $s_i \leq u_i$ for $i=1,2,..., d$. We define $S[a,b]=\{x \in \mathbb{Z}^d|a \leq x \leq b\}$ and $S\langle a,b]=S[a,b]\backslash \{a\}$ for $a,b \in \mathbb{Z}^d,$ such that $a\leq b$ and $a\neq b$. 

A random field $(Y_s)_{s \in \mathbb{Z}^d}$ is said to be a spatial ARMA of order $p,q\in \mathbb{Z}^d$ if it is weakly stationary and satisfies the equation
\begin{equation}\label{arma_model}
Y_s-\sum_{j\in S\langle 0,p]}\phi_jY_{s-j}=\varepsilon_s+\sum_{k\in
S\langle 0,q]}\theta_k\varepsilon_{s-k},
\end{equation}
where $(\phi_j)_{j\in S\langle 0,p]}$ and $(\theta_k)_{k\in S\langle 0,q]}$ denote the autoregressive and moving average parameters respectively with $\phi_0=\theta_0=1,$ and $(\varepsilon_s)_{s \in \mathbb{Z}^d}$ denotes a sequence of i.i.d. random variables with variance $\sigma^2.$ Note that if $q=0$ the process is called spatial autoregressive AR($p$) random field. An ARMA random field is called causal if it can be represent for the following equation:
\begin{equation}
Y_s=\sum_{j\in S[ 0,\infty]}\phi_j\varepsilon_{s-j},
\label{MAinf}
\end{equation}
with $\sum_j |\phi_j|<\infty.$
Similar to the time series case, there are conditions on the (AR or MA) polynomials for them to be stationary and invertible respectively (\cite{Bas}). As an example, consider a first-order autoregressive process as in (\ref{arma_model}) with $d=2$, $p=(1,1)$ and $q=(0,0)$. Then $S=\langle (0,0), (1,1)]=\{(1,0), (0,1), (1,1)\}$ and the model is of the form
\begin{equation}
\label{modeloAR}
Y_{i,j}=\phi_{1}Y_{i-1,j}+\phi_{2}Y_{i,j-1}+\phi_{3}Y_{i-1,j-1}+\varepsilon_{i,j}
\end{equation}
where to simplify the notation it took $\phi_1=\phi_{1,0}$, $\phi_2=\phi_{0,1}$ and $\phi_3=\phi_{1,1}$. In equivalent form, (\ref{modeloAR}) can be expressed as
\begin{equation}
\label{modeloARresum}
\Phi(B_1,B_2)Y_{i,j}=\varepsilon_{i,j}
\end{equation}
where $B_1$ and $B_2$ are the backward operators given by $B_1Y_{i,j}=Y_{i-1,j}$, $B_2Y_{i,j}=Y_{i,j-1}$ and in (\ref{modeloARresum}), $\Phi(B_1,B_2)=(1-\phi_{1} B_{1}-\phi_{2}B_{2}-\phi_{3} B_{1}B_{2})$. In the case that $\Phi(B_1,B_2)$ has inverse, we can write equation (\ref{MAinf}) as
\begin{equation*}
Y_{i,j}=\Phi(B_1,B_2)^{-1}\varepsilon_{i,j}
\end{equation*}
\cite{Bas} studied the correlation structure of a process like (\ref{modeloAR}). They obtained conditions to guarantee the existence of the stationary representation of the model (\ref{modeloAR}) as in (\ref{MAinf}). In that case, the use of a multinomial expansion for $\Phi(B_1,B_2)^{-1}$ implies the convergent representation
\begin{equation}
\label{MAinf2}
	Y_{i,j}=\sum_{k=0}^{\infty}\sum_{l=0}^{\infty}\sum_{r=0}^{\infty}\lambda_{klr}\varepsilon_{i-k-r,j-l-r}
\end{equation}
where $\lambda_{klr}=\frac{(k+l+r)!}{k!l!r!}\phi_{1}^k\phi_{2}^l\phi_{3}^r$ with $k,l,r\in \mathbb{N}\cup\{0\}$ are the coefficients of this multinomial expansion. The increase in the number of parameters in the model also increases the diversity of possible textures but in contrast, the calculations become more complex. In this paper we worked with the AR 2D model with three parameters as in (\ref{modeloAR}).\\

\subsection{Types of contamination in AR-2D processes}

\cite{Maro} described (Chapter 8) some probability models for time series outliers, including additive outliers (AOs), replacement outliers (ROs) and innovation outliers (IOs). 

In this section we generalized the notion of replacement outliers for spatial processes. Let $Y$ be a stationary process, for example an AR-2D process as (\ref{modeloAR}) and let $Z$ be the observed process. It is said that $Z$ process behaves like a two-dimensional Replacement Outlier model (RO) if it is given by
\begin{equation}
\label{eqContam}
Z_{i,j}=(1-\xi_{i,j}^{\alpha})Y_{i,j}+\xi_{i,j}^{\alpha}W_{i,j}
\end{equation}
where $\xi^{\alpha}$ is a zero-one process such that $P(\xi_{i,j}^{\alpha}=1)=\alpha$  and $P(\xi_{i,j}^{\alpha}=0)=1-\alpha$ and $W$ is a replacement process that is not necessarily independent of $Y$. The fraction $\alpha$ is positive and small.\\
A particular case of the RO models is the two-dimensional Additive Outlier model (AO), in which 
\begin{equation*}
W_{i,j}=Y_{i,j}+\nu_{i,j},
\end{equation*}
$\nu$ is a stationary process independent of $Y$ and $\xi^{\alpha}$ is a Bernoulli process. This type of contamination is very important for satellite image processing; for example, it is present in optical images such as those from Landsat satellites. When $W$ does not follow the pattern of AO, we say that the contamination process follows a Strictly Replacement Outlier model (SRO).

\subsection{Robust Parametric Estimation}

Because LS estimators are very sensitive to the presence of atypical values (Martin, 1980), several alternative estimators arise to mitigate the impact of contaminated observations on estimates. Most of these proposals are natural extensions of robust estimators studied in time series.

Robust estimators have been defined for models with a small amount of  parameters. Here, we summarize the well-known robust estimators for the model (\ref{modeloAR}); however, a more general development can be found for the AR and MA models in \cite{Kas}, \cite{All4}, \cite{Oje4}, \cite{Vall3} and \cite{Bus5}.

Note that model (\ref{modeloAR}) can be rewritten in the linear model form:
\begin{equation*}
Y_{i,j} = {\bm \phi}^T {Z}_{i, j} + \epsilon_{i,j},
\end{equation*}
where $\mbox{\boldmath{$\phi$}} ^{T}=( \phi_{1},\phi_{2},\phi_{3})$ is a parameter vector and $ {Z}_{i,j}^T = (Y_{i-1,j}; Y_{i,j-1}; Y_{i-1,j-1}).$
To obtain the LS estimator of $\bm \phi$, we minimize the following function:
\begin{equation*}
\sum_{i,j}\left[ \varepsilon_{i,j}({\bm \phi})\right]^2,
\end{equation*}
where
\begin{equation}
\label{resAR}
\varepsilon_{i,j}({\bm \phi}) = Y_{i,j}-{\bm \phi}^T{Z}_{i, j} = Y_{i,j}-\phi_{1}Y_{i-1,j}-\phi_{2}Y_{i,j-1}-\phi_{3}Y_{i-1,j-1}
\end{equation}
Similarly, the class of M estimators for 2D autoregressive processes (\cite{Kas}) is defined by the minimization of the function 

\begin{equation}
M_{nm}({\bm \phi})=\frac{1}{(n-1)(m-1)}\sum_{i=2}^{n}\sum_{j=2}^{m}\rho\left(\frac{\varepsilon_{i,j}({\bm \phi})}{\hat{\sigma}}\right) 
\label{FuncObjAR}
\end{equation}
Because the M estimators are very sensitive when the process is contaminated with additive outliers is that other robust estimators arise to reduce the effects of additive outliers. Alternatively, \cite{All4} developed the generalized M estimators (GM) for spatial AR processes. A GM estimator of $\bm \phi$ is the solution to the problem of minimizing the non-quadratic function defined by:
\begin{equation*}
Q(\phi,\sigma)=\sum_{i,j}l_{ij}t_{ij}\left[\rho\left(\frac{Y_{i,j}-{\bm\phi}^T{Z}_{i, j}}{l_{ij}\sigma}\right)+\frac{1}{2}\right]\sigma,
\end{equation*}
where $\rho$ is as a function like the one that was defined in (\ref{FuncObjAR}), $t_{ij}$ and $l_{ij}$ are weights corresponding to the respective $Z_{i,j}.$

In \cite{Oje4}, the authors presented the robust autocovariance (RA) estimator for AR-2D processes which was first defined for time series models by \cite{Bus3}. This estimator is determined by the following equations

\begin{equation*}
\label{RA}
\sum_{k,l,r=0}^\infty  p_{ \bm \phi}(k,l,r)  f(k,l,r) = 0 
\end{equation*}
\begin{equation*}
\label{RA2}
\sum_{(m,n) \in (W_M \setminus S\langle 0,(1,1)])}\psi\left(\frac{r(m,n)}{\hat{\sigma}}\right)=0, 
\end{equation*}

\noindent where $f$ is a function that depends of the residues, $p_{\bm \phi}$ are coefficients that depend on the parameters  and $\hat{\sigma}$ is a independent estimate

\section{A new approach}
\label{sec:BMM}
\subsection{BIP-AR 2D models}
A new class of bounded nonlinear AR-2D models is presented in this work: the bounded innovation propagation AR-2D model (BIP-AR 2D). This model arises from the need to estimate the best possible parameters of an autoregressive central model when a contaminated process is observed. The BIP-AR 2D model is a two-dimension generalization of the model presented for time series by \cite{Mull}.\\

Given a stationary and invertible AR-2D model like in (\ref{modeloAR}), it supports a stationary representation as in (\ref{MAinf2}). We define the BIP-AR 2D auxiliary model as:
\begin{equation}
\label{modeloBIPgen}
	Y_{i,j}=\sum_{(k,l,r)\in D}\lambda_{klr}\sigma\eta \left( \frac{\varepsilon_{i-k-r,j-l-r}}{\sigma}\right) +\varepsilon_{i,j}	
\end{equation}
with $D=\{(k,l,r)\in \mathbb{N}^3_0\}\setminus \{(0,0,0)\}$, $\varepsilon_{i,j}$'s are i.i.d. random variables with symmetric distribution and $\eta(x)$ is an odd and bounded function. Besides, $\sigma$ is a robust M-scale of $\varepsilon_{i,j}$, which coincides with the standard deviation if $\varepsilon_{i,j}$  are normal, and is defined as the solution of the equation $E(\rho(\varepsilon_{i,j}/\sigma))=b$.\\

\noindent Note that (\ref{modeloBIPgen}) can also be written as
%\begin{equation}
\begin{align*}
	Y_{i,j} &= \sum_{\{k, l, r\geq 0\}}\lambda_{klr}\sigma\eta \left( \frac{\varepsilon_{i-k-r,j-l-r}}{\sigma} \right) -\sigma\eta \left( \frac{\varepsilon_{i,j}}{\sigma} \right) +\varepsilon_{i,j}\notag \\
	&= \sigma\Phi(B_1,B_2)^{-1}\eta \left( \frac{\varepsilon_{i,j}}{\sigma} \right) -\sigma\eta \left( \frac{\varepsilon_{i,j}}{\sigma} \right) +\varepsilon_{i,j}
\end{align*}	
%\end{equation}
and multiplying both members by $\Phi(B_1,B_2)$, we get
\begin{equation*}
\Phi(B_1,B_2)Y_{i,j} = \sigma\eta \left( \frac{\varepsilon_{i,j}}{\sigma} \right) -\sigma\Phi(B_1,B_2)\eta \left( \frac{\varepsilon_{i,j}}{\sigma} \right) +\Phi(B_1,B_2)\varepsilon_{i,j}
\end{equation*}
\noindent which is equivalent to
\begin{align*}
Y_{i,j} &=\phi_{1}Y_{i-1,j}+\phi_{2}Y_{i,j-1}+\phi_{3}Y_{i-1,j-1}+ \sigma\phi_{1}\eta \left( \frac{\varepsilon_{i-1,j}}{\sigma} \right) \nonumber\\
& \quad+\sigma\phi_{2}\eta \left( \frac{\varepsilon_{i,j-1}}{\sigma} \right) +\sigma\phi_{3}\eta \left( \frac{\varepsilon_{i-1,j-1}}{\sigma} \right)+\varepsilon_{i,j}\nonumber\\
& \quad -\phi_{1}\varepsilon_{i-1,j}-\phi_{2}\varepsilon_{i,j-1}-\phi_{3}\varepsilon_{i-1,j-1}
\end{align*}

\subsection{BMM estimator for AR-2D processes}

In time series, \cite{Mull} introduced the MM- estimators for ARMA models based in the definition of MM-estimate for regression where the residuals are calculated as in the BIP-ARMA model instead of just as in the pure ARMA model. The idea of MM-estimators in regression is to compute a highly robust estimator of the error scale in a first stage, and this estimated scale is used to calculate an M-estimator of the regression parameters in a second stage. However, in time series this differs somewhat because an MM-estimate is not enough to guarantee robustness.\\
In the same way that residues of AR-2D model exist (\ref{resAR}), there are residues obtained from BIP-AR 2D model:
\begin{align}
\label{resBIP}
\varepsilon_{i,j}^b({\bm \phi},\sigma) & = Y_{i,j}-\phi_{1}Y_{i-1,j}-\phi_{2}Y_{i,j-1}-\phi_{3}Y_{i-1,j-1}\nonumber\\ 
						 & \quad - \sigma \phi_{1}\eta \left( \frac{\varepsilon_{i-1,j}^b({\bm \phi},\sigma)}{\sigma} \right) - \sigma \phi_{2}\eta \left( \frac{\varepsilon_{i,j-1}^b({\bm \phi},\sigma)}{\sigma} \right) \nonumber\\
						& \quad - \sigma \phi_{3}\eta \left( \frac{\varepsilon_{i-1,j-1}^b({\bm \phi},\sigma)}{\sigma}\right) + \phi_{1}\varepsilon_{i-1,j}^b({\bm \phi},\sigma) \nonumber\\
						& \quad +\phi_{2}\varepsilon_{i,j-1}^b({\bm \phi},\sigma)+\phi_{3}\varepsilon_{i-1,j-1}^b({\bm \phi},\sigma)
\end{align}
for all $i,j\geq 2$. With this residue the objective function that must be minimized to obtain the M-estimator of the parameters under a model BIP-AR 2D is defined:

\begin{equation}
M_{nm}^b({\bm \phi})=\frac{1}{(n-1)(m-1)}\sum_{i=2}^{n}\sum_{j=2}^{m}\rho\left(\frac{\varepsilon_{i,j}^b({\bm \phi},\hat{\sigma})}{\hat{\sigma}}\right) 
\label{FuncObjBIP}
\end{equation}

\noindent where $\hat{\sigma}$ is a robust estimate of $\sigma$.\\
One way of robustly estimating the scale was introduced in 1964 by \cite{Hub} as follows: Given a sample $\bm{u}=(u_1,...,u_n)$, with $u_i\in \mathbb{R}$, an M-estimate of scale $S_n(\bm{u})$ is defined by any value $s\in(0,\infty)$ satisfying 

\begin{equation}
\frac{1}{n}\sum_{i=1}^n \rho\left( \frac{u_i}{s}\right)=b 
\label{eqScale}
\end{equation}
where $\rho$ is a continuous and non- constant function, non-decreasing in $|x|$ and symmetric around zero as well. To make the M-scale estimate consistent with the standard deviation when the data are normal, it requires that $E(\rho(x))=b$ under the standard normal distribution. Taking $b=\max(\rho)/2$, we get a maximum breakdown point of 0.5. With all this, we can define the new BMM-2D estimator by following the two steps given below:\\

\vspace{0.5cm}

\textbf{\textsl{First Step:}}
At this stage, an estimate of $\sigma$ is obtained. For this purpose, two $\sigma$ estimates are considered: one using an AR-2D model, another using a BIP-AR 2D model; then we choose the smallest of them.\\
Let $\rho_1$ a continue, non-constant, non-decreasing in $|x|$, bounded and symmetric function and such that:  $b=E(\rho_1(u)) \Rightarrow b/max(\rho_1)=0.5$. This guarantees that for a normal random sample, the M-scale estimator $s$ based on $\rho_1$ converges to the standard deviation and the breakdown point of $s$ is 0.5. Put
\begin{equation*}
\mathcal{B}=\{\bm{\phi}\in\mathbb{R}^3 : \ \text{if} \ (z_1,z_2)\in\mathbb{C}^2 \ \text{is such that} \ \bm{\Phi}(z_1,z_2)=0 \ \text{then} \ |z_1|\geq 1+\zeta \wedge |z_2|\geq 1+\zeta \}
\end{equation*}
\noindent for some $\zeta>0$. Then, we defined an estimate of ${\bm \phi}\in\mathcal{B}$:
\begin{equation*}
\hat{{\bm \phi}}_S=arg\min_{{\bm \phi}\in\mathcal{B}}S_{nm}({\boldsymbol\varepsilon}_{nm}({\bm \phi}))
\end{equation*}
and the corresponding estimate of $\sigma$:
\begin{equation}
s_{nm}=S_{nm}({\boldsymbol\varepsilon}_{nm}(\hat{{\bm \phi}}_S))
\label{eqDesvio}
\end{equation}
\noindent where ${\boldsymbol\varepsilon}_{nm}({\bm \phi})=(\varepsilon_{22}({\bm \phi}),...,\varepsilon_{n2}({\bm \phi}),...,\varepsilon_{2m}({\bm \phi}),...,\varepsilon_{nm}({\bm \phi}))$,\\
with $\varepsilon_{ij}({\bm \phi})$ as in (\ref{resAR}) and $S_{nm}$ is the M-estimate of scale based on $\rho_1$ and $b$ defined as in (\ref{eqScale}). Later, we described the estimate corresponding to the BIP-AR model. Define $\hat{{\bm \phi}}_S^b$ by the minimization of $S_{nm}({\boldsymbol\varepsilon}_{nm}^b({\bm \phi},\hat{\sigma}({\bm \phi})))$ over $\mathcal{B}$.
The value $\hat{\sigma}({\bm \phi})$ is an estimate of $\sigma$ computed as if ${\bm \phi}$ were the true parameters and the $\varepsilon_{i,j}$'s were normal. Then, from (\ref{modeloBIPgen}), because in this case the M-scale $\sigma$ coincides with the
standard deviation of $\varepsilon_{i,j}$, we had:
\begin{equation*}
\sigma^2=\frac{\sigma^2_Y}{1+\kappa^2\sum_{k,l,r\geq0}\lambda_{klr}^2}
\end{equation*}
 where $\kappa^2=Var(\eta(\frac{\varepsilon_{i,j}}{\sigma}))$ and $\sigma^2_Y=Var(Y_{i,j})$. Let $\hat{\sigma}^2_Y$ a robust estimate of $\sigma^2_Y$ and $\kappa^2=Var(\eta(Z))$ where $Z\sim N(0,1)$. Then, we defined
\begin{equation*}
\hat{\sigma}^2(\bm \phi)=\frac{\hat{\sigma}^2_Y}{1+\kappa^2\sum_{k,l,r\geq0}\lambda_{klr}^2({\bm \phi})}
\end{equation*}
\noindent The scale estimate $s_{nm}^b$ corresponding to the BIP-AR-2D model is defined by
\begin{equation*}
\hat{{\bm \phi}}_S^b=arg\min_{{\bm \phi}\in\mathcal{B}}S_{nm}({\boldsymbol\varepsilon}_{nm}^b({\bm \phi},\hat{\sigma}({\bm \phi})))
\end{equation*}
and
\begin{equation*} 
s_{nm}^b=S_{nm}({\boldsymbol\varepsilon}_{nm}^b(\hat{{\bm \phi}}_S^b,\hat{\sigma}(\hat{{\bm \phi}}_S^b)))
\end{equation*}
\noindent where, for simplify, we denoted $\tilde{\sigma}=\hat{\sigma} (\bm \phi),$ and

\begin{equation*}
{\boldsymbol\varepsilon}_{nm}^b({\bm \phi},\tilde{\sigma})=(\varepsilon_{22}^b({\bm \phi},\tilde{\sigma}),.,\varepsilon_{n2}^b({\bm \phi},\tilde{\sigma}),.,\varepsilon_{2m}^b({\bm \phi},\tilde{\sigma}),.,\varepsilon_{nm}^b({\bm \phi},\tilde{\sigma}))
\end{equation*}
\noindent with $\varepsilon_{ij}^b({\bm \phi},\tilde{\sigma})$ defined as in (\ref{resBIP}). Finally, our $\sigma$ estimate was
\begin{equation*}
s_{nm}^*=\min(s_{nm},s_{nm}^b)
\end{equation*}

\vspace{0.5cm}
\textbf{\textsl{Second Step:}}
We considered a bounded function $\rho_2$ that satisfies the same properties as $\rho_1$ but also $\rho_2\leq\rho_1$. This function was chosen such that the corresponding M-estimator is highly efficient under normal innovations. Given the objective functions defined in (\ref{FuncObjAR}) and (\ref{FuncObjBIP}) with the scale obtained in the first step ($s_{nm}^*$):

\begin{equation*}
\label{funcObjBMM1}
M_{nm}({\bm \phi})=\frac{1}{(n-1)(m-1)}\sum_{i=2}^{n}\sum_{j=2}^{m}\rho_2 \left( \frac{\varepsilon_{i,j}({\bm \phi})}{s_{nm}^*} \right)
\end{equation*}
and 
\begin{equation*}
\label{funcObjBMM2}
M_{nm}^b({\bm \phi})=\frac{1}{(n-1)(m-1)}\sum_{i=2}^{n}\sum_{j=2}^{m}\rho_2 \left(\frac{\varepsilon_{i,j}^b({\bm \phi},s_{nm}^*) }{s_{nm}^*} \right)
\end{equation*}

\noindent The corresponding M-estimators of the parameters for each function are:
\begin{equation*}
\hat{{\bm \phi}}_M=arg\min_{{\bm \phi}\in\mathcal{B}}M_{nm}({\bm \phi})$$ and $$\hat{{\bm \phi}}_M^b=arg\min_{{\bm \phi}\in\mathcal{B}}M_{nm}^b({\bm \phi})
\end{equation*}

\noindent Then, we defined the BMM-estimator 2D as:
\[\hat{{\bm \phi}}^*_M=
       \begin{cases}
       \hat{{\bm \phi}}_M, & \text{if } M_{nm}(\hat{{\bm \phi}}_M)\le M^b_{nm}(\hat{{\bm \phi}}^b_M)\\
       \hat{{\bm \phi}}^b_M, & \text{if } M_{nm}(\hat{{\bm \phi}}_M)> M^b_{nm}(\hat{{\bm \phi}}^b_M)\\
       \end{cases}\]

%%%%%%%%%%%%%%%%%%%%%%%%%%%%%%%%%%%%%%%%
\section{Monte Carlo Results}
\label{sec:MC}
The aim of this section is to analyze the performance of the new BMM estimator to estimate the parameters in the model (\ref{modeloAR}) compared to the LS, M, GM and RA estimators. We performed several experiments. Each experiment is based on  different Monte Carlo studies. We set the parameter values of 
(\ref{modeloAR}) as:
\begin{equation}
\label{modeloARE}
Y_{i,j}=0.15 Y_{i-1,j}+ 0.17Y_{i,j-1}+0.2Y_{i-1,j-1}+\varepsilon_{i,j}
\end{equation}
It is important to mention that the set parameters were chosen at random satisfying the conditions of invertibility of \cite{Bas}.
  We performed our study under five different conditions of the model (Cases); in Case I, the model was non-contaminated, while in Cases II, III, IV and V, the model was contaminated according to (\ref{eqContam}): 

\begin{itemize}
\item Case I) Non-contaminated model like in (\ref{modeloARE}), where $\varepsilon$  is a normal distribution process with $Var(\varepsilon_{i,j})=1$ and $E(\varepsilon_{i,j})=0$ for all $i,j$.
\item Case II) AO, where the $\nu$ process is independent of the $Y$ process and follows a normal distribution with zero mean and variance 50. 
\item Case III) SRO, where the replacement process $W$  follows a t-student distribution with 2.3 f.d.
\item Case IV) SRO, where the replacement process $W$ is another autoregressive process, independent of the $Y$process, with parameters $\tilde{\phi_1}=0.1$, $\tilde{\phi_2}=0.2$ and $\tilde{\phi_3}=0.3$.
\item Case V) SRO, where the replacement process $W$ is a white noise process with normal distribution with zero mean and variance 50.
\end{itemize}

In each of the five variants of the model (\ref{modeloARE}), the parameters $\phi_1, \phi_2$ and $\phi_3$ were estimated by the five procedures presented in the previous sections. In each experiment, 500 simulations of the model were generated, and the mean value, the mean square error (MSE) and the sample variance were computed. For the contaminated models we considered four  levels of contamination: 5\%, 10\%, 15\% and 20\%. Besides, we performed our study considering different window sizes: $8\times8$, $16\times16$, $32\times32$, and $57\times57$. For the calculation of the BMM estimator, a robust estimator of the scale was obtained as in  (\ref{eqDesvio}). $\rho_1(x)=\rho_2(\frac{x}{0.405})$ was selected according to the same criteria that was taken for the definition of the BMM estimator for time series present in \cite{Mull}, where the function $\rho_2$ is given by:
\vspace{0.5cm}
\begin{equation*}
\label{funcRho2}
\rho_2(x)=
\begin{cases}
   0.5x^2, & \text{if } |x|\leq 2;\\
   0.002x^8-0.052x^6+0.432x^4-0.972x^2+1.792, & \text{if } 2<|x|\leq 3;\\
   3.25, & \text{if } 3<|x|
\end{cases}
\end{equation*}
\vspace{0.5cm}
The same $\rho_2$ function was used to calculate the M estimators. In addition, for the implementation of the GM estimator the weights were set:
\begin{equation*}
\vspace{0.5cm}
l_{i,j}=1 \ \ \forall \ i,j
\end{equation*}
\begin{equation*}
t_{i,j}=\frac{\psi_H((Y_{i-1,j}^2+Y_{i,j-1}^2+Y_{i-1,j-1}^2)/3}{(Y_{i-1,j}^2+Y_{i,j-1}^2+Y_{i-1,j-1}^2)/3}
\end{equation*}
\vspace{0.5cm}
where $\psi_H$ is the following version of the Huber function:
\begin{equation*}
\label{funcPsiH}
\psi_H(x)=
\begin{cases}
   x, & \text{if } |x|\leq 1.5;\\
   1.5, & \text{if } 1.5<x;\\
   -1.5, & \text{if } x<-1.5
\end{cases}
\end{equation*}
Finally the RA estimators were implemented according to the details formulated in \cite{Oje}.
To facilitate the paper reading, only the boxplots of the simulations have been included in the body of the work; the numerical Monte Carlo results are shown in the Appendix.

%%%%%%%%%%%%%%%%%%%%%%%%%%%%%%%%% 

\subsection{Experiments}

In a first experiment, we studied the performance of the BMM estimator for the non-contaminated model (Case I).  All the methods estimated the parameters quite well. Table \ref{tabSinCont} shows the results obtained for the four different window sizes considered. In Figure \ref{figBoxSinCont}, the corresponding boxplots are shown. In this case, it is convenient to use the LS method due to its simplicity and calculation speed.\\
The second experiment was developed in the context of Case II. We analyzed the ability of the BMM method to estimate the parameters of the model, considering a 10\% of additive contamination, and for window sizes $8\times 8$, $16\times 16$, $32\times 32$ and $57\times 57$, in comparison with the LS, M, GM and RA methods. Table \ref{tabContAdit} shows the estimated values for $\phi_1$, $\phi_2$ and $\phi_3$,  using the different five procedures analyzed. Figure \ref{figBoxContAdit} exhibits the corresponding boxplots. For window size $32\times 32$ and $57\times 57$, it can be seen that the BMM estimator is the best  because its values are closer to the parameter than the estimates produced by the other methods mentioned. In addition, BMM estimator has the lowest variance and the lowest MSE. When the window size was $8\times 8$ or $16\times 16$, the best performance corresponded to the GM and RA estimators; however, the BMM estimation values were similar to the RA and GM estimations. An analogue affirmation is valid to the sample variance and MSE of BMM. We also noted that for any window size, the M estimator had a very small sample variance but their estimations were wrong when compared to the ones in the other methods.\\
\clearpage
\begin{multicols}{2}
%\begin{figure}%[h!]
%\begin{center}
%\begin{tabular}[c]{p{8cm} p{10cm}}
%\begin{minipage}[c]{5cm}
%label va SIEMPRE dps del caption para que haga referencia al num de imagen o tabla, sino da la seccion en la que esta.
%\end{minipage} &
{\centering
  \begin{minipage}[c]{7.75cm}
  \includegraphics[width=7.75cm, height=5cm]{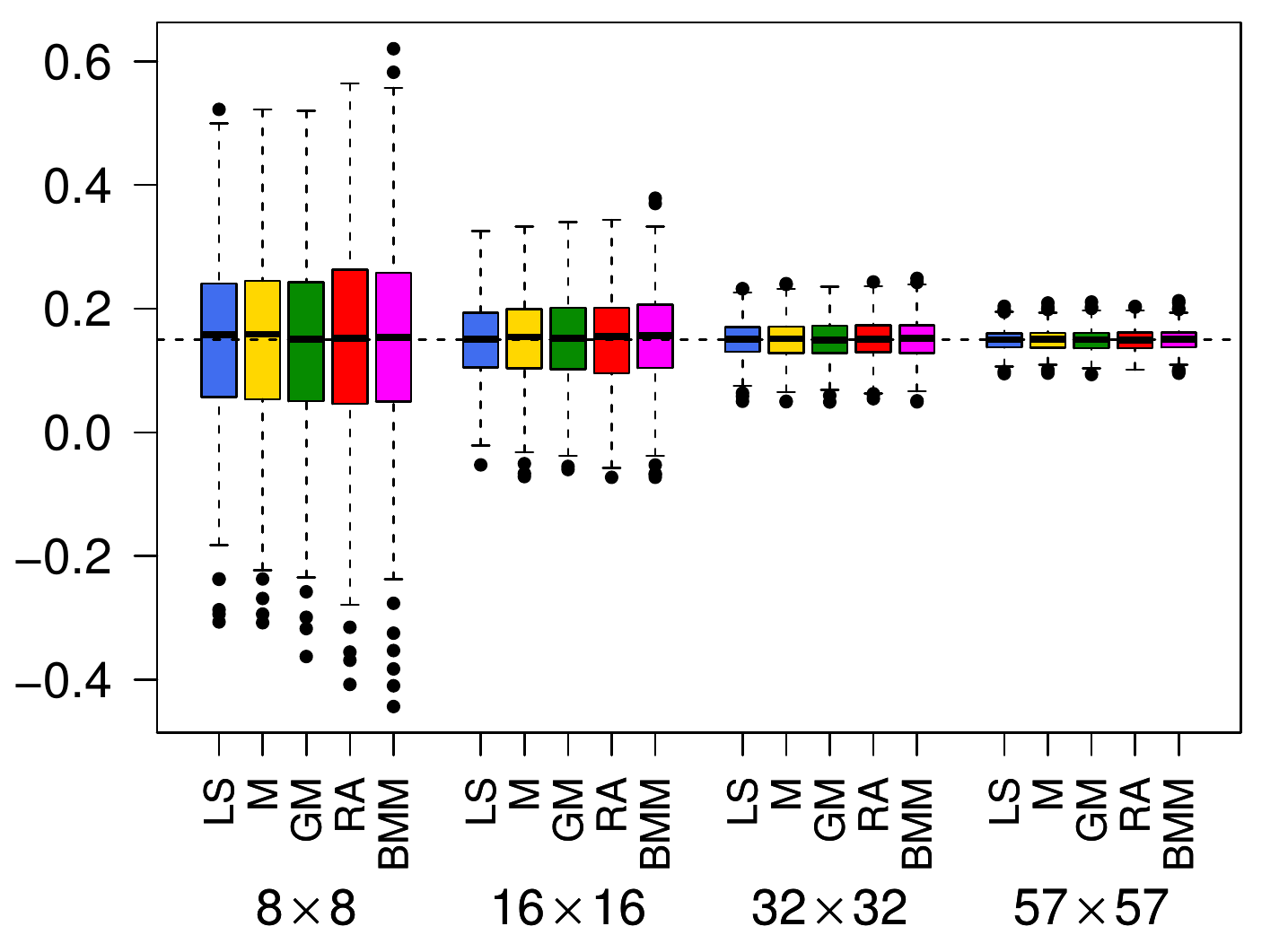}
  \centerline{(a)}
	\smallskip
  \end{minipage}\\
  \begin{minipage}[c]{7.75cm}
   \includegraphics[width=7.75cm, height=5cm]{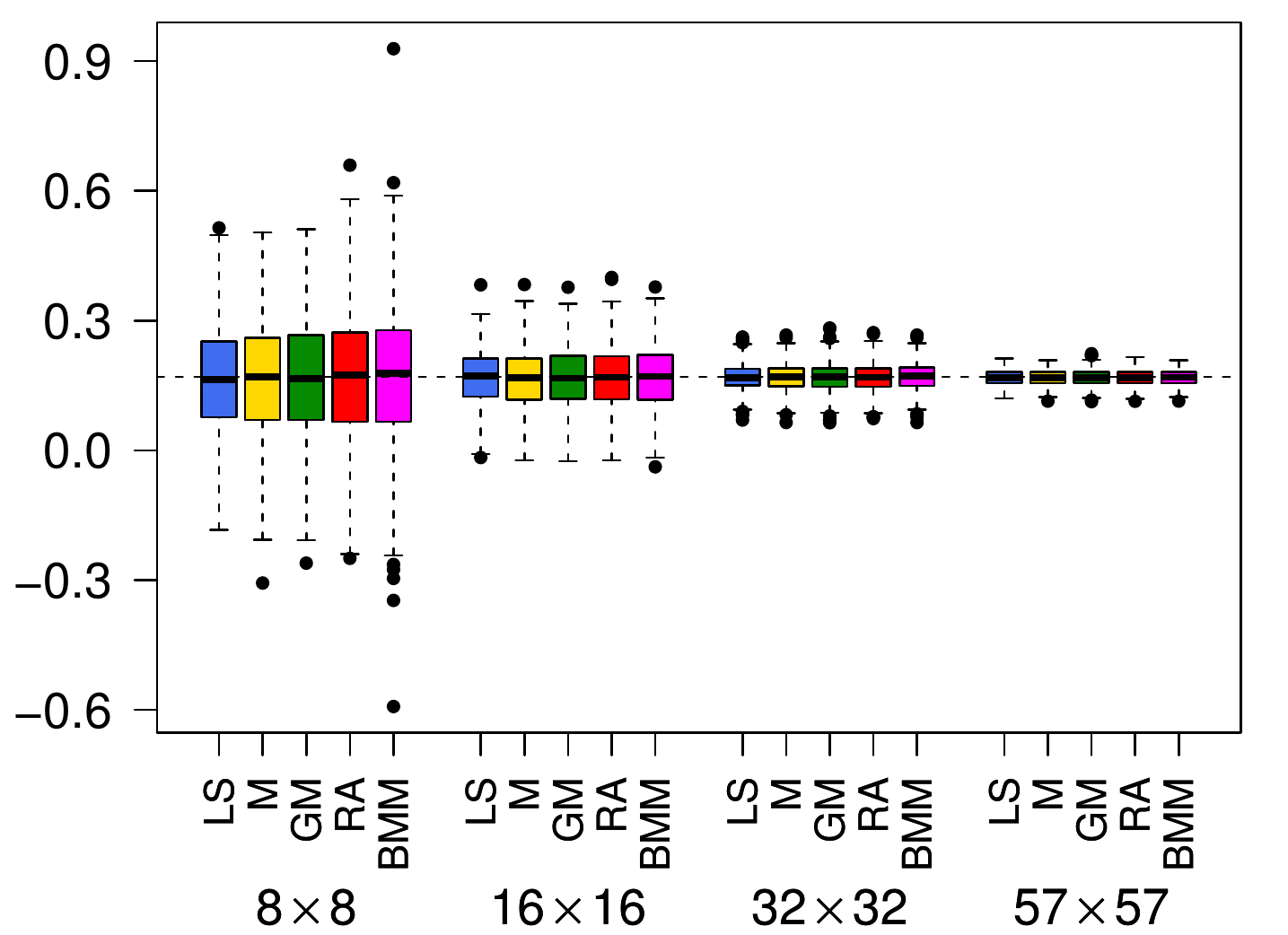}
   \centerline{(b)}
	\smallskip
   \end{minipage}\\
  \begin{minipage}[c]{7.75cm}
   \includegraphics[width=7.75cm, height=5cm]{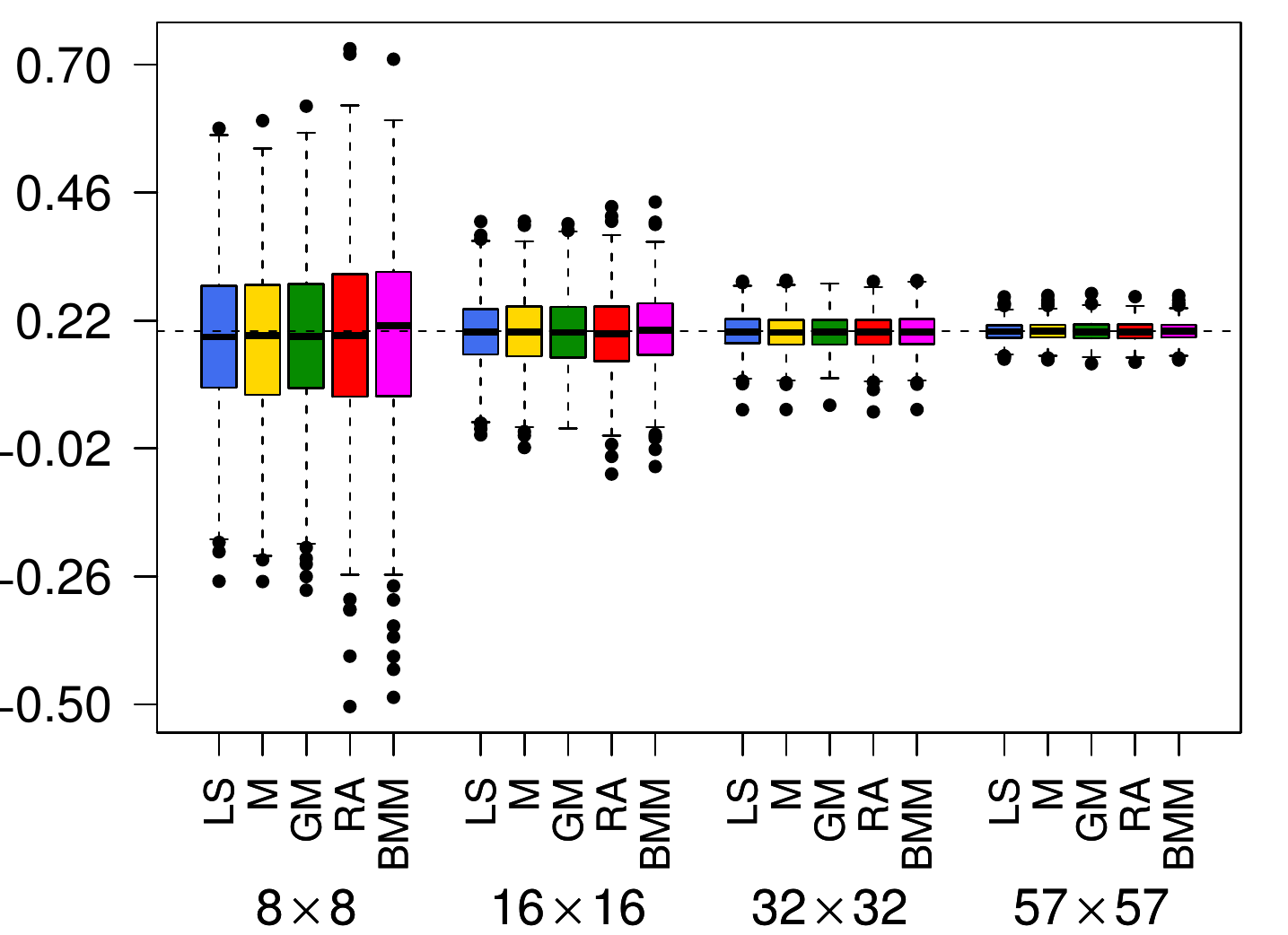}
   \centerline{(c)}
	\smallskip
   \end{minipage}\\
%\end{tabular}
\begin{minipage}[c]{7.75cm}
\captionof{figure}{Case (I): LS, M, GM, RA and BMM estimation boxplots for (a) $\phi_1=0.15$, (b) $\phi_2=0.17$ and (c) $\phi_3=0.2$; model (\ref{modeloARE}) without contamination, varying the window sizes.}\label{figBoxSinCont}
\end{minipage}
%\end{center}
%\end{figure}
}

%\vfill\null
%\columnbreak
%%%%%%%%%%%%%%%%%%%%%%%%%%%%
%\begin{figure}%[h!]
%\begin{center}
%\begin{tabular}[c]{p{5cm} p{13cm}}
%\begin{minipage}[c]{5cm}
%\end{minipage} &
{\centering
  \begin{minipage}[c]{7.75cm}
  \includegraphics[width=7.75cm, height=5cm]{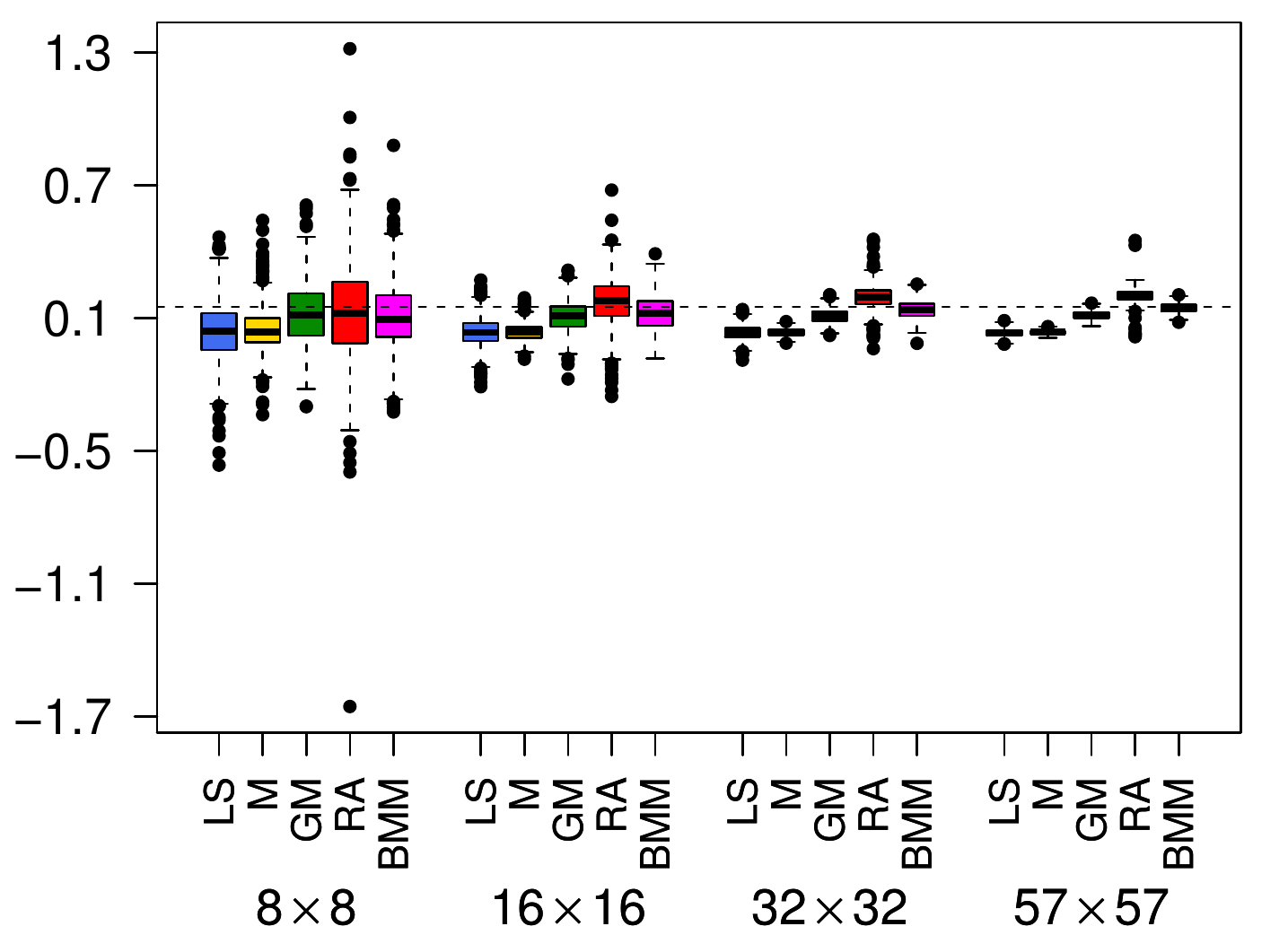}
  \centerline{(a)}
	\smallskip
  \end{minipage}\\
 \begin{minipage}[c]{7.75cm}
   \includegraphics[width=7.75cm, height=5cm]{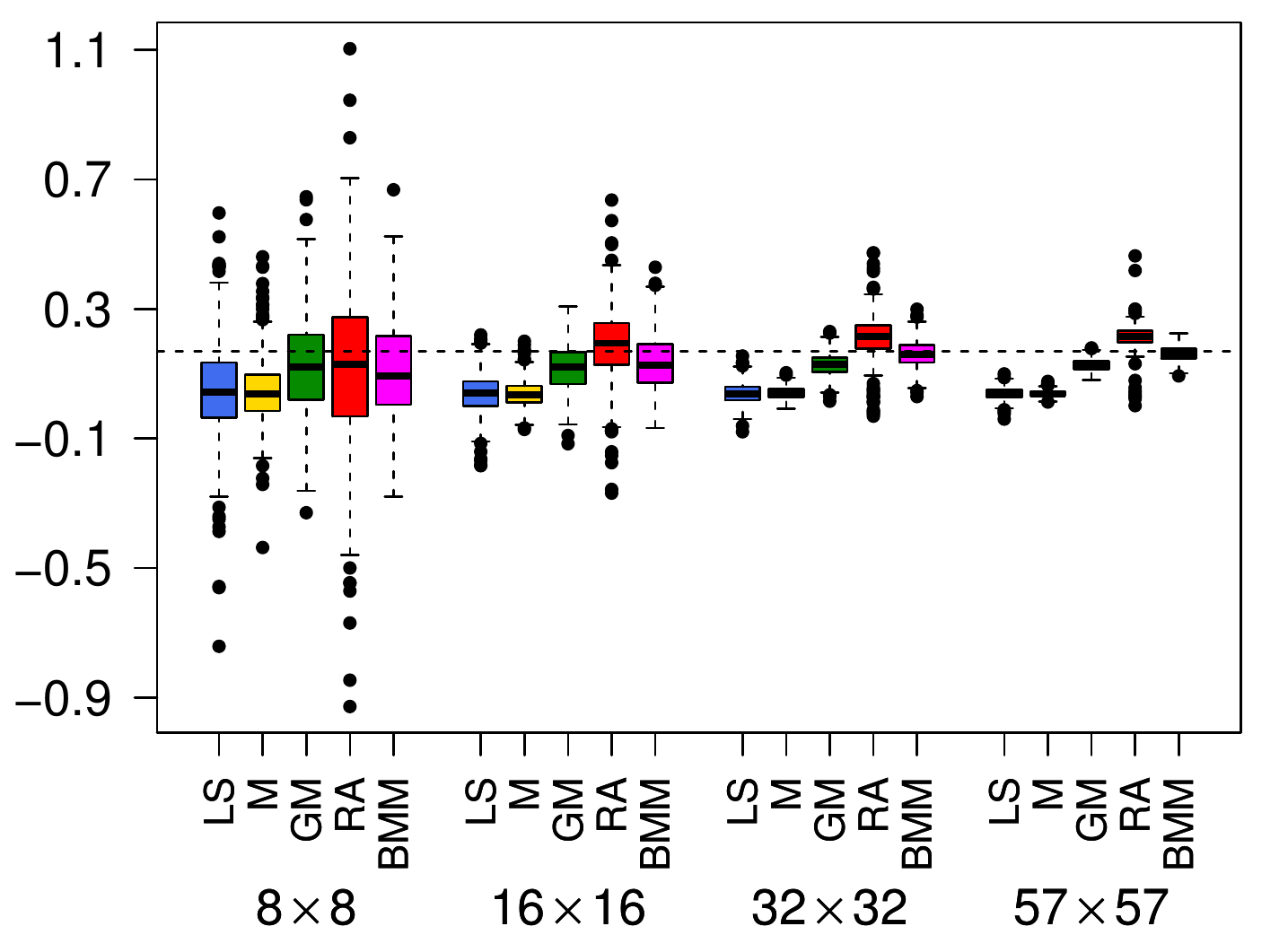}
   \centerline{(b)}
	\smallskip
   \end{minipage}\\
  \begin{minipage}[c]{7.75cm}
   \includegraphics[width=7.75cm, height=5cm]{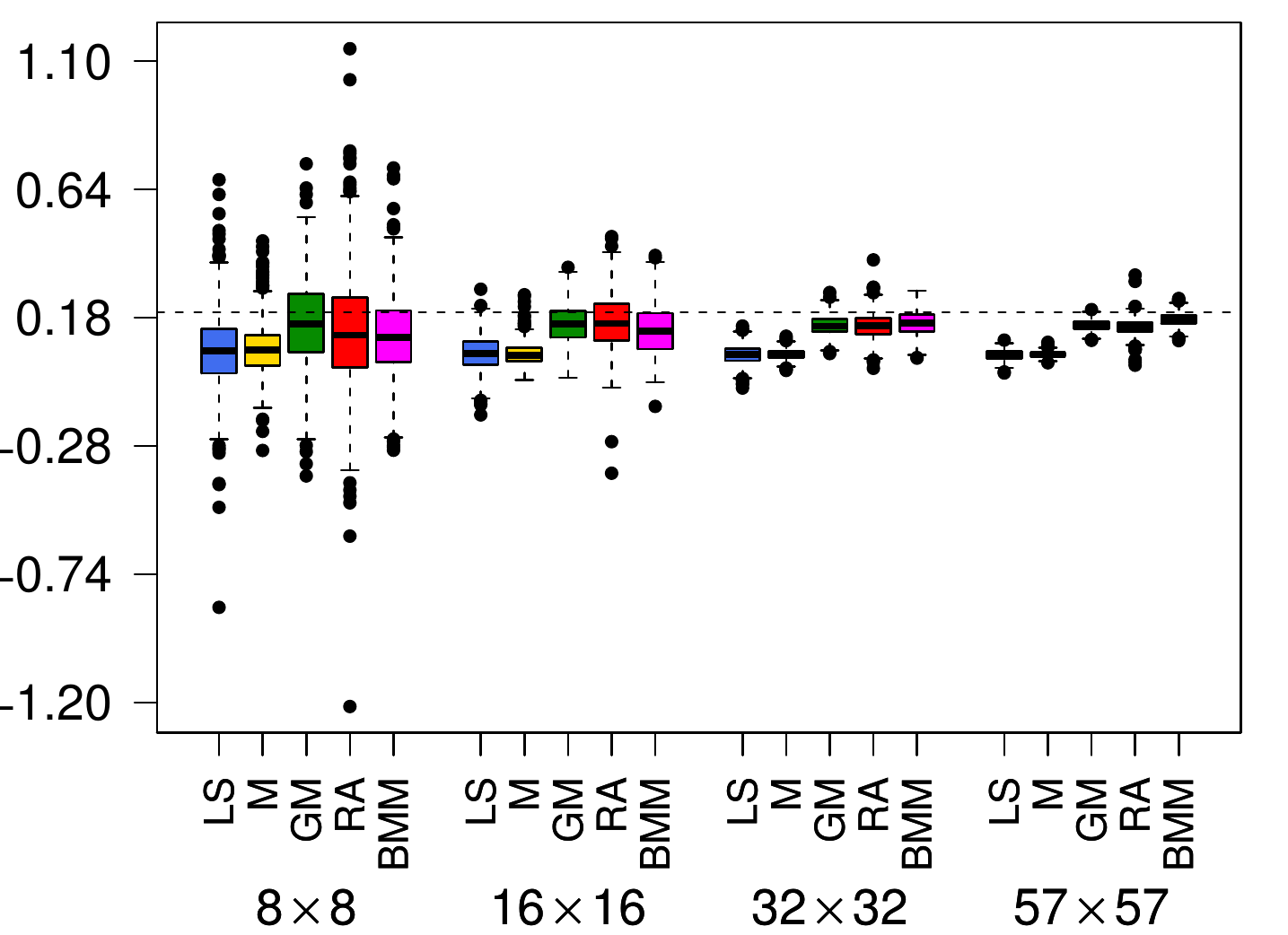}
   \centerline{(c)}
	\smallskip
   \end{minipage}\\
%\end{tabular}
\begin{minipage}[c]{7.75cm}
\captionof{figure}{Case (II): LS, M, GM, RA and BMM estimation boxplots for (a) $\phi_1=0.15$, (b) $\phi_2=0.17$ and (c) $\phi_3=0.2$, varying the window sizes. Model (\ref{modeloARE}) with additive contamination $10\%$ level, with a normal noise.}\label{figBoxContAdit} 
\end{minipage}
%\end{center}
%\end{figure}
}
\end{multicols}
\clearpage
\begin{multicols}{2}
%\begin{figure}%[h!]
%\begin{center}
%\begin{tabular}[c]{p{5cm} p{13cm}}
%\begin{minipage}[c]{5cm}
%\end{minipage} &
{\centering
  \begin{minipage}[c]{7.75cm}
  \includegraphics[width=7.75cm, height=5cm]{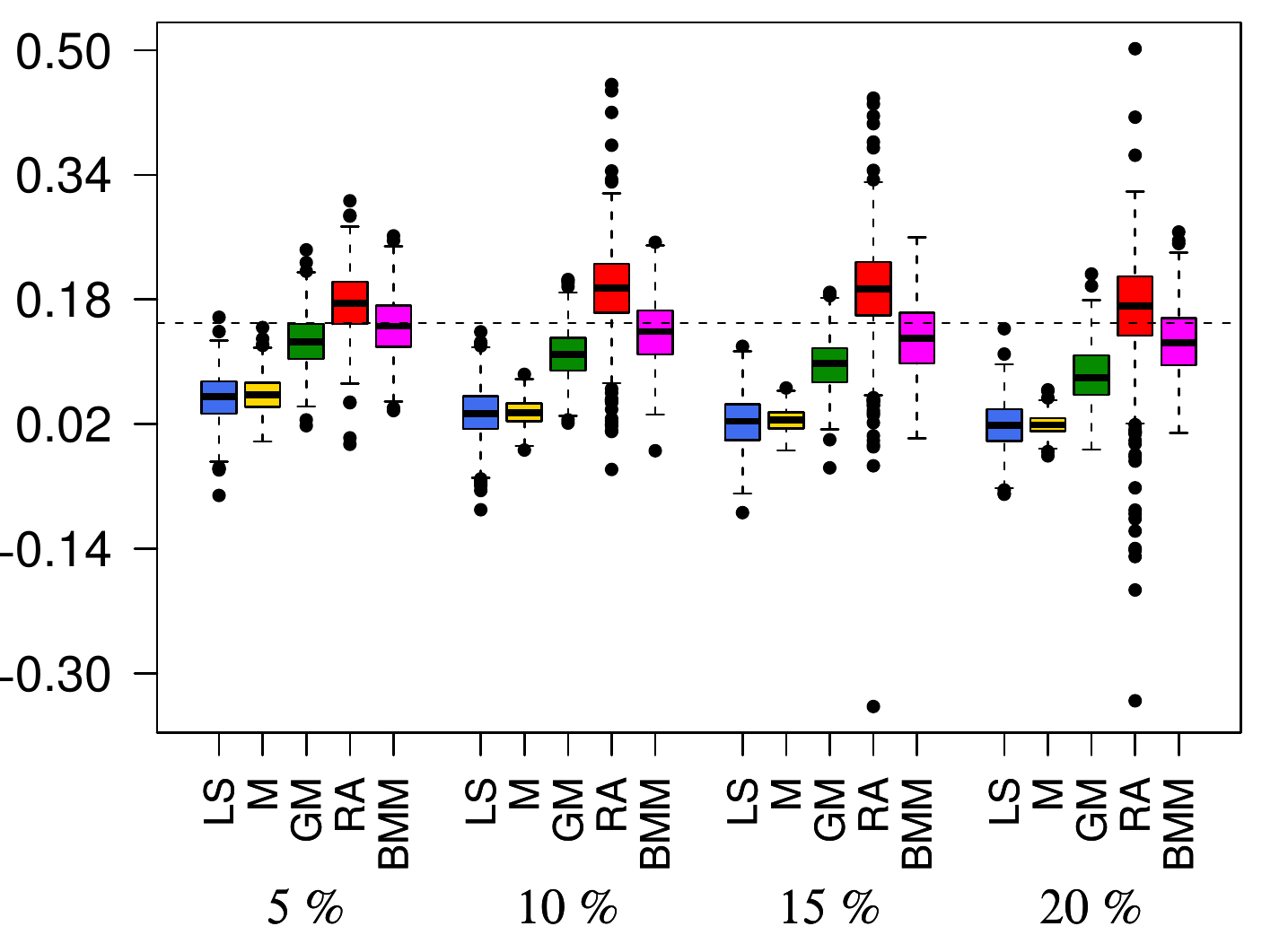}
  \centerline{(a)}
	\smallskip
  \end{minipage}\\
 \begin{minipage}[c]{7.75cm}
   \includegraphics[width=7.75cm, height=5cm]{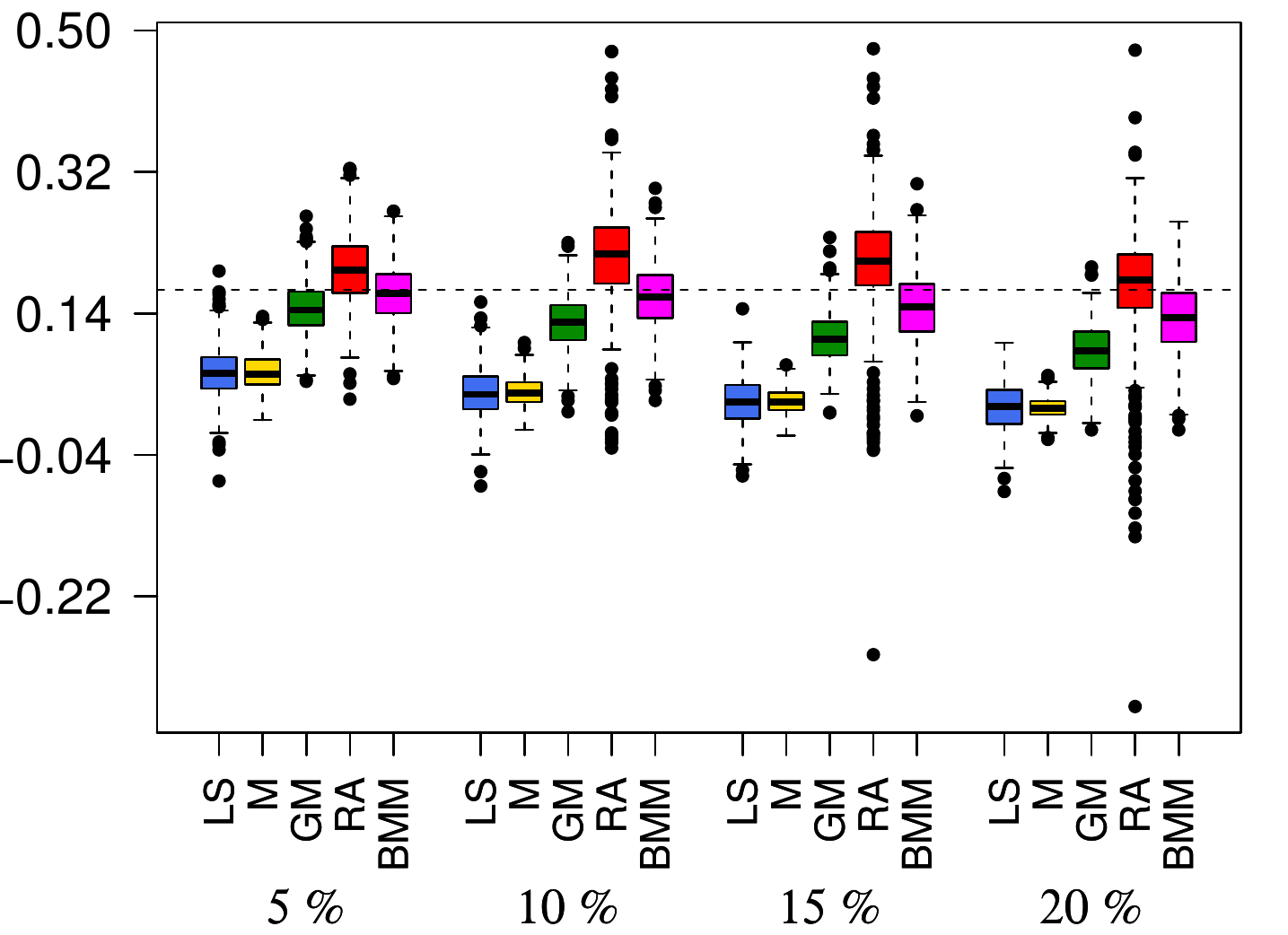}
   \centerline{(b)}
	\smallskip
   \end{minipage}\\
 \begin{minipage}[c]{7.75cm}
   \includegraphics[width=7.75cm, height=5cm]{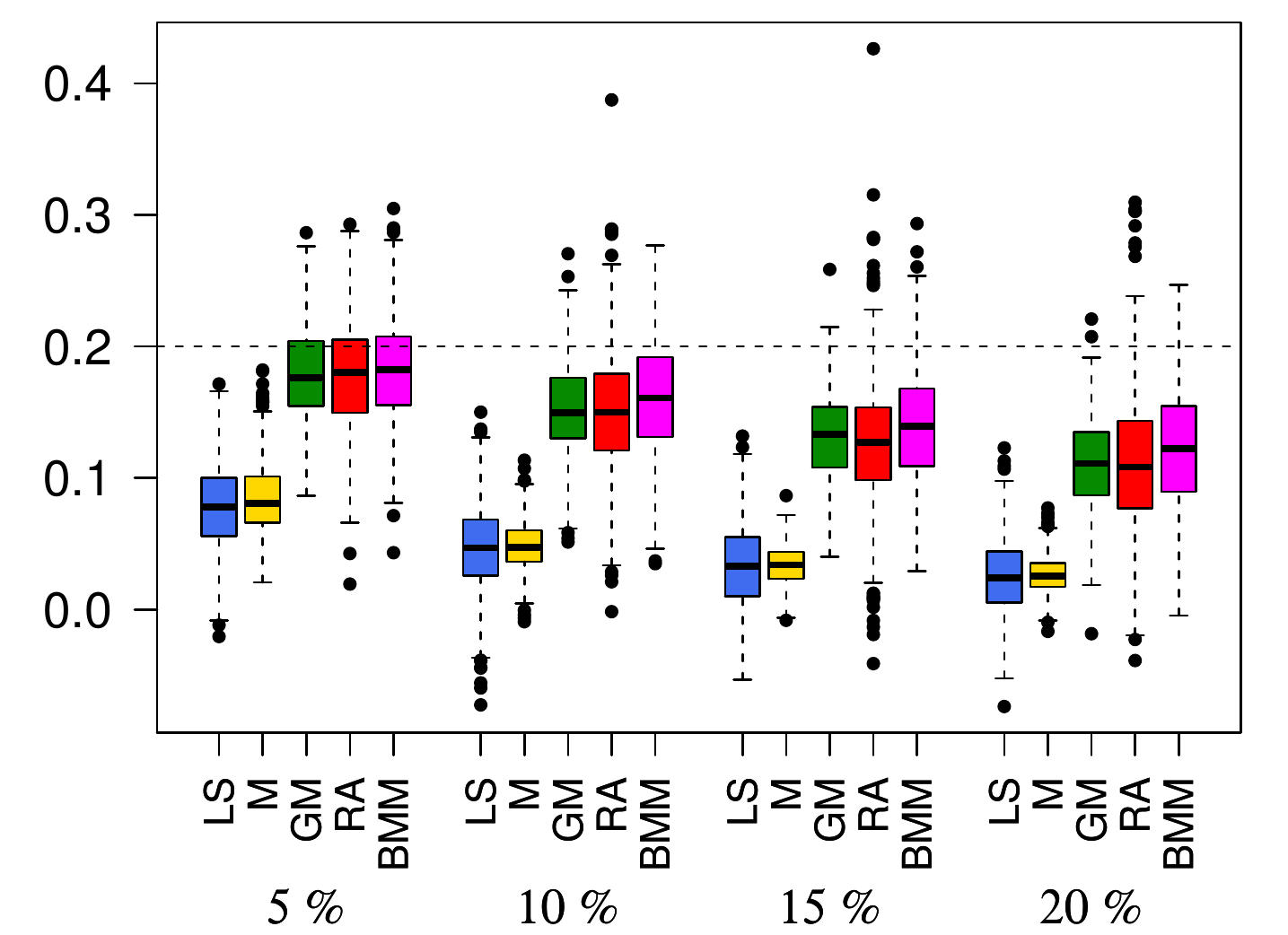}
   \centerline{(c)}
	\smallskip
   \end{minipage}\\
%   \end{tabular}
\begin{minipage}[c]{7.75cm}
\captionof{figure}{Case (II): LS, M, GM, RA and BMM estimation boxplots, for (a) $\phi_1=0.15$, (b) $\phi_2=0.17$ and (c) $\phi_3=0.2$ in model \ref{modeloARE}, with additive contamination, varying the contamination level, with a $32\times32$ window size. The contamination process is a normal noise, with $\sigma^2=50$. }\label{figBoxContAditPorc}
\end{minipage}
}
%\end{center}
%\end{figure}

%%%%%%%%%%%%%%%%%%%%%%%%%%%%%
%\begin{figure}%[h!]
%\begin{center}
%\begin{tabular}[c]{p{5cm} p{13cm}}
%\begin{minipage}[c]{5cm}
%\end{minipage} &
{\centering
  \begin{minipage}[c]{7.75cm}
  \includegraphics[width=7.75cm, height=5cm]{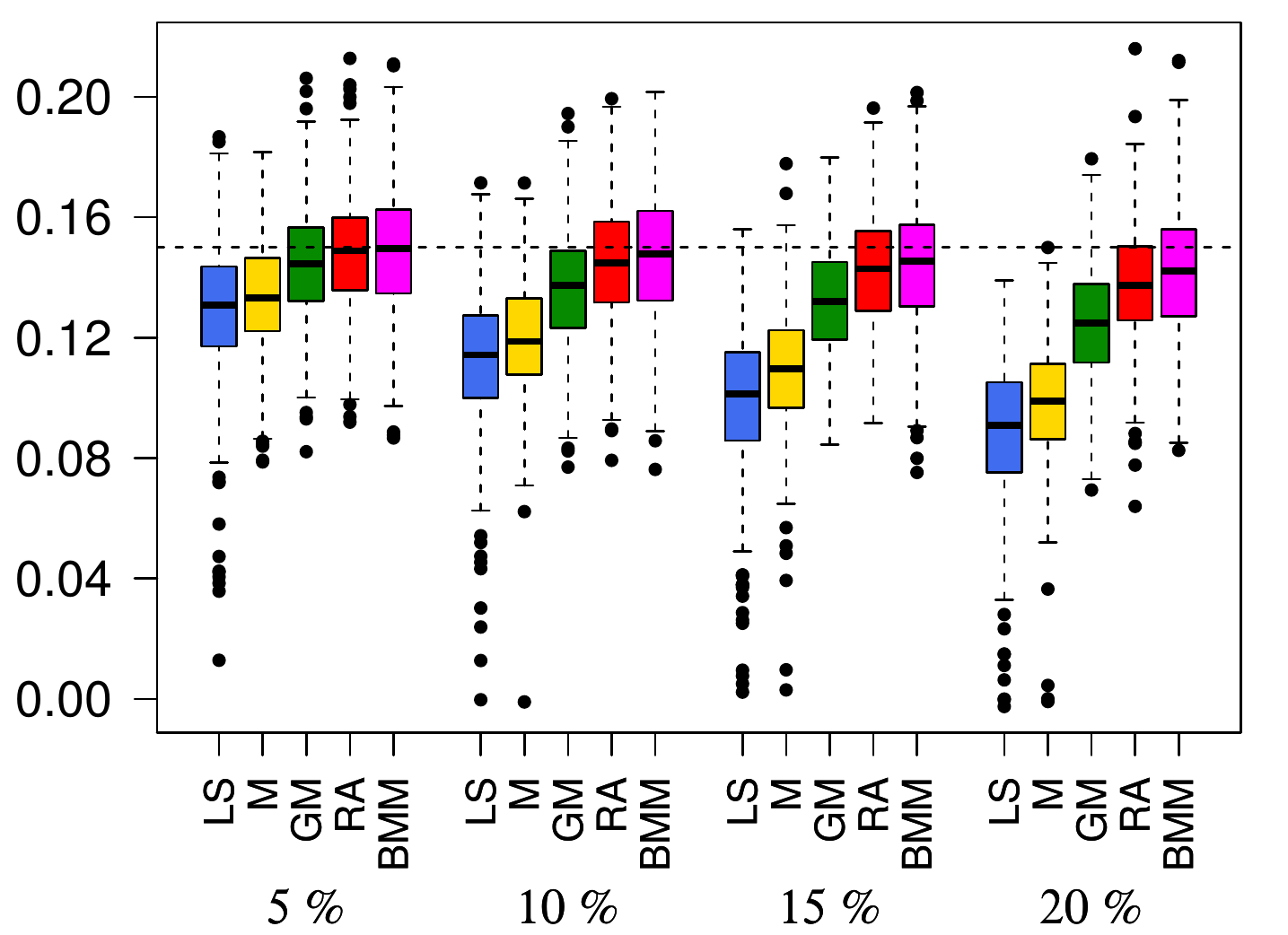}
  \centerline{(a)}
	\smallskip
  \end{minipage}\\
  \begin{minipage}[c]{7.75cm}
   \includegraphics[width=7.75cm, height=5cm]{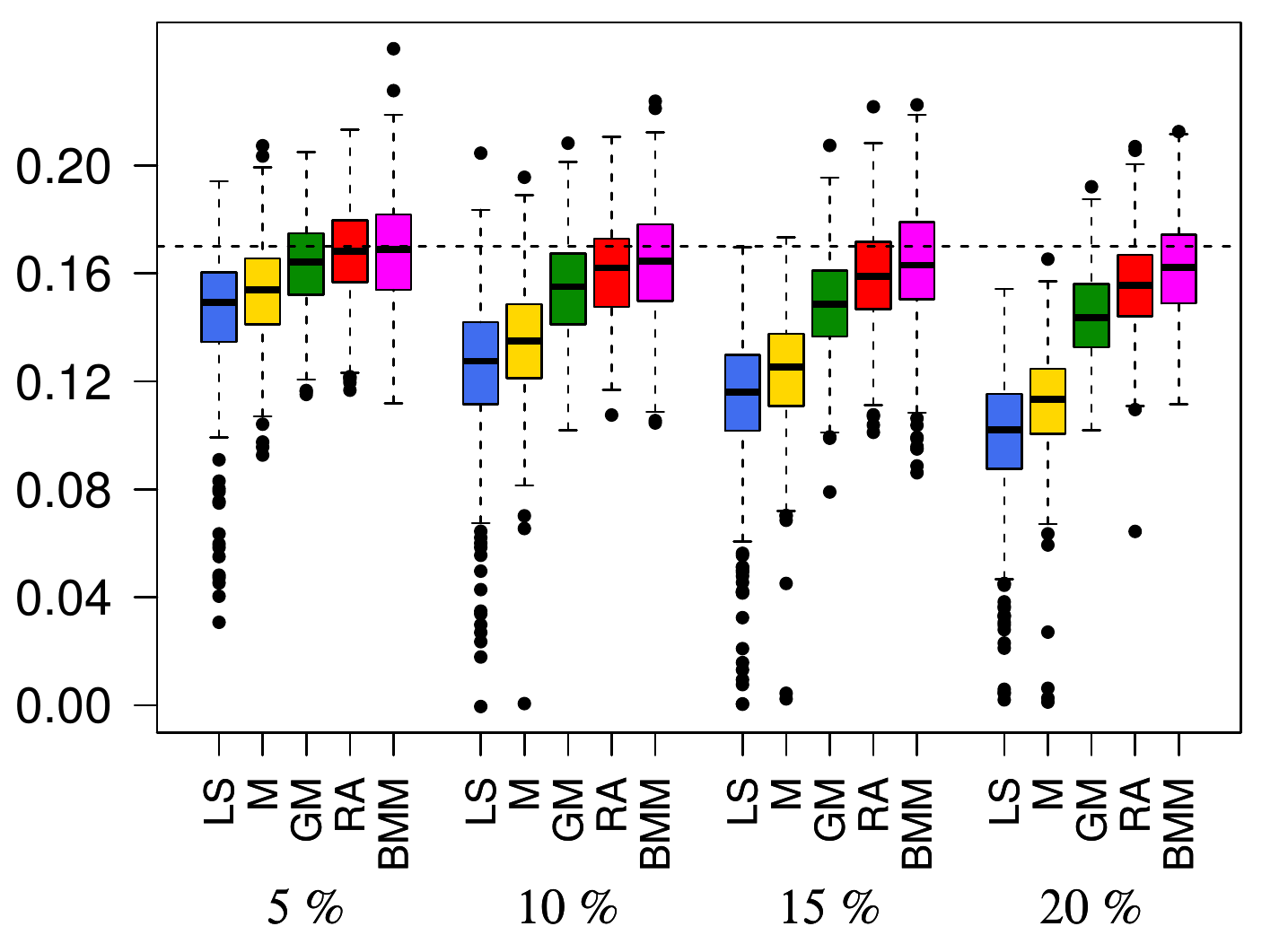}
   \centerline{(b)}
	\smallskip
   \end{minipage}\\
  \begin{minipage}[c]{7.75cm}
   \includegraphics[width=7.75cm, height=5cm]{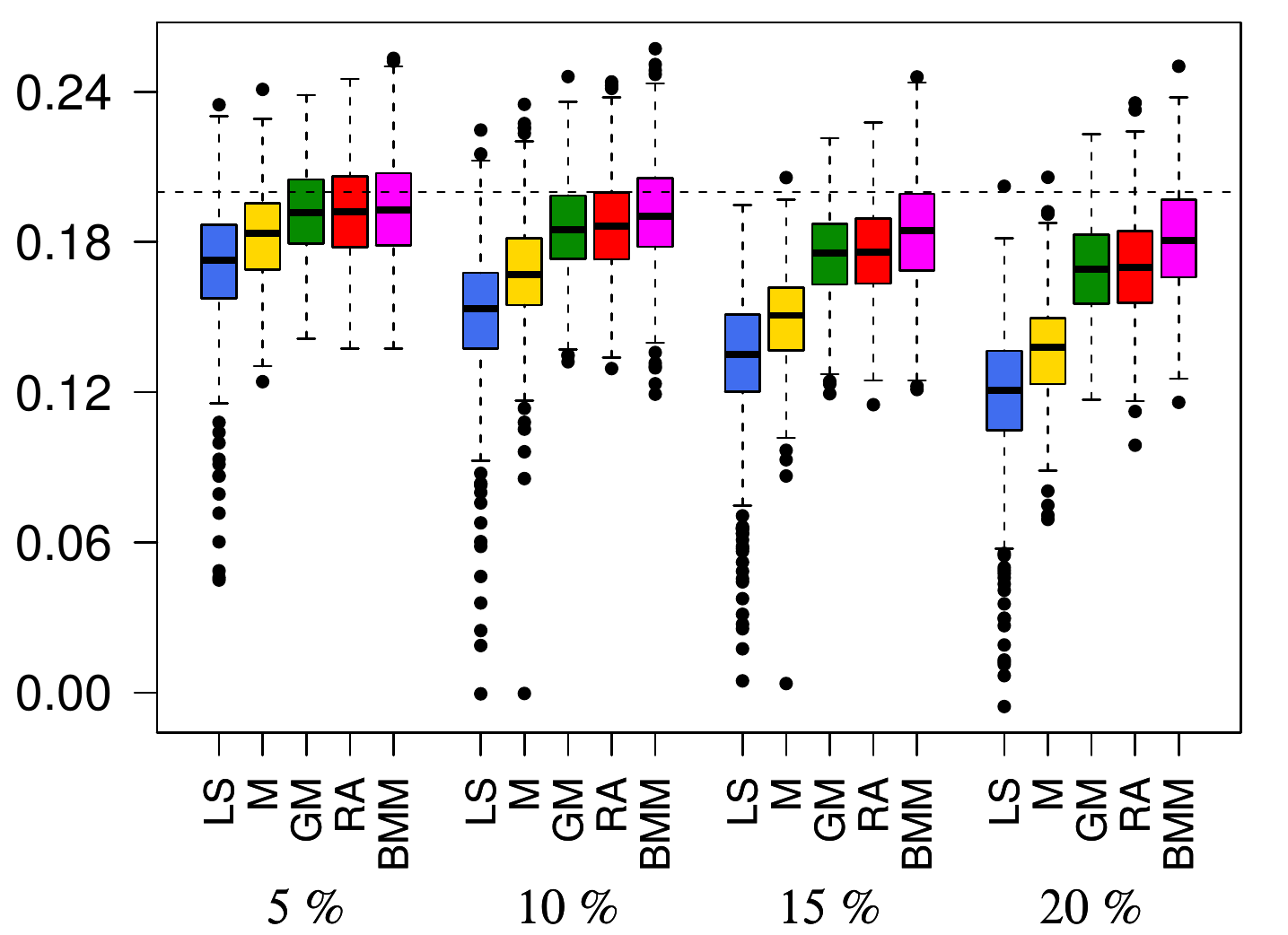}
   \centerline{(c)}
	\smallskip
   \end{minipage}\\
%\end{tabular}
\begin{minipage}[c]{7.75cm}
\captionof{figure}{Case (III): LS, M, GM, RA and BMM estimation boxplots, for (a) $\phi_1=0.15$, (b) $\phi_2=0.17$ and (c) $\phi_3=0.2$ in model \ref{modeloARE}, with contamination of replacement, varying the contamination level, with a $57\times 57$ window size. The process of contamination follows a t-Student distribution with 2.3 d.f.}\label{figBoxContStudPorc}
\end{minipage}
%\end{center}
%\end{figure}
}
\end{multicols}
\clearpage
The third experiment also refers to Case II. We set the window size at $32\times32$ and varied the additive contamination level, considering four levels: $5\%$, $10\%$, $15\%$ and $20\%$.
The BMM method was the best in most of the cases studied, followed by the RA estimator. This behavior is deduced from the comparison of the values estimated by the BMM method with the respective estimates obtained by the other procedures. The values of the dispersion measures also point out that the BMM estimator is the most accurate methodology. These results can be seen in Table \ref{tabContAditPorc}. In addition, from Figure \ref{figBoxContAditPorc} we can note that using any of the five estimators, the parameter $\phi_3$ was estimated with less precision than $\phi_1 $ and $\phi_2$ as the percentage of contamination increases. Besides, while $\phi_3 $ was underestimated by all methods, for all levels of contamination, the RA estimator was the only method that overestimated $\phi_1$ and $\phi_2,$ independently of the contamination level.
The fourth experiment is related to Case III. The process of contamination is a replacement contamination where the replacement process $W$ follows a t-student distribution with 2.3 f.d.. The simulations were performed for a $57 \times 57$ window. Table \ref{tabContStudent} and plot \ref{figBoxContStudPorc} show the results. Boxplots show that the BMM method is the best performing estimator, followed by the RA, GM, M and LS methods, in that order. We also noted that in all methods the estimates deteriorate as the level of contamination increases. Additionally, the classic LS estimator presented greater dispersion of the data.
The fifth experiment was performed in the context of Case IV, where the replacement process $W$ was a autoregressive process. We set the window size at $32\times 32$, varying the level of contamination ($5\%,$ $10\%,$ $15\%$ and $20\%$). Table \ref{tabContReemARPorc} shows these results. Besides, Figure \ref{figBoxContReemARPorc} displays the boxplots corresponding to these tables. 
We can see a pattern similar to the fourth experiment; except that in this case, the variances  seem very much alike.  
Finally, the sixth experiment was carried out according to Case V. The replacement process of the contamination was a white noise with variance 50. As in the fifth experiment, we set the window size at $32\times 32$, varying the level of contamination. Table \ref{tabContReemRBPorc} presents the estimated values obtained. The corresponding boxplots are shown in Figure \ref{figBoxContReemRBPorc}. The parameter values $\phi_1,$ $\phi_2$ and $\phi_3$ were underestimated for all methods, excluding the RA estimator that overestimated the values of $\phi_1$ and $\phi_2$ parameters. The BMM estimator was less affected by the contamination process. The LS and M estimators are less accurate than the GM, RA and BMM estimators. Comparatively, the RA estimator presented the highest variance, whereas the GM estimator, although quite accurate, deteriorates more than the BMM estimator as the level of contamination increases.
\vspace{-0.6cm}
\subsection{Computational Time Evaluation}
\begin{figure}[h!]
\begin{center}
\includegraphics[width=8cm, height=4.7cm]{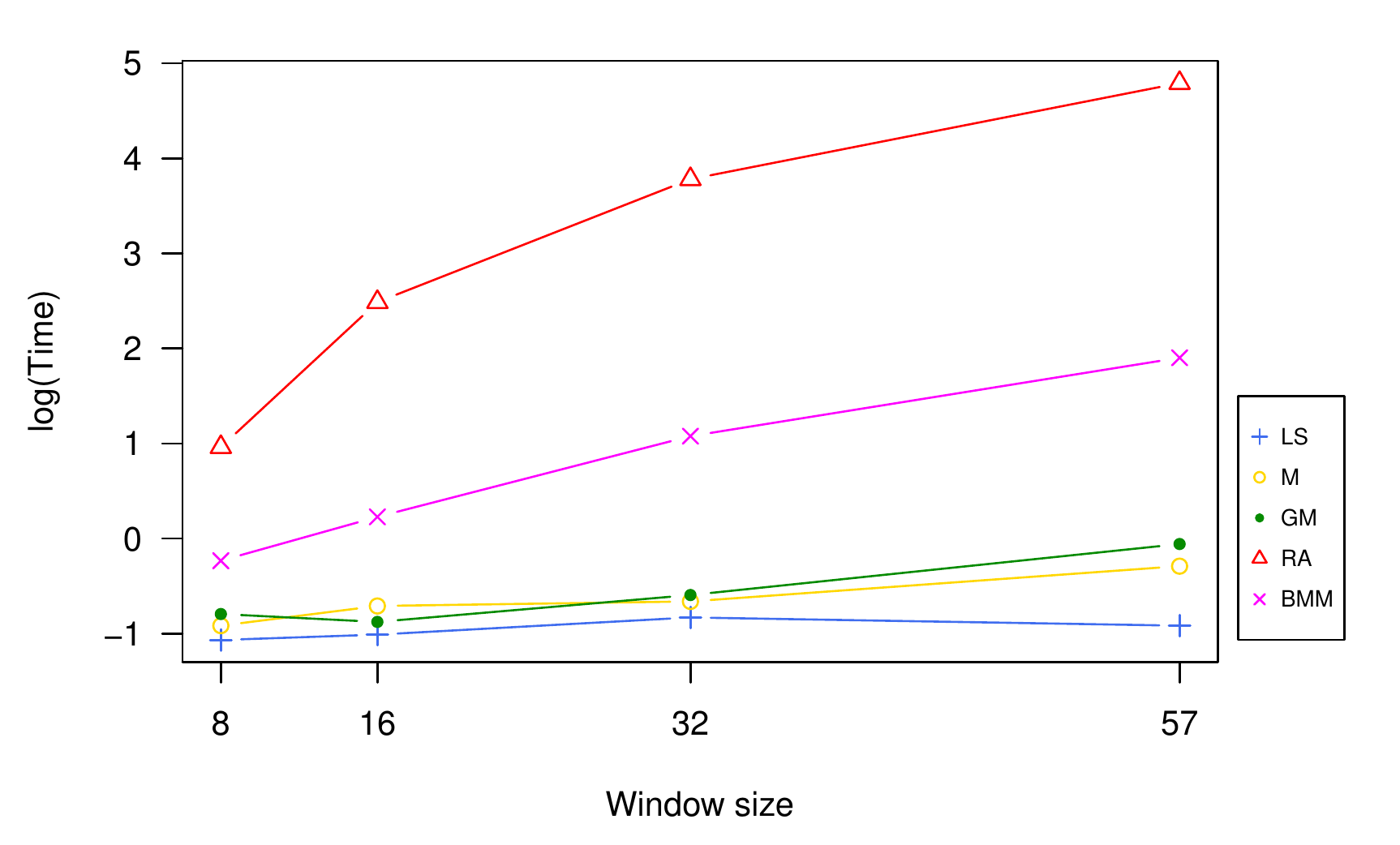}%
\caption{Logarithm of the estimation time (in seconds) when the process has additive contamination of $\sigma^2=50$ according to the window size.}
\label{figTiempos}
\end{center}
\end{figure}

\vspace{-0.3cm}
All the computational routines were developed in R and were carried out on the server JupiterAce of FaMAF-UNC. It has 12-cores 2.40GHz Intel Xeon E5-2620v3 processor, with 128 GiB 2133MHz of available DDR4 RAM. Running time as time logarithm of a single simulation to each estimator vs the window size in Case I is presented in Figure \ref{figTiempos}. Time was expressed in seconds. The graph shows that the computational cost of the RA estimator is the highest; for example, in a $32\times32$ window size, the RA running time was 43.812 seconds, while  BMM, GM, M and LS computational costs were 2.936, 0.552,  0.516 and 0.436 seconds, respectively. This results show that, although RA estimator is one of the major competitors of BMM estimator, due to its accuracy and good asymptotic properties, it exhibits its computational cost as a disadvantage. This makes RA an unattractive estimator for the processing of big size images.
\vspace{-0.3cm}
%\newpage
%%%%%%%%%%%%%%%%%%%%%%%

\clearpage

%%%%%%%%%%%%%%%%%%%%%%%%%%%%%

%\clearpage
%%%%%%%%%%%%%%%%%%%%%%%%%%%%%
\begin{multicols}{2}
%\begin{figure}%[h!]
%\begin{center}
%\begin{tabular}[c]{p{5cm} p{13cm}}
%\begin{minipage}[c]{5cm}

%\end{minipage} &
{\centering
  \begin{minipage}[c]{7.75cm}
  \includegraphics[width=7.75cm, height=5cm]{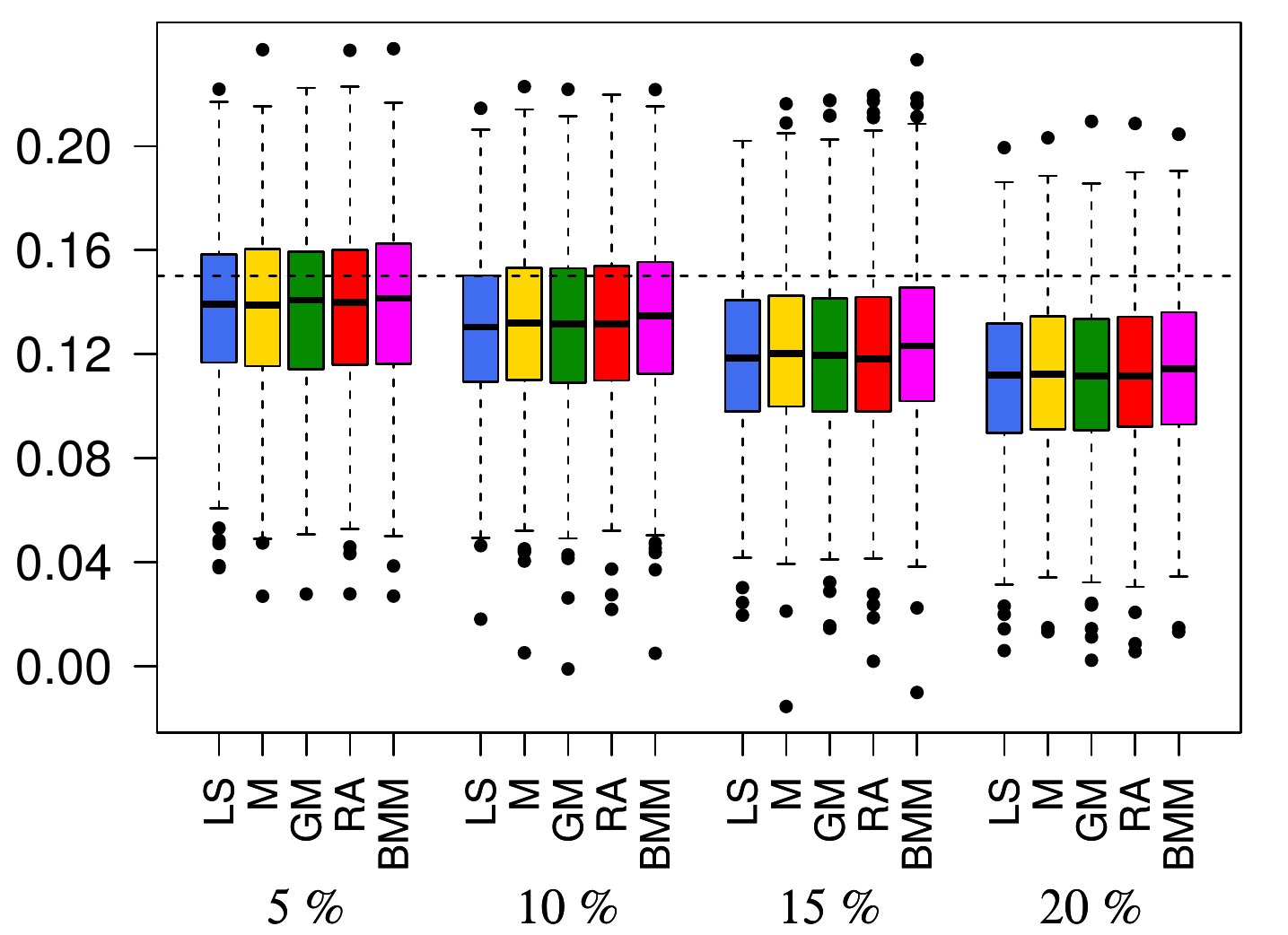}
  \centerline{(a)}
	\smallskip
  \end{minipage}\\
  \begin{minipage}[c]{7.75cm}
   \includegraphics[width=7.75cm, height=5cm]{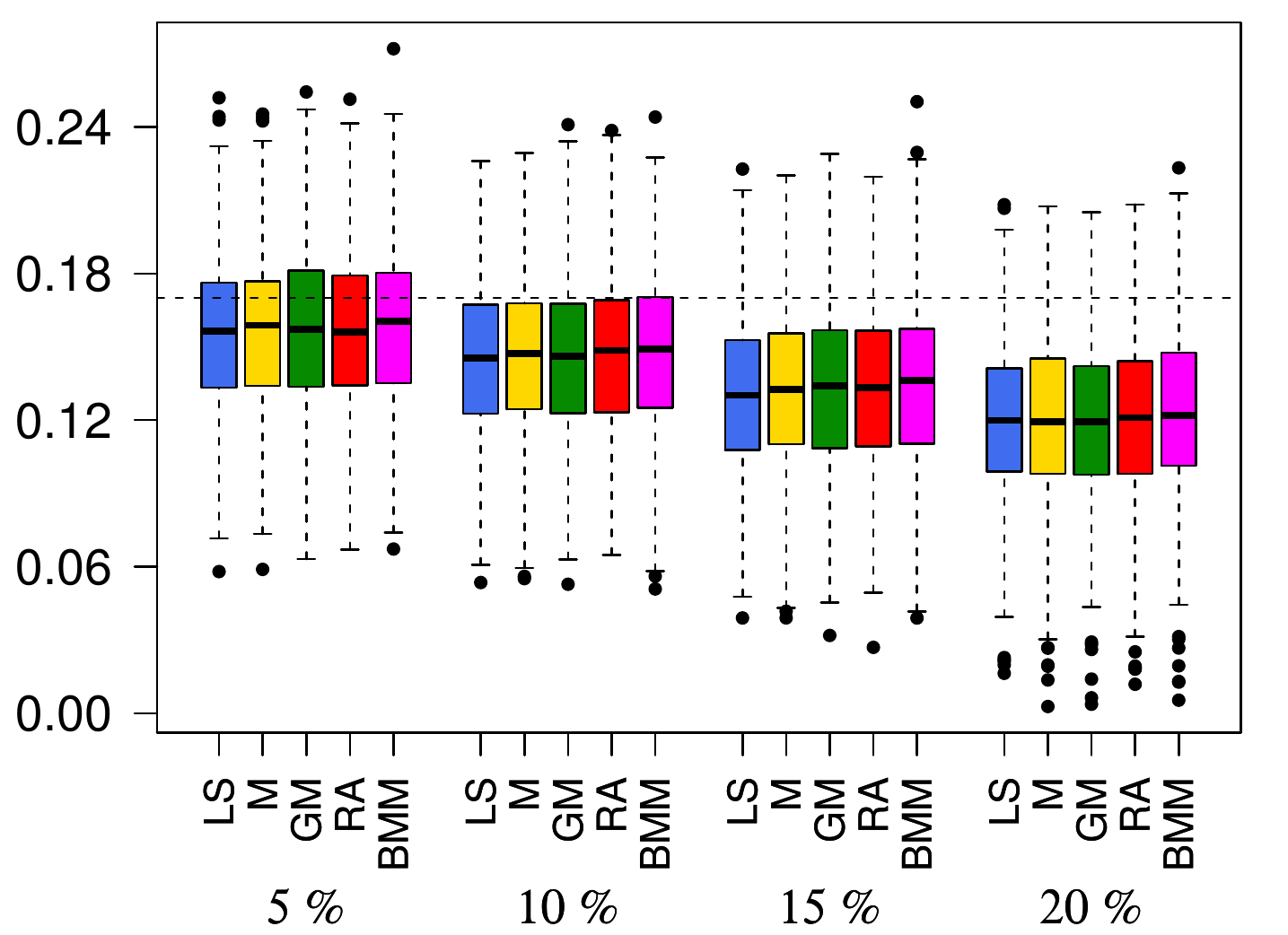}
   \centerline{(b)}
	\smallskip
   \end{minipage}\\
  \begin{minipage}[c]{7.75cm}
   \includegraphics[width=7.75cm, height=5cm]{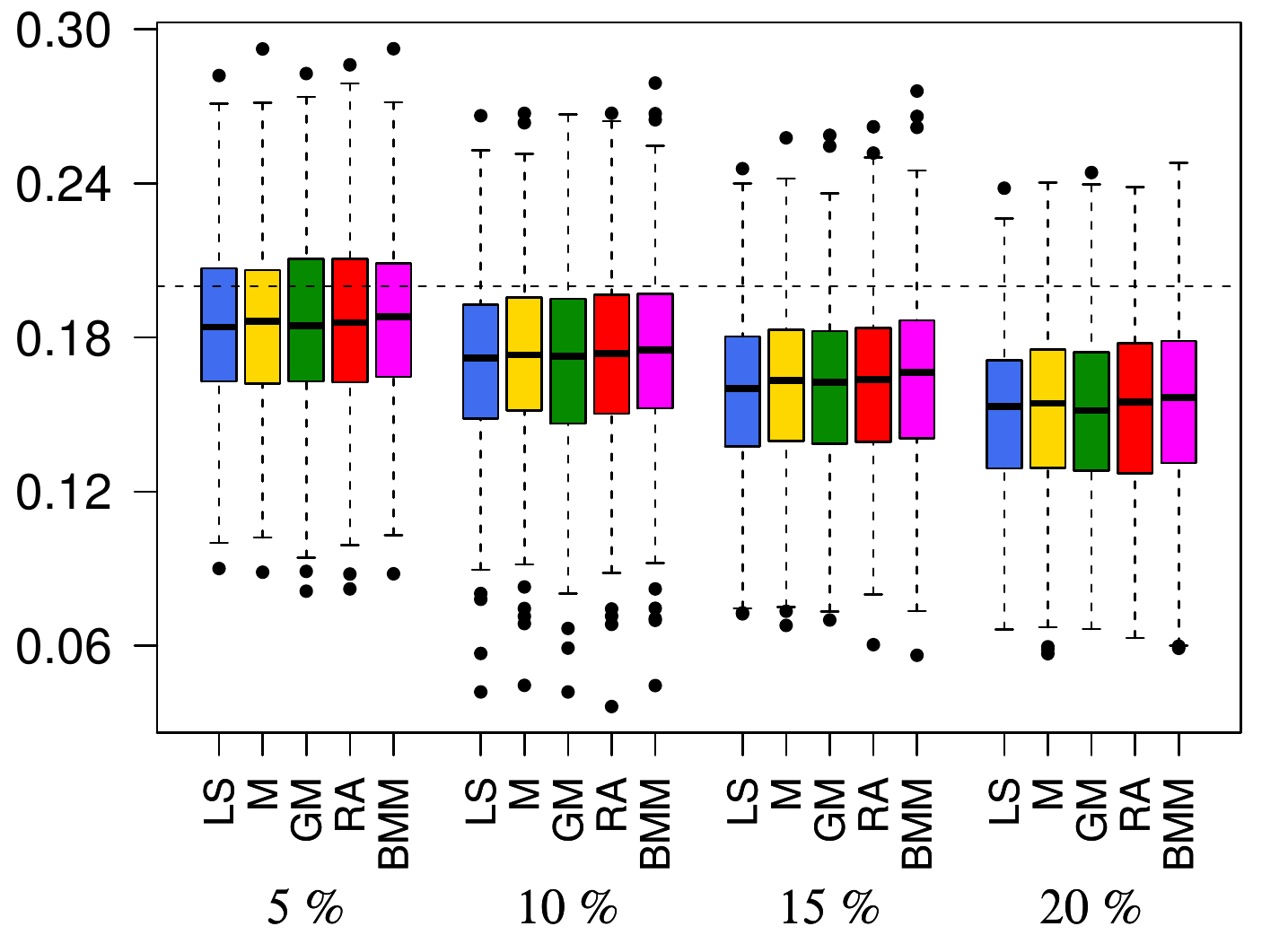}
   \centerline{(c)}
	\smallskip
   \end{minipage}\\
%\end{tabular}
\begin{minipage}[c]{7.75cm}
\captionof{figure}{Case (IV): LS, M, GM, RA and BMM estimation boxplots, for (a) $\phi_1=0.15$, (b) $\phi_2=0.17$ and (c) $\phi_3=0.2$ in model \ref{modeloARE}, varying the contamination level, with a $32\times 32$ window size. The contamination process is of replacement type, by an AR process with $\tilde{\phi}_1=0.1$, $\tilde{\phi}_2=0.2$ and $\tilde{\phi}_3=0.3$ parameters. }\label{figBoxContReemARPorc}
\end{minipage}

%\end{center}
%\end{figure}
}
%%%%%%%%%%%%%%%%%%%%%%%%%%%%%
{\centering
%\begin{figure}%[h!]
%\begin{center}
%\begin{tabular}[c]{p{5cm} p{13cm}}
%\begin{minipage}[c]{5cm}
%\end{minipage} &
  \begin{minipage}[c]{7.75cm}
  \includegraphics[width=7.75cm, height=5cm]{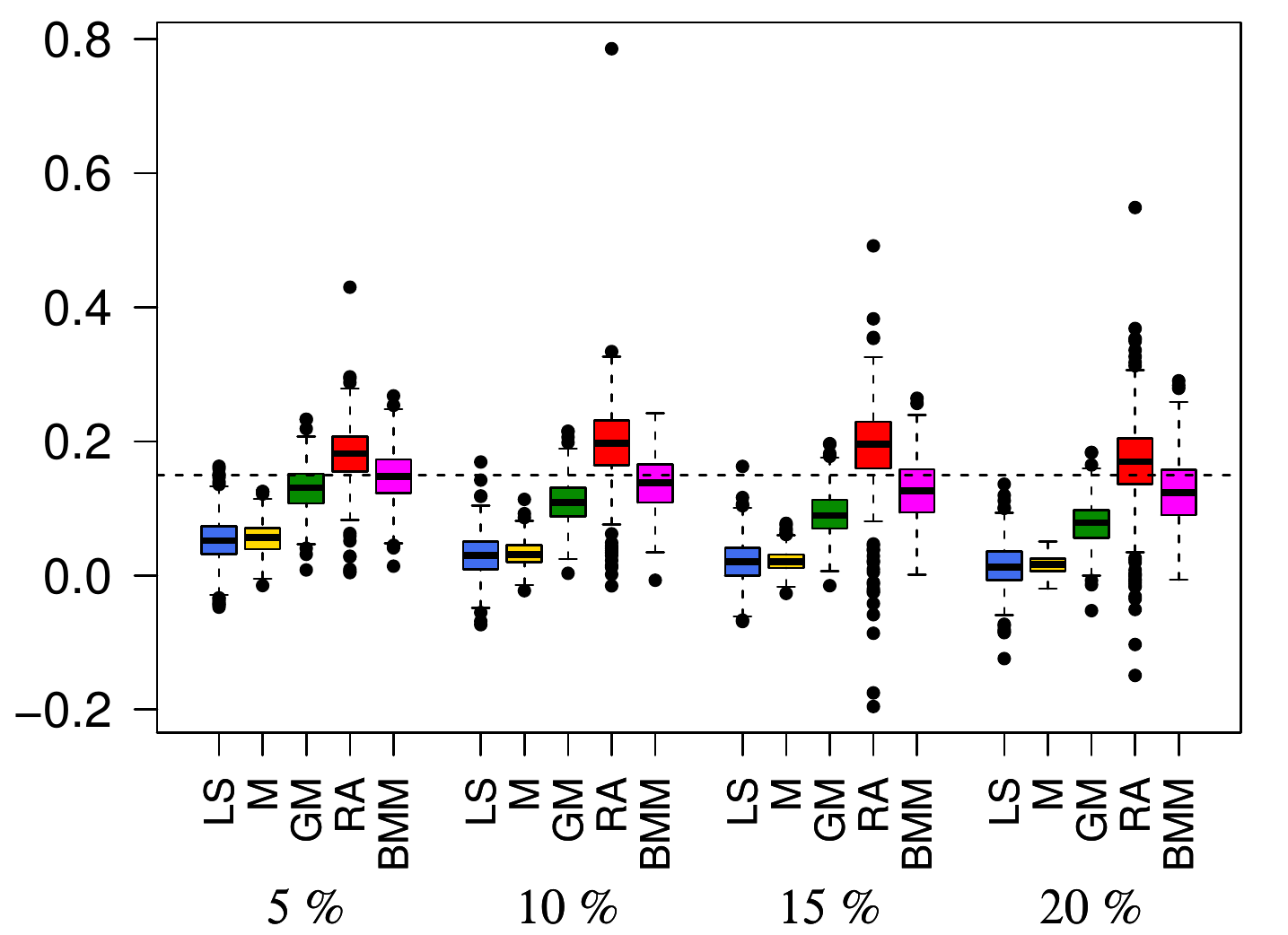}
  \centerline{(a)}
	\smallskip
  \end{minipage}\\
  \begin{minipage}[c]{7.75cm}
   \includegraphics[width=7.75cm, height=5cm]{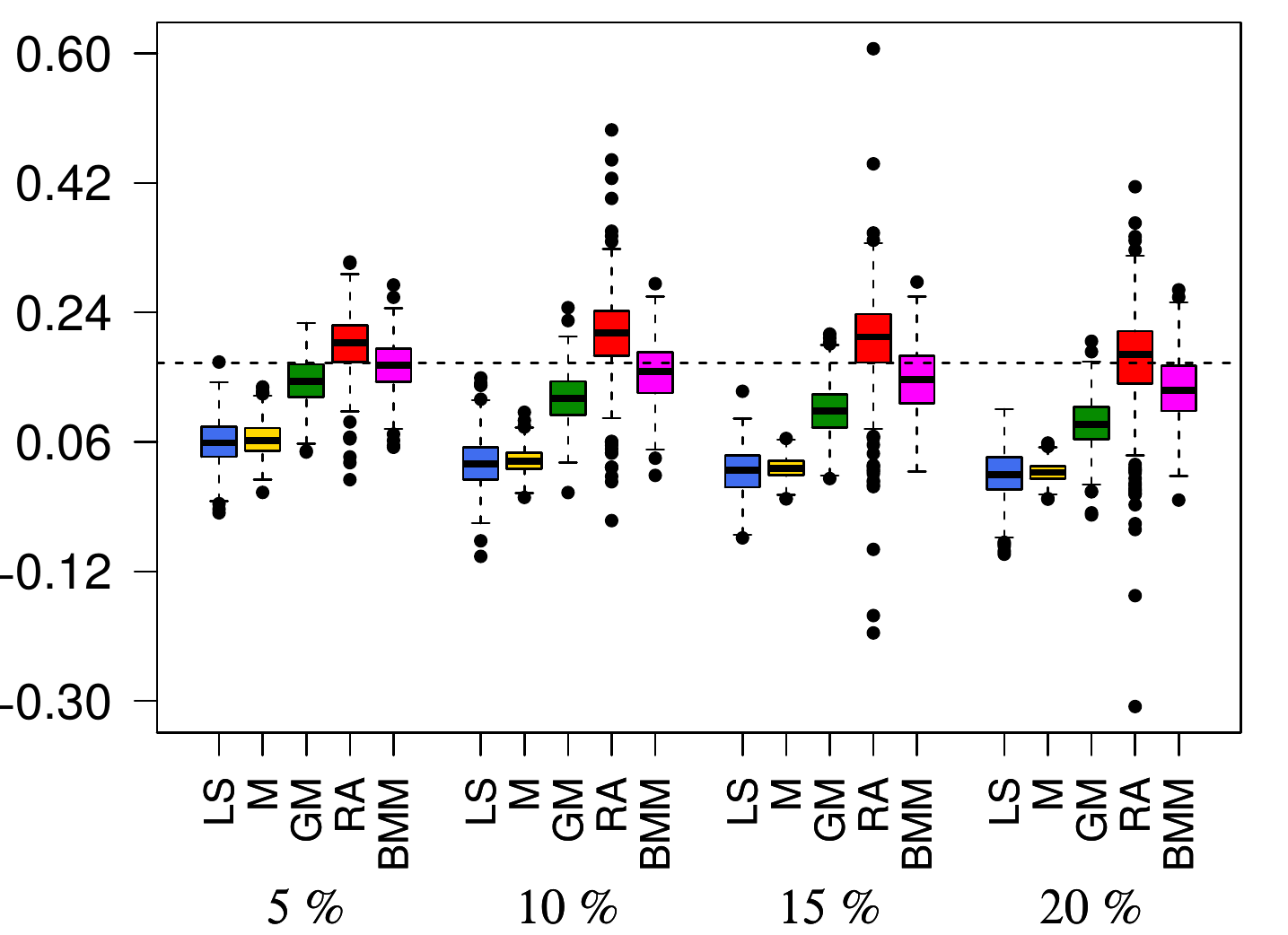}
   \centerline{(b)}
	\smallskip
   \end{minipage}\\
  \begin{minipage}[c]{7.75cm}
   \includegraphics[width=7.75cm, height=5cm]{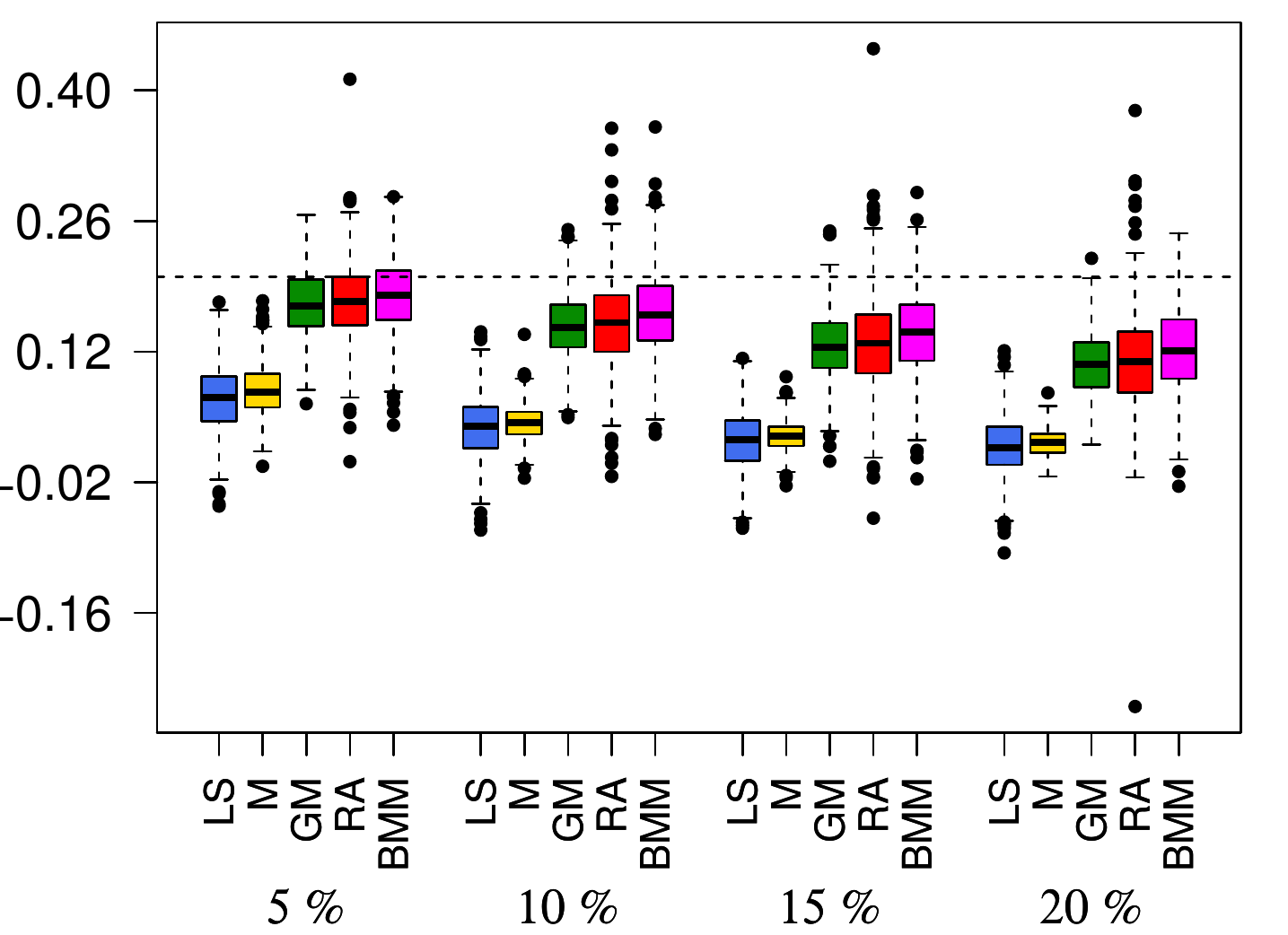}
   \centerline{(c)}
	\smallskip
   \end{minipage}\\
%\end{tabular}
\begin{minipage}[c]{7.75cm}
\captionof{figure}{Case (V): LS, M, GM, RA and BMM estimation boxplots, for (a) $\phi_1=0.15$, (b) $\phi_2=0.17$ and (c) $\phi_3=0.2$ in model \ref{modeloARE}, varying the contamination level, with a $32\times 32$ window size. The contamination process is of replacement type, with a white noise, of variance 50.}\label{figBoxContReemRBPorc}
\end{minipage}
}
%\end{center}
%\end{figure}
\end{multicols}
%%%%%%%%%%%%%%%%%%%%%%%%%%%%%%%%%%%%%%%%%%%%%%%%%%%%%%%%
%%%%%%%%%%%%%%

%\newpage
%\clearpage
%%%%%%%%%%%%%%%%%%%%%%%%%%%%%%%%%%%%%%%%%%%%%%%%%%%%%%%%%%%%%%%%%%%%%%%%%%%%%%%%%%%%%%%%%%%%%%%%%%%
\section{Application to real images}
\label{sec:apli}
\begin{figure}[h!]
\begin{center}
\begin{tabular}[c]{p{3.5cm} p{3.5cm} p{3.5cm} p{3.5cm}}
\begin{minipage}[c]{3.5cm}
\includegraphics[width=3.5cm, height=3.5cm]{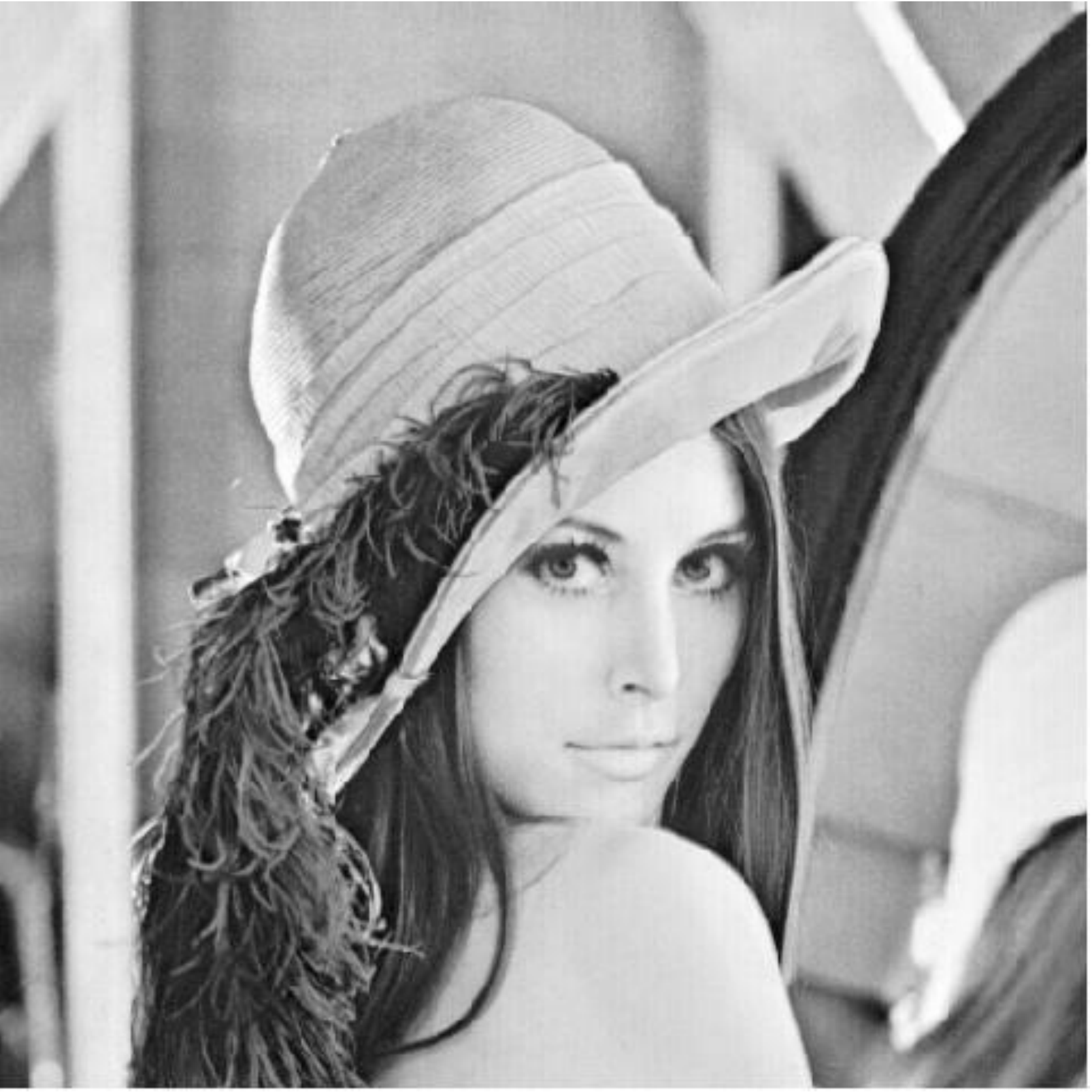}
\centerline{(I)}
\end{minipage} &
\begin{minipage}[h]{10.5cm}
\caption{Image (I) Lena original image. (Right) The first row has the reconstructions done for BMM estimator adjusting window sizes $8\times 8$, $16\times 16$, $32\times 32$ and $57\times 57$ respectively ((a) to (d)). The second row has the respective differences ((e) to (h)) with respect to the original image (I).}\label{figLenRecVent}
\end{minipage}\\
\begin{minipage}[c]{3.5cm}
\includegraphics[width=3.5cm, height=3.5cm]{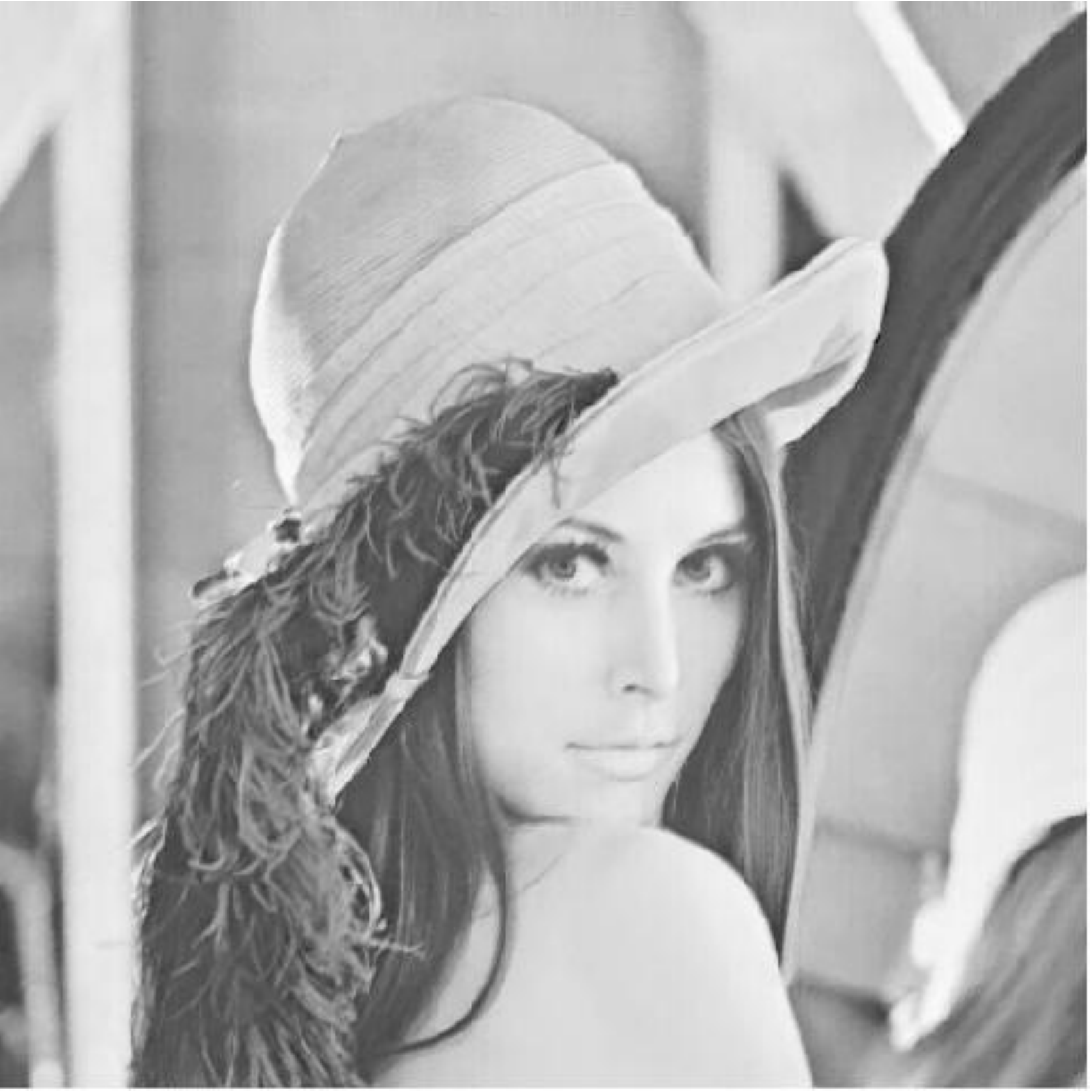}
\centerline{(a)}
\end{minipage} &
 \begin{minipage}[c]{3.5cm}
\includegraphics[width=3.5cm, height=3.5cm]{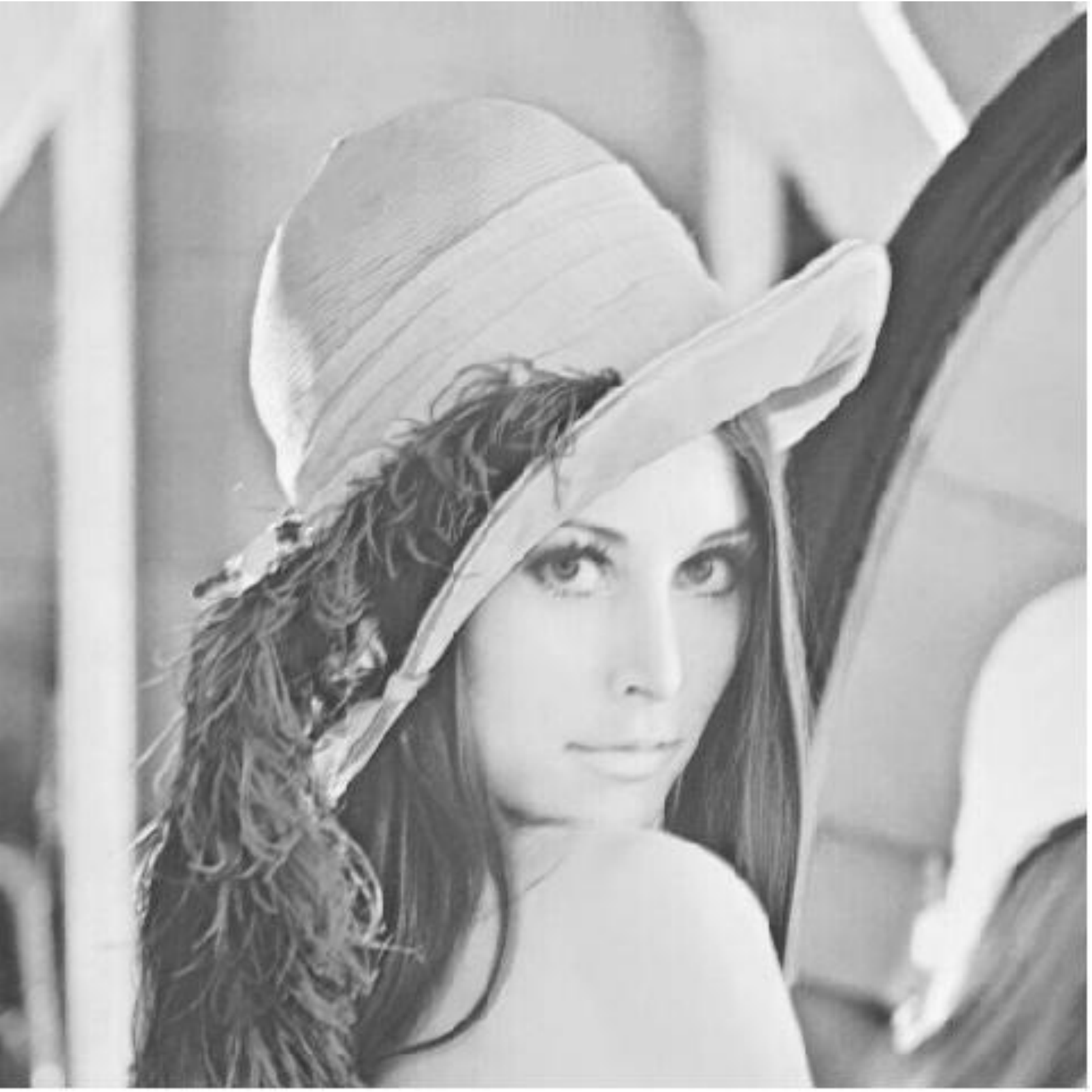}
\centerline{(b)}
\end{minipage} &
\begin{minipage}[c]{3.5cm}
\includegraphics[width=3.5cm, height=3.5cm]{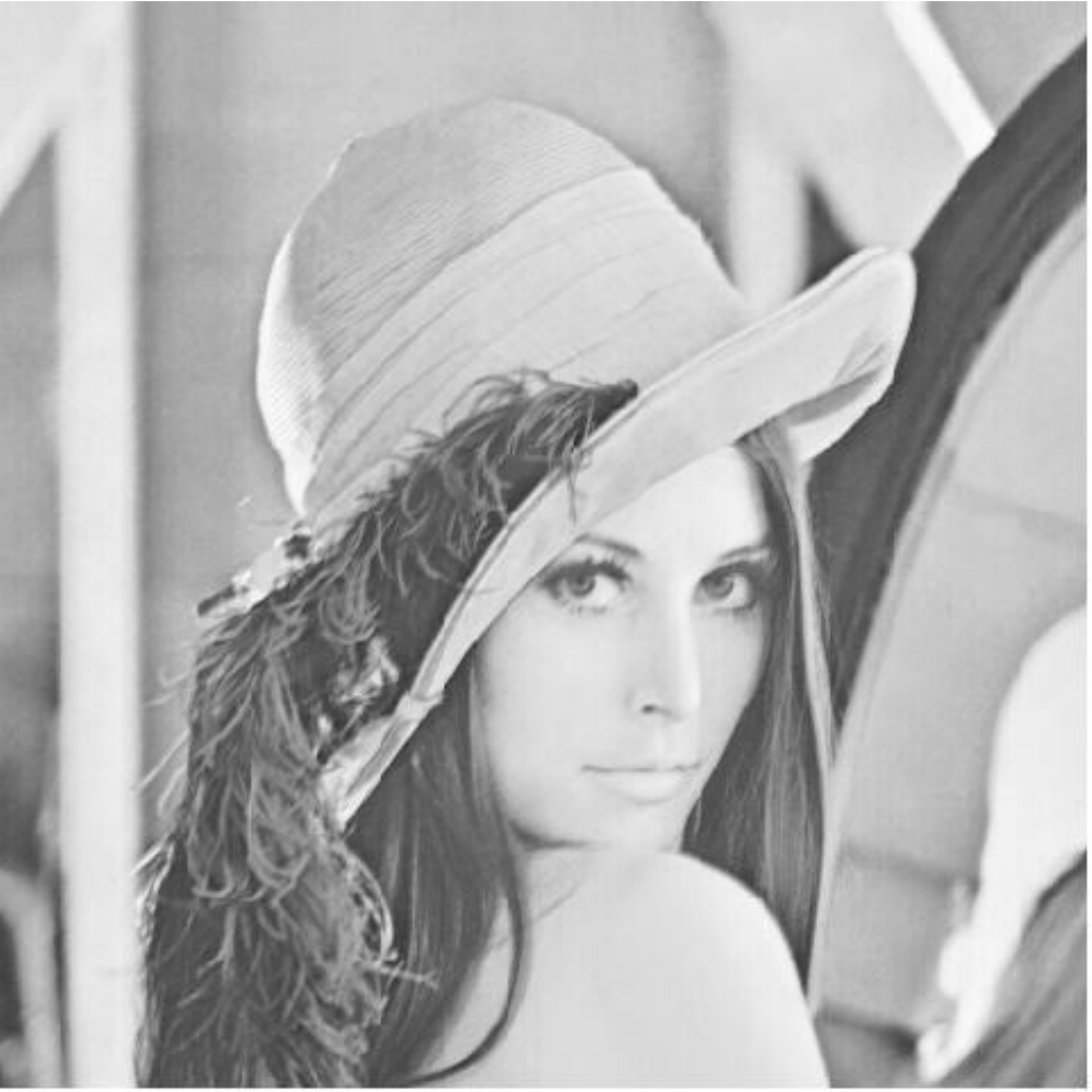}
\centerline{(c)}
\end{minipage} & 
\begin{minipage}[c]{3.5cm}
\includegraphics[width=3.5cm, height=3.5cm]{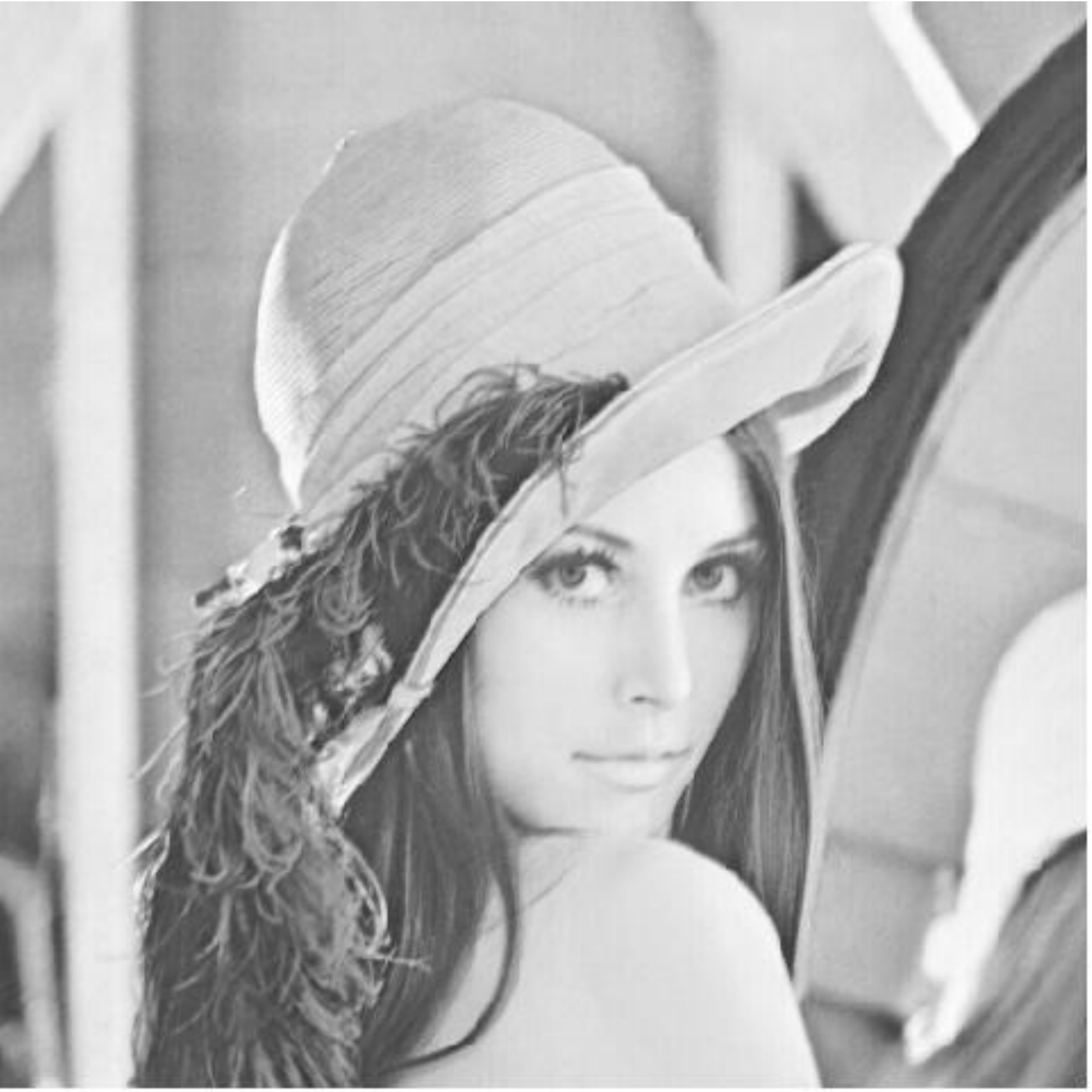}
\centerline{(d)}
\end{minipage}\\
\begin{minipage}[c]{3.5cm}
\includegraphics[width=3.5cm, height=3.5cm]{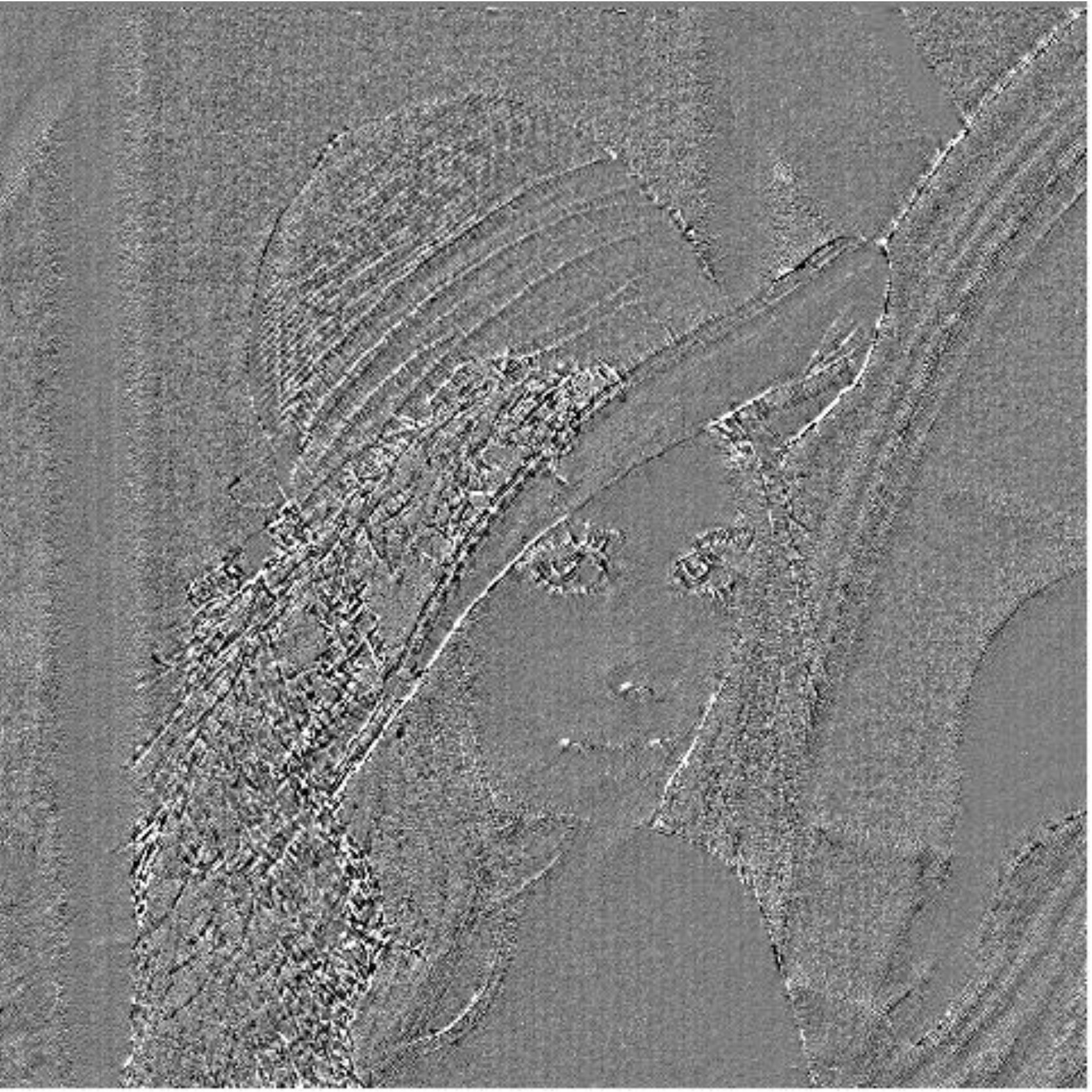}
\centerline{(e)}
\end{minipage} &
\begin{minipage}[c]{3.5cm}
\includegraphics[width=3.5cm, height=3.5cm]{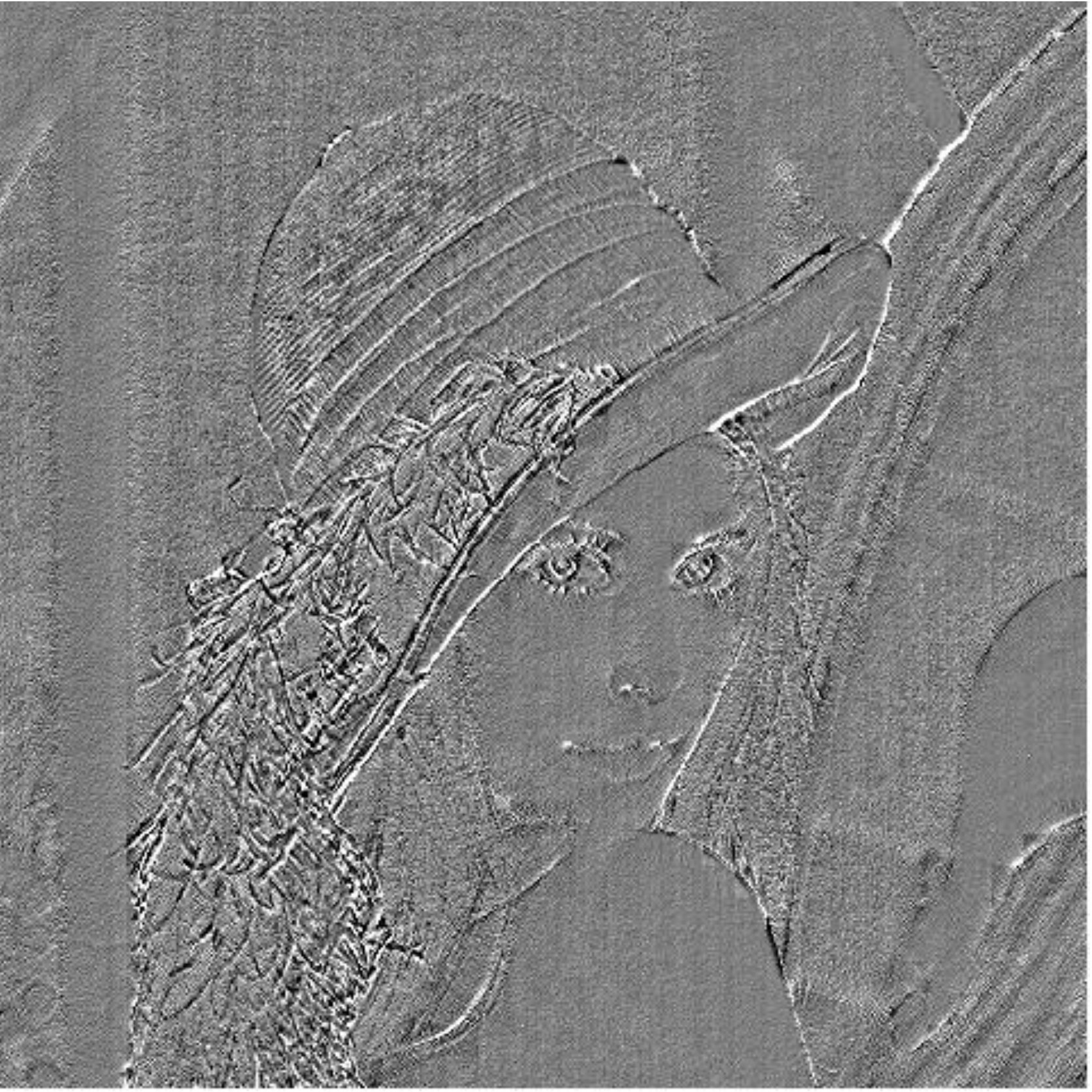}
\centerline{(f)}
\end{minipage} &
\begin{minipage}[c]{3.5cm}
\includegraphics[width=3.5cm, height=3.5cm]{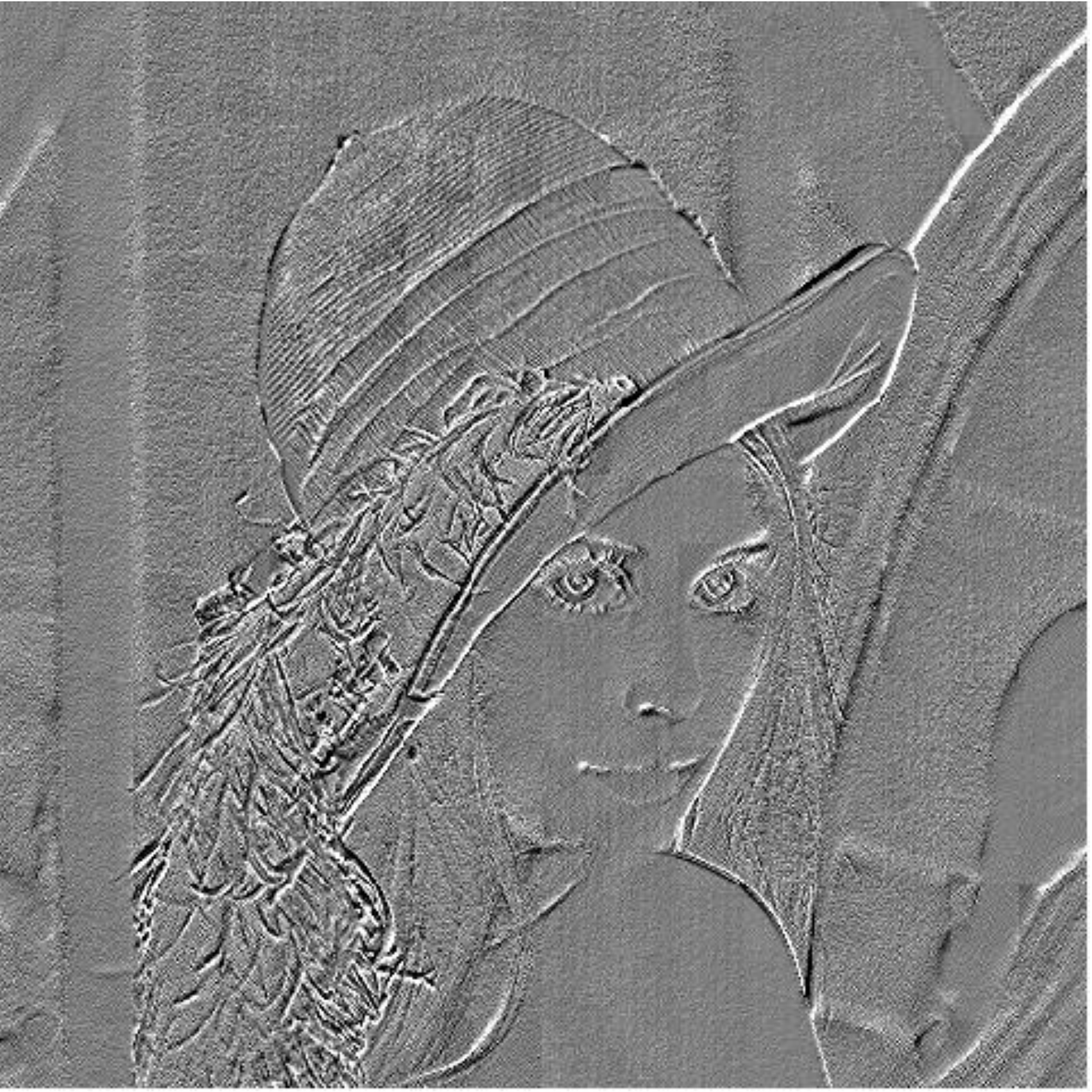}
\centerline{(g)}
\end{minipage} & 
\begin{minipage}[c]{3.5cm}
\includegraphics[width=3.5cm, height=3.5cm]{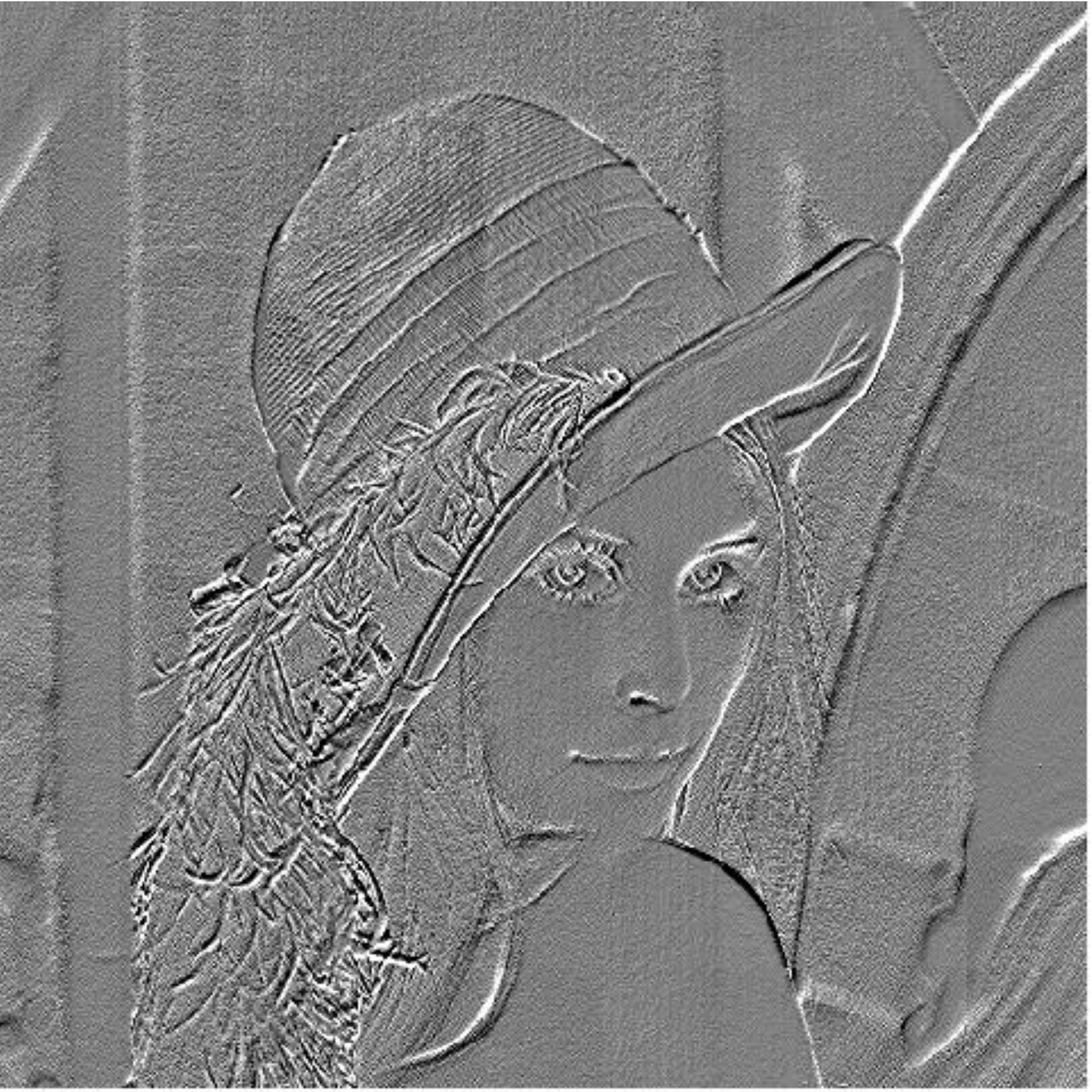}
\centerline{(h)}
\end{minipage}\\
\end{tabular}
\end{center}
\end{figure}

The analysis of contaminated images is of great interest in several areas of research. For example, the reconstruction of contaminated images is relevant in modeling of images (\cite{All2}; \cite{Vall2}), and, in general, the reduction of the noise produced by interferences taking place in the processes of obtaining the physical image and transmitting it electronically plays an important role in the literature (\cite{Bus1}).\\ 
In \cite{Oje2}, two algorithms for image processing based on the unilateral AR-2D model with two parameters were presented. The foundations of the algorithms are random field theory and robustness for spatial autoregressive processes. The first one produces a local approximation of images, and the second one, is a segmentation algorithm. In this work, we proposed to use a variant of these algorithms using a unilateral AR-2D process with three parameters (model (\ref{modeloAR})), instead of two parameters as it was originally proposed. We called the modified algorithms as Algorithm 1 and Algorithm 2. We applied them to reconstruction and segmentation of images using the LS, GM and BMM estimators of the parameters in the model (\ref{modeloAR}). Later, we inspected and compared the performance of these estimators in Algorithms 1 and 2 on contaminated images. To compare the images generated by the algorithms and, therefore, the performance of the different estimators, we calculated three indexes used in the literature; the SSIM index (\cite{Wan}), the CQ index (\cite{Oje3}), and  CQmax index (\cite{Pis}). Next, we present two numerical experiments using the image ``Lenna'', which was taken from the USC-SIPI image database http://sipi.usc.edu/database/. In Figure \ref{figLenRecVent}-(I), the original $512\times512$ image is shown.\\ 
In the following we present Algorithms \ref{alg1} and \ref{alg2}, for more details about the notation you can see the work \cite{Oje2}.

\begin{algorithm}[h!]
\begin{algorithmic}[1]
\REQUIRE Original image $Z$.
\ENSURE  Approximated image $\hat{Z}$ of the original image $Z$\\
\STATE Define $X$ as $X=Z-\overline{Z}$
\STATE Generate block $B_X(i_b,j_b)$
\STATE Compute the estimations $\hat{\phi}_1^{(i_b,j_b)}$, $\hat{\phi}_2^{(i_b,j_b)}$, $\hat{\phi}_3^{(i_b,j_b)}$ of $\phi_1$, $\phi_2$ and $\phi_3$ corresponding to the block $B_X(i_b,j_b)$ extended to $B'_X(i_b,j_b)=[X_{r,s}]_{(k-1)(i_b-1)\leq r\leq (k-1)i_b, (k-1)(j_b-1)\leq s\leq (k-1)j_b}$
\STATE Define $\hat{X}$ on the block $B_X(i_b,j_b)$ by\\
 $$\hat{X}_{r,s}=\hat{\phi}_1^{(i_b,j_b)}X_{r-1,s}+\hat{\phi}_2^{(i_b,j_b)}X_{r,s-1}+\hat{\phi}_3^{(i_b,j_b)}X_{r-1,s-1}$$
where $(k-1)(i_b-1)+1\leq r\leq (k-1)i_b$ and $(k-1)(j_b-1)+1\leq s\leq (k-1)j_b$\\
\STATE Define $\hat{Z}$ as $\hat{Z}=\hat{X}-\overline{Z}$\\
\end{algorithmic}
\caption{Local approximation of images by using AR-2D processes}\label{alg1}
\end{algorithm}

\begin{algorithm}[h]
\begin{algorithmic}[1]
\REQUIRE Original image $Z$
\ENSURE Segmentated image $W$
\STATE Generate an approximated image $\hat{Z}$ of $Z$ with the Algorithm 1.\\
\STATE Compute the residual image $W$ defined as $W=Z-\hat{Z}$.
\end{algorithmic}
\caption{Segmentation}\label{alg2}
\end{algorithm}

In the first experiment, Algorithm 1 was applied to image representation. We locally adjusted an AR-2D process to the original image for different window sizes, and estimated the parameters of the model with the BMM estimator. Fig. \ref{figLenRecVent}, (a), (b), (c) and (d) exhibits the BMM reconstructed images obtained respectively using the window sizes $8\times 8$, $16\times 16$, $32\times 32$ and $57\times 57$. For all window sizes, the BMM recostructed images are visually good; although a quantitative analysis of the similarity between each BMM reconstructed image and the original image showed differences. We calculated the SSIM, CQ (1,1) and CQ$_{\max}$ index, between each reconstructed image and the original image. The three indexes revealed that the similarity decreases as the size of the window increases (Table \ref{tabSimilLenavsRec}); so the best fits were obtained with small window sizes. This result reflects the assumption that the two-dimensional autoregressive model is a local adjustment model. Next, we applied Algorithm 2 and generated four difference images (e), (f), (g) and (h) shown in Fig. \ref{figLenRecVent}. We observed  that the difference image (h) highlights the edges more than the others do. This shows that when we performed the reconstruction with a $57\times 57$ window size, (Fig. \ref{figLenRecVent} (d)), a lot of information got lost and this is reflected by  the difference image (Fig. \ref{figLenRecVent} (h)).

%%%%%%%%%%%%%%%%%%%%%%%%%
\begin{table}[h!]
\begin{center}
\caption{SSIM, CQ and CQ$_{max}$ index between the original image and each one of the BMM reconstructed images (a), (b), (c) and (d) in Figure \ref{figLenRecVent}.}
\begin{tabular}[c]{c c c c}
\hline
Window size & SSIM & CQ(1,1) & CQ$_{\max}$ \\
\hline
$8\times 8$   & 0.9948914 & 0.8582201 & 0.9706984 \\ 
$16\times 16$ & 0.9827996 & 0.8309626 & 0.9544317 \\ 
$32\times 32$ & 0.9779204 & 0.8151581 & 0.9462133 \\ 
$57\times 57$ & 0.9762065 & 0.8073910 & 0.9423786 \\ \hline
\end{tabular}
\label{tabSimilLenavsRec}
\end{center}
\end{table}

\newpage

\begin{figure}[h!]
\begin{center}
\begin{tabular}[c]{p{3.5cm} p{3.5cm} p{3.5cm} p{3.5cm}}
\begin{minipage}[c]{3.5cm}
\includegraphics[width=3.5cm, height=3.5cm]{2lenna_sin_cont_vent8-eps-converted-to.pdf}
\centerline{(I)}
\end{minipage} &
\begin{minipage}[c]{3.5cm}
\includegraphics[width=3.5cm, height=3.5cm]{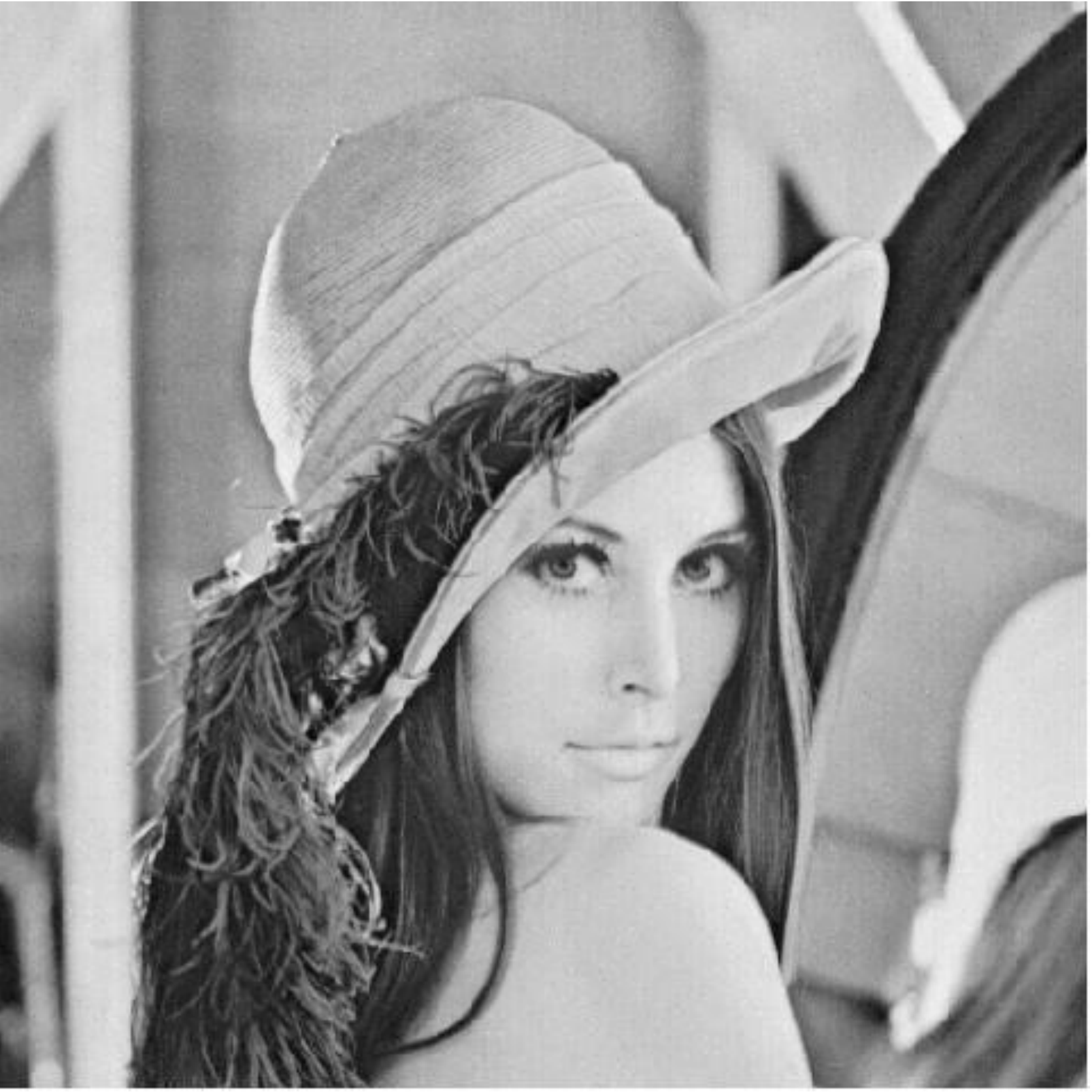}
\centerline{(II)}
\end{minipage} & 
\begin{minipage}[c]{7cm}
\caption{Image (I) Lena original image; Image (II) image with 10\% additive contamination and $\sigma^2=50$. In the first row, adjustments made with LS; in the second row, with GM; and in the third row, with BMM. Columns 1 and 2 correspond to adjustments with $8\times 8$ windows, and columns 3 and 4 to $57\times 57$ windows. Columns 1 and 3 are the reconstruction for Algorithm 1 and columns 2 and 4 are the segmented images for Algorithm 2.}\label{figLenaRec8y57}
 \
\end{minipage}\\
\begin{minipage}[c]{3.5cm}
\includegraphics[width=3.5cm, height=3.5cm]{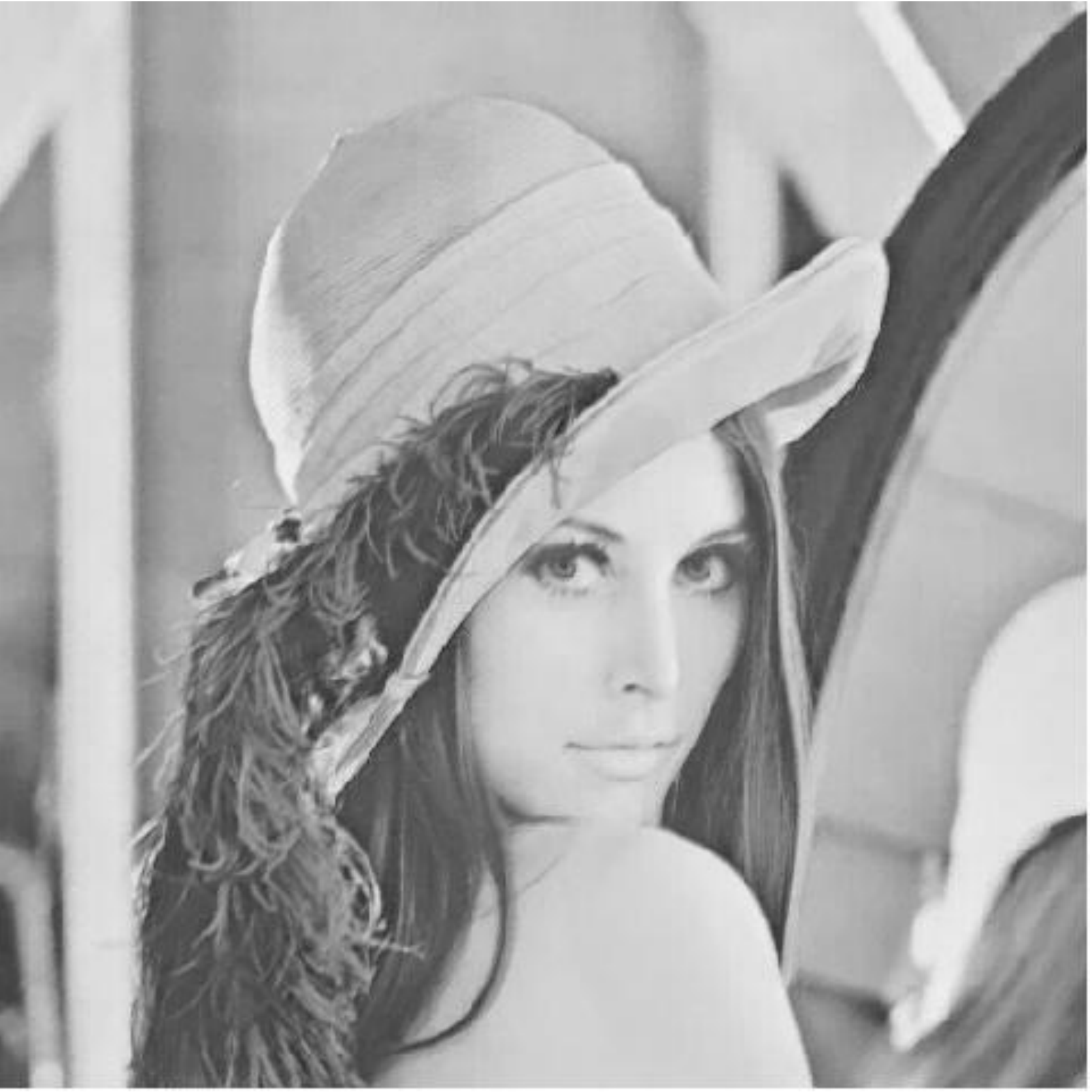}
\centerline{(a)}
\end{minipage} & 
\begin{minipage}[c]{3.5cm}
\includegraphics[width=3.5cm, height=3.5cm]{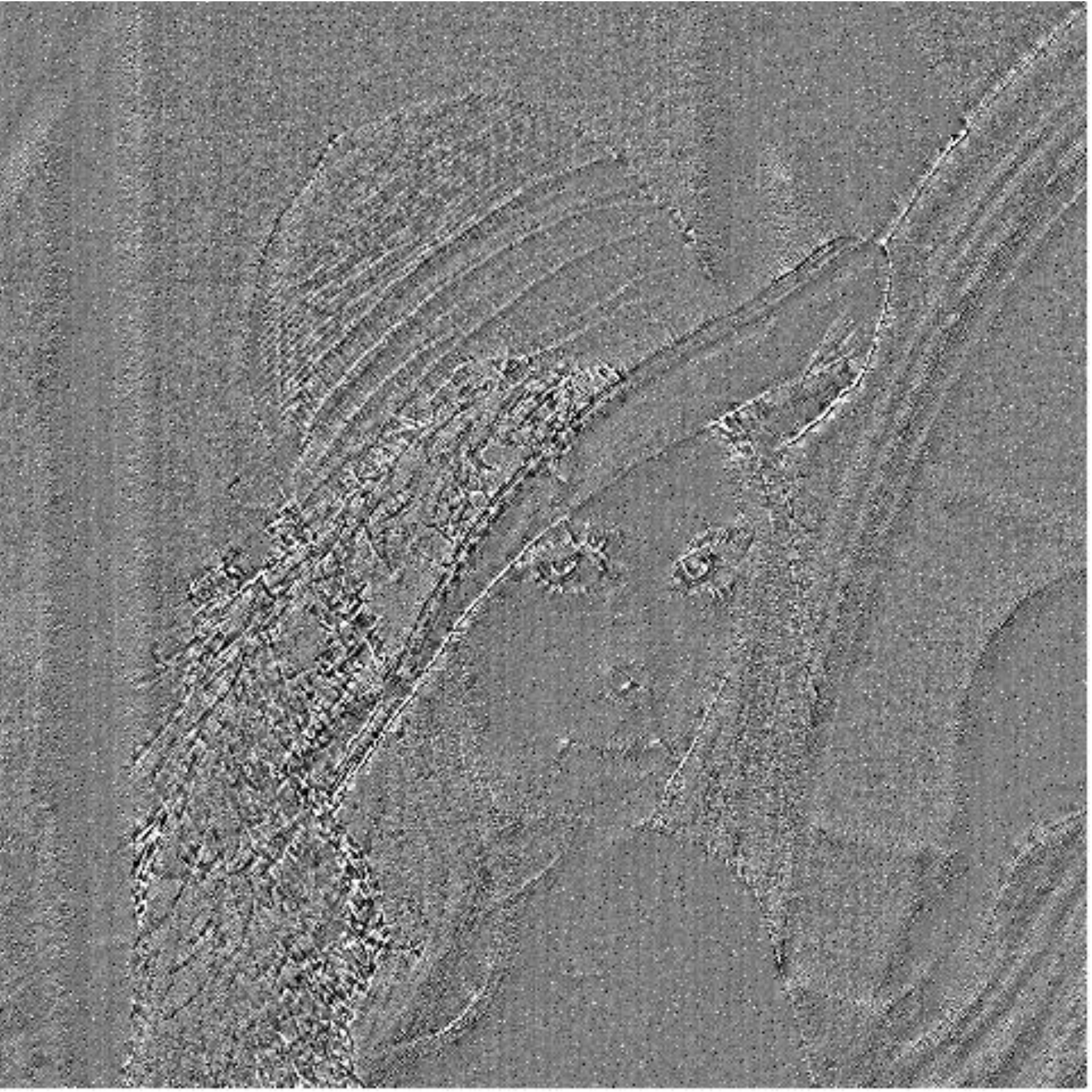}
\centerline{(d)}
\end{minipage} &
\begin{minipage}[c]{3.5cm}
\includegraphics[width=3.5cm, height=3.5cm]{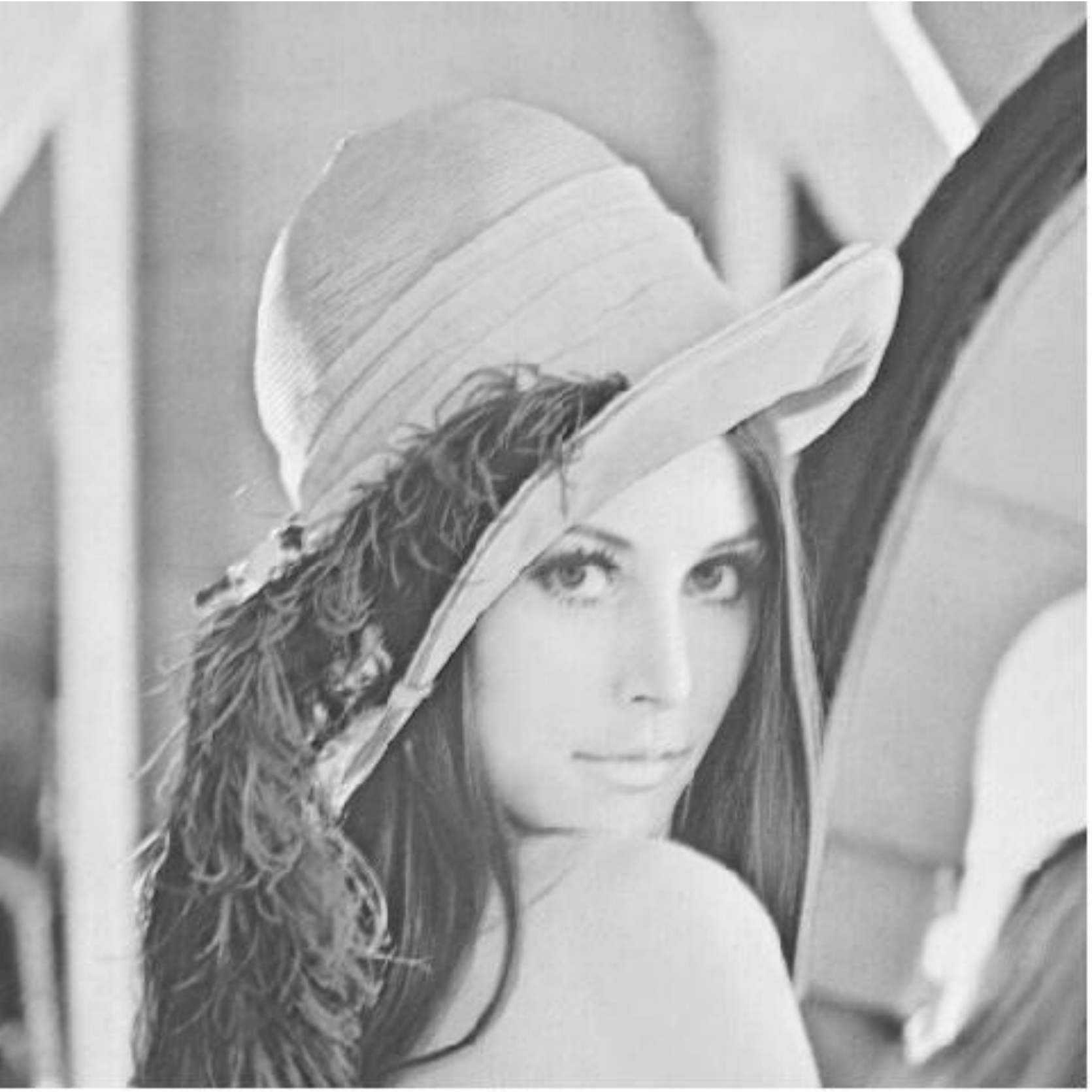}
\centerline{(g)}
\end{minipage} &
\begin{minipage}[c]{3.5cm}
\includegraphics[width=3.5cm, height=3.5cm]{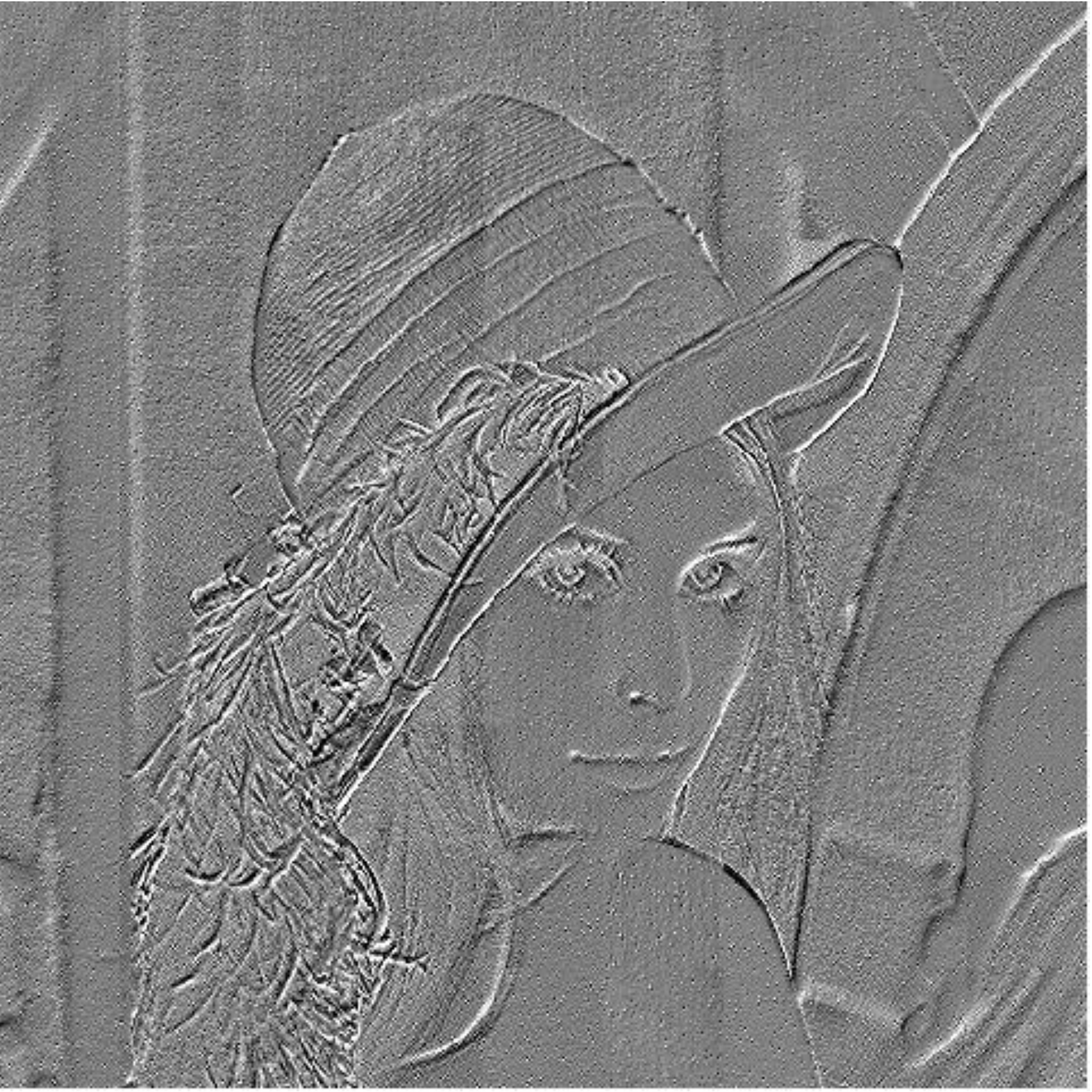}
\centerline{(j)}
\end{minipage}\\
\begin{minipage}[c]{3.5cm}
\includegraphics[width=3.5cm, height=3.5cm]{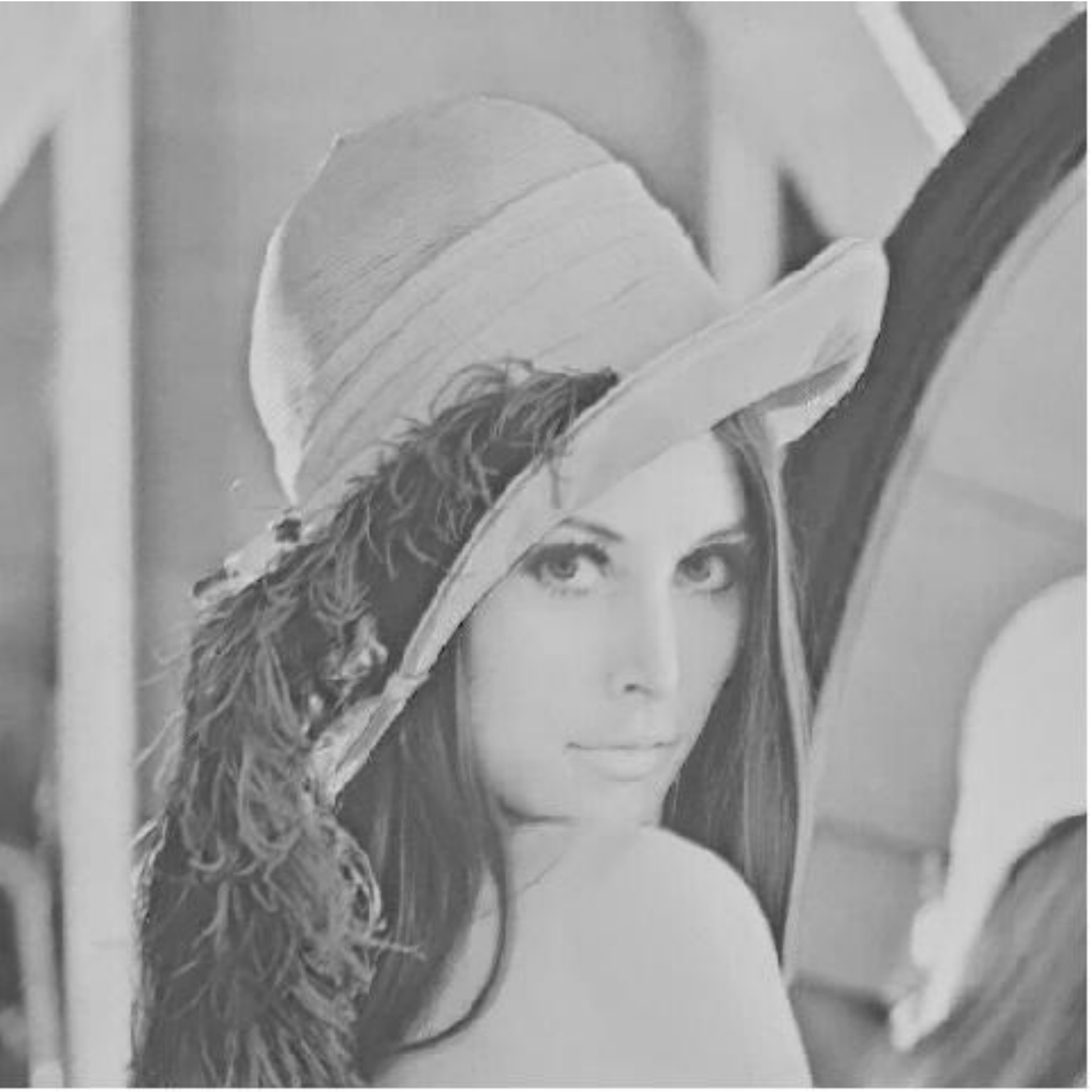}
\centerline{(b)}
\end{minipage} &
 \begin{minipage}[c]{3.5cm}
\includegraphics[width=3.5cm, height=3.5cm]{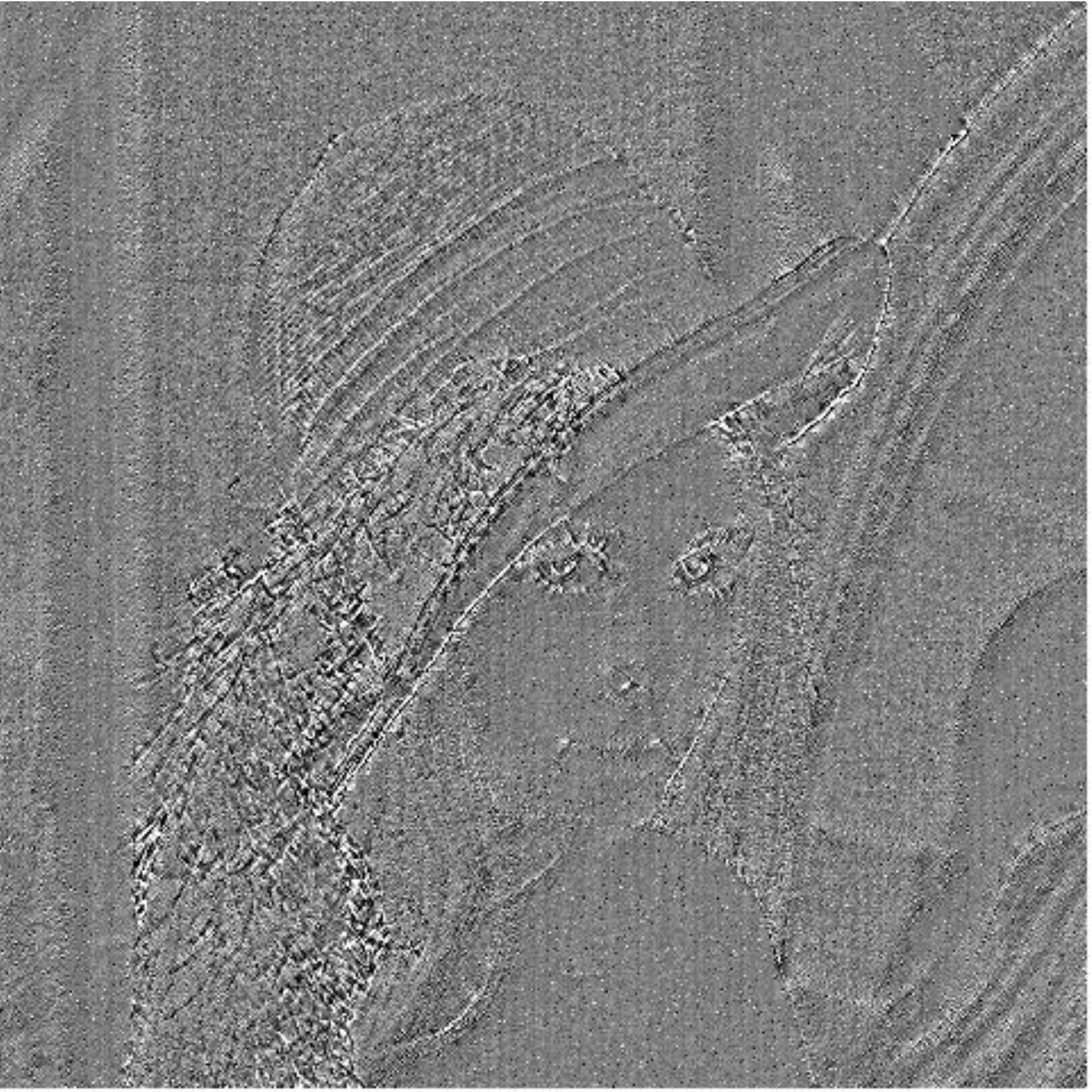}
\centerline{(e)}
\end{minipage} &
\begin{minipage}[c]{3.5cm}
\includegraphics[width=3.5cm, height=3.5cm]{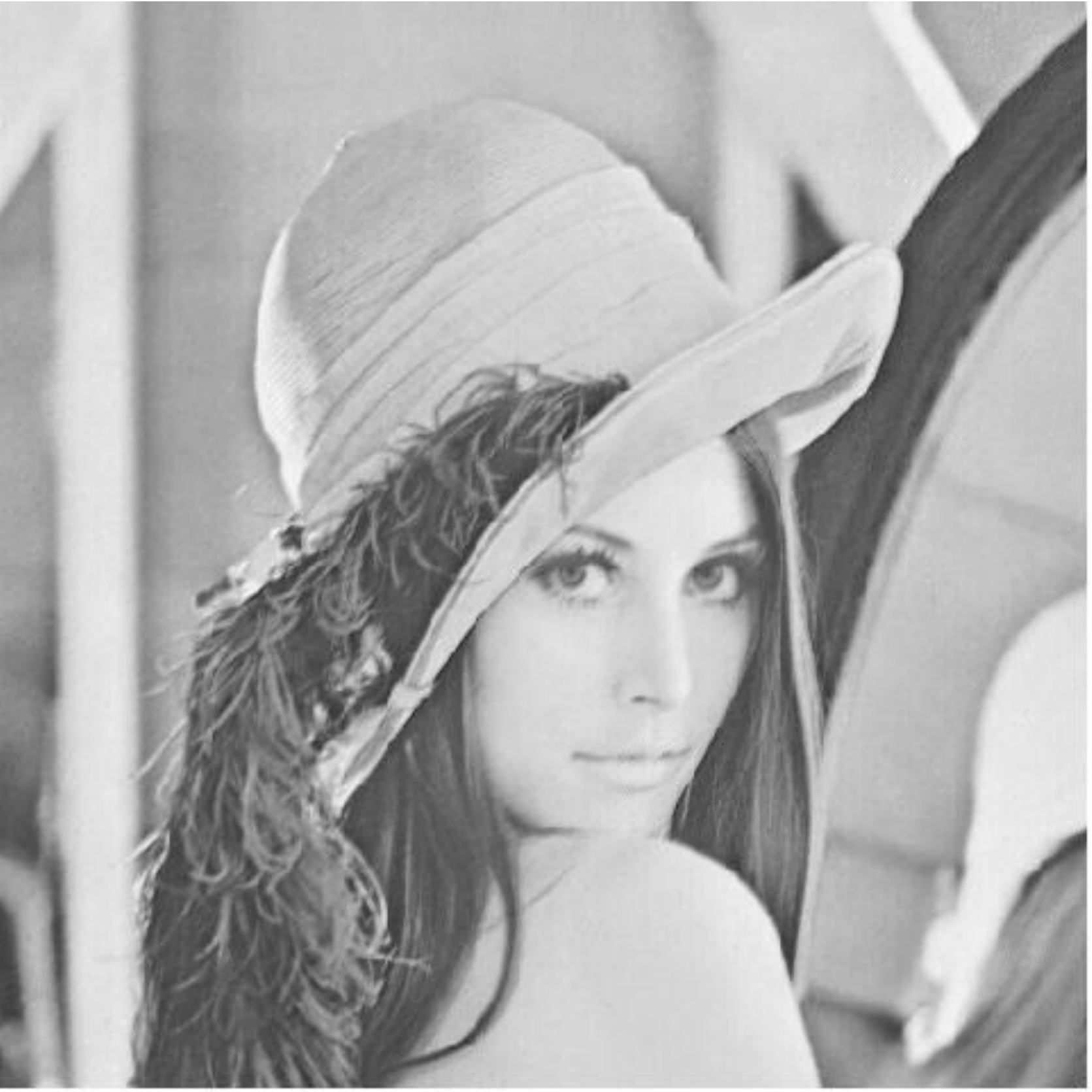}
\centerline{(h)} 
\end{minipage} &
\begin{minipage}[c]{3.5cm}
\includegraphics[width=3.5cm, height=3.5cm]{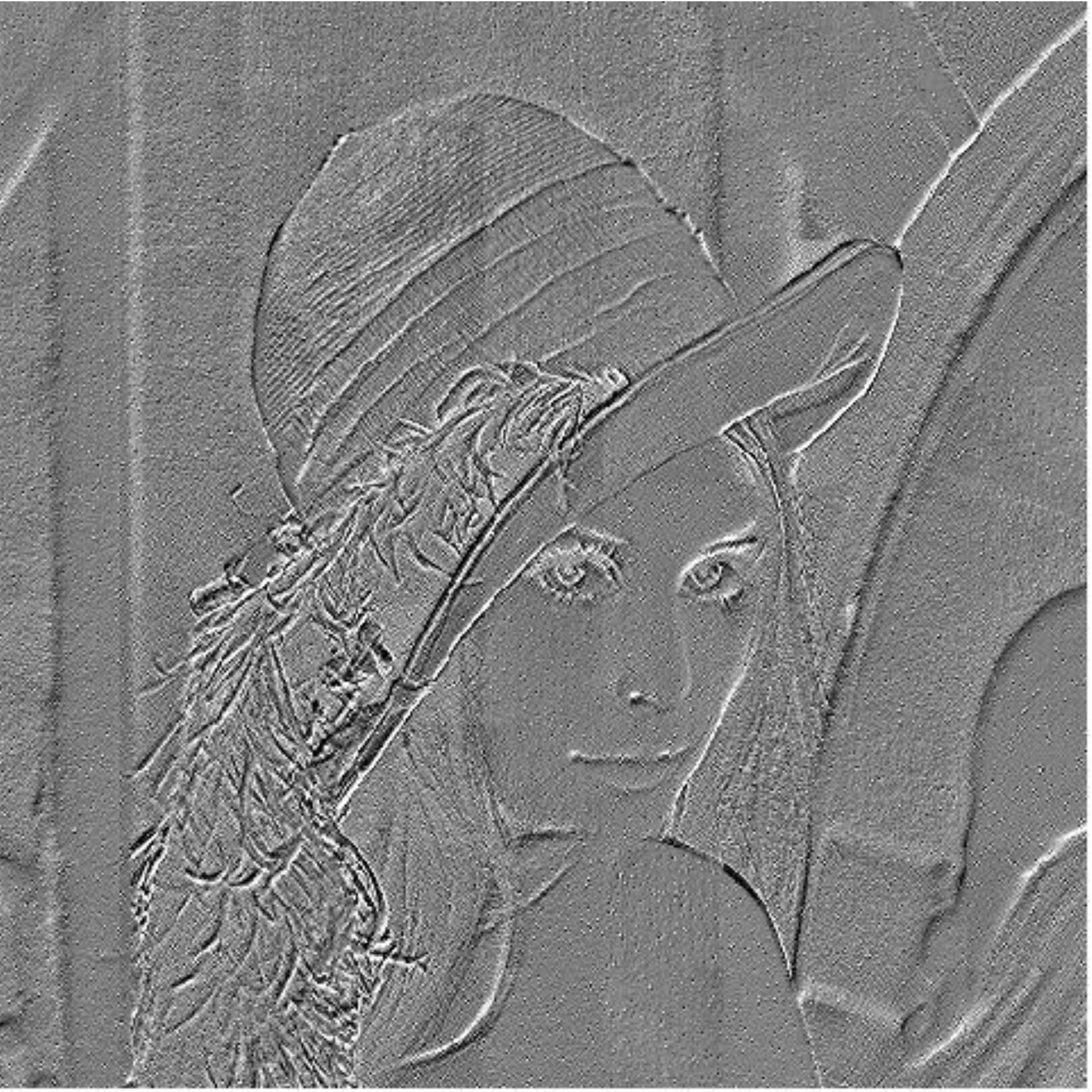}
\centerline{(k)}
\end{minipage}\\
\begin{minipage}[c]{3.5cm}
\includegraphics[width=3.5cm, height=3.5cm]{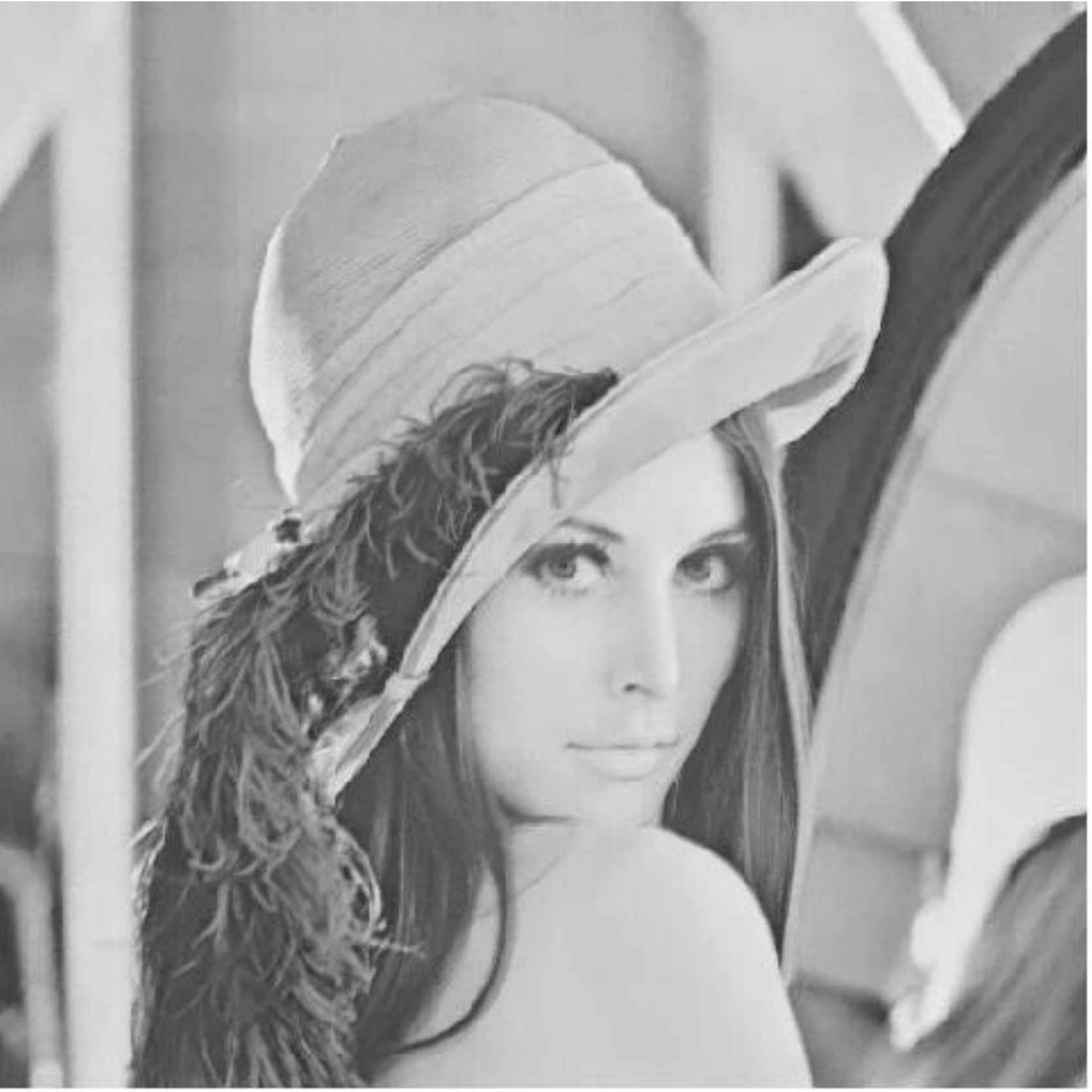}
\centerline{(c)}
\end{minipage} & 
\begin{minipage}[c]{3.5cm}
\includegraphics[width=3.5cm, height=3.5cm]{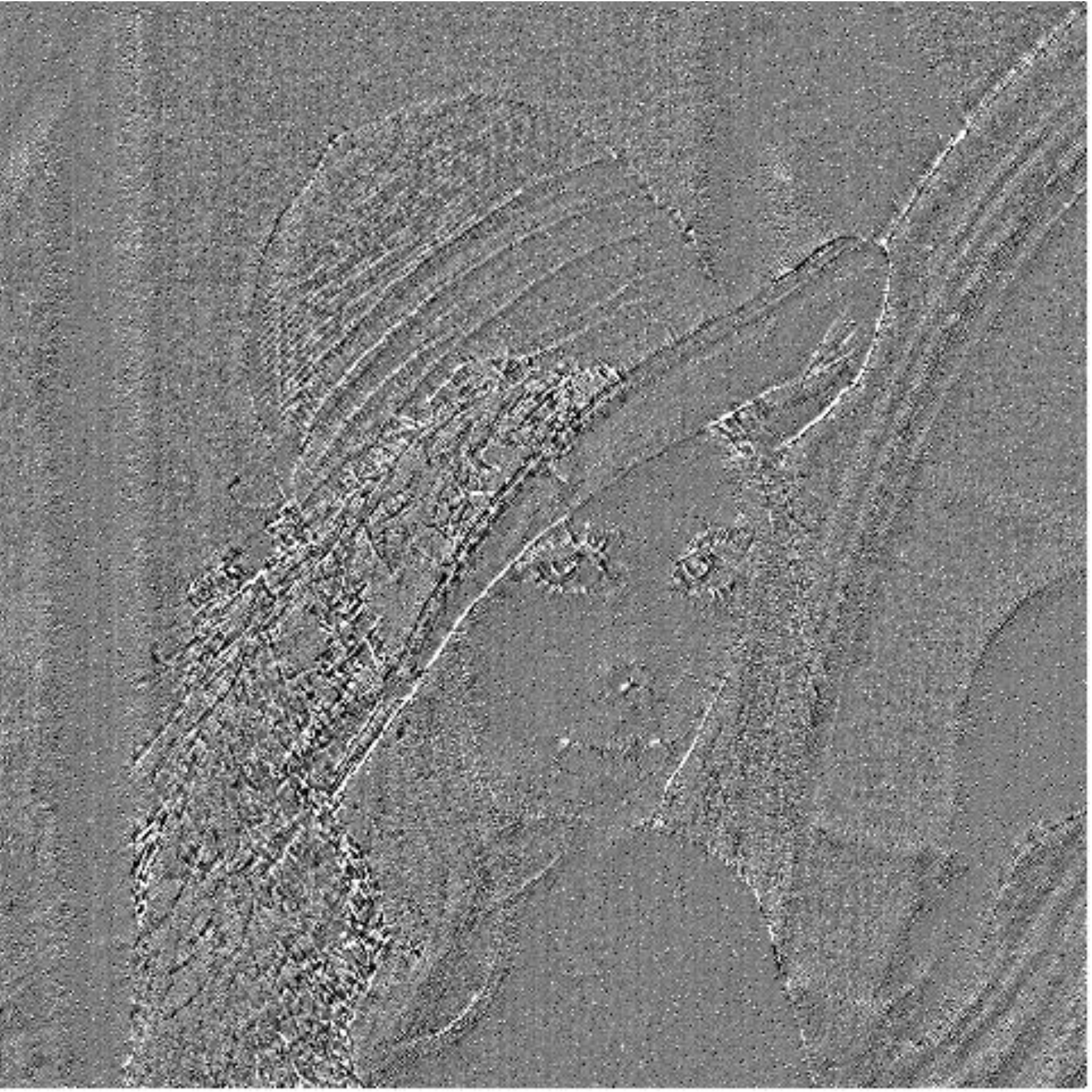}
\centerline{(f)}
\end{minipage} &
\begin{minipage}[c]{3.5cm}
\includegraphics[width=3.5cm, height=3.5cm]{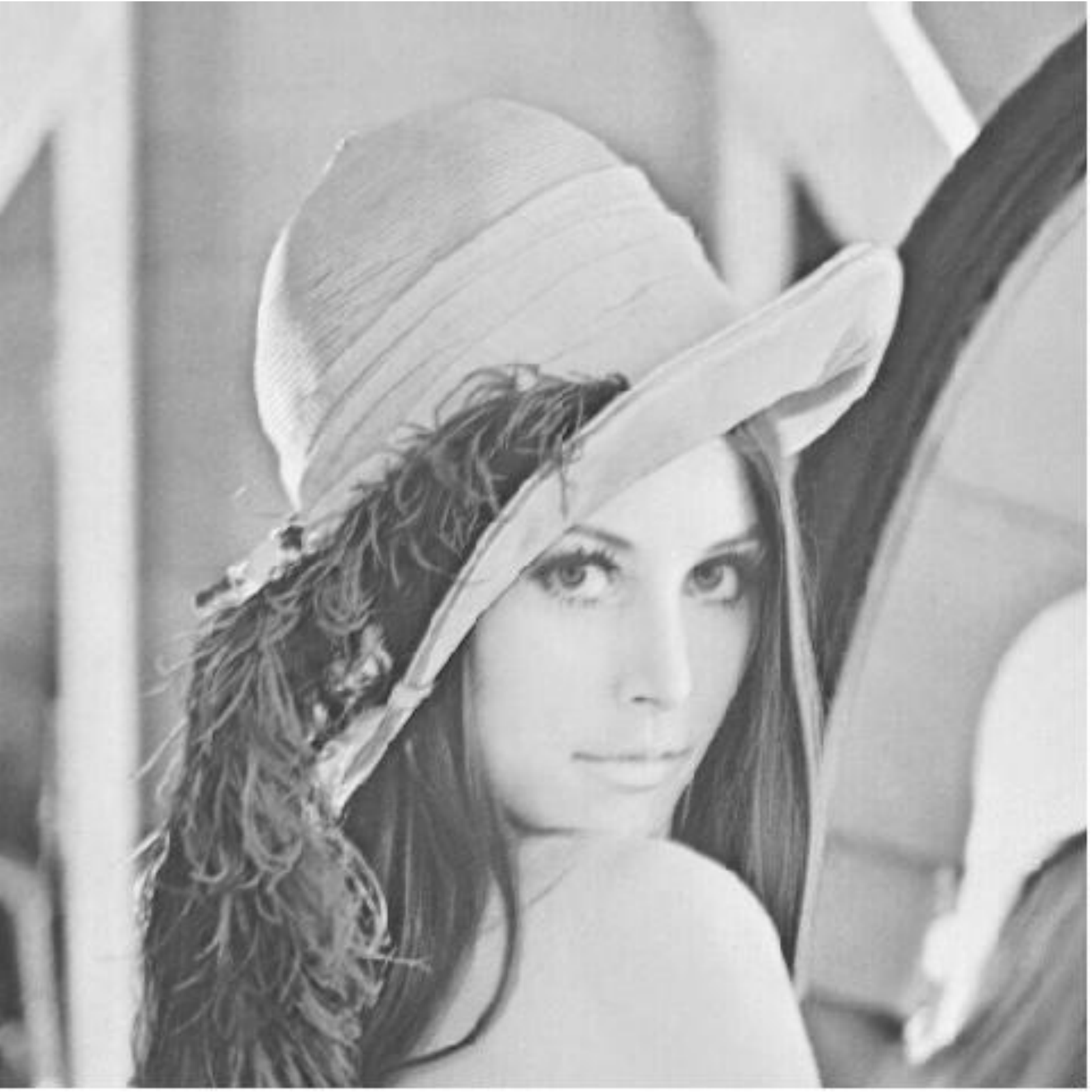}
\centerline{(i)}
\end{minipage} & 
\begin{minipage}[c]{3.5cm}
\includegraphics[width=3.5cm, height=3.5cm]{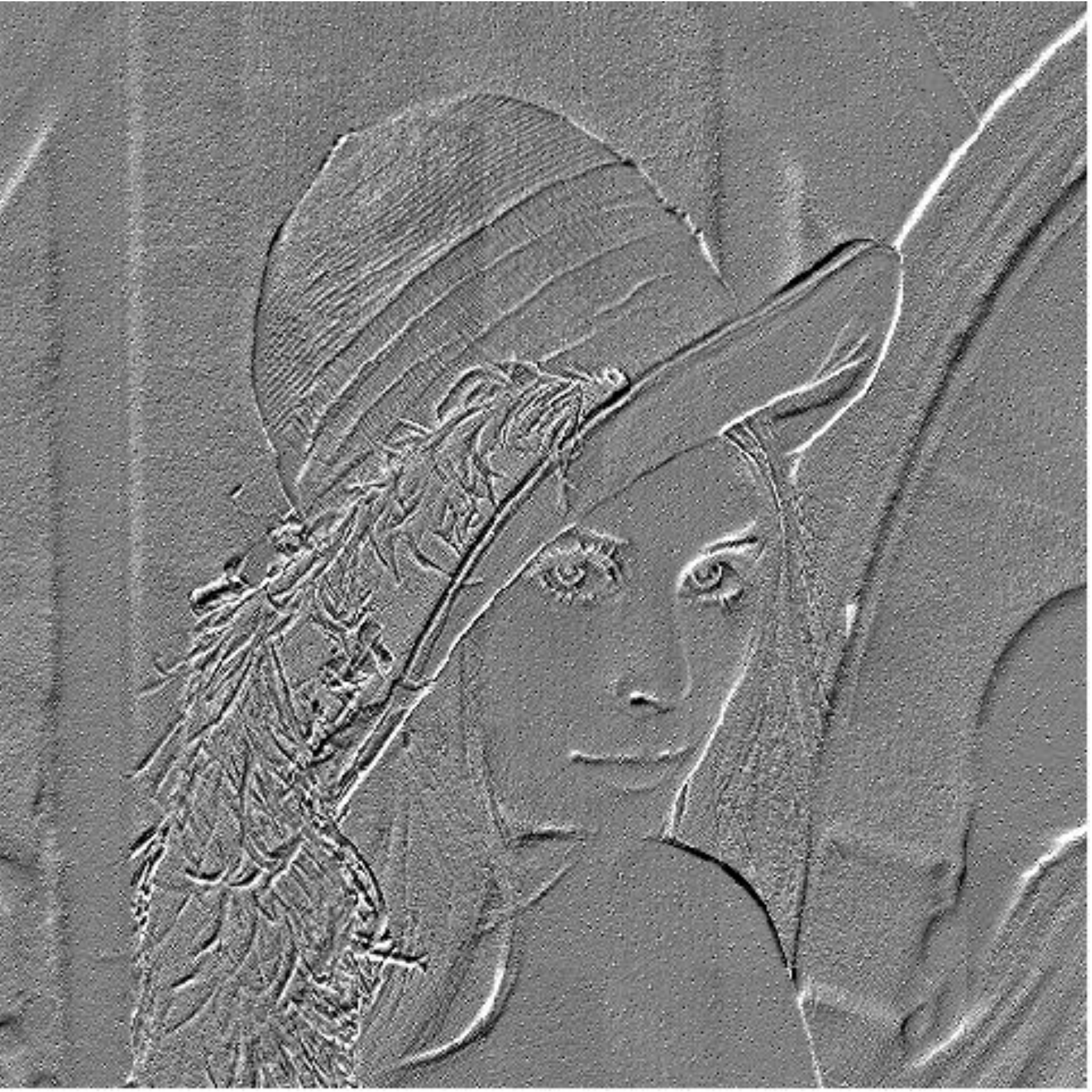}
\centerline{(l)}
\end{minipage}\\
\end{tabular}
\end{center}
\end{figure}

In the second experiment, the original image was 10\% additively contaminated (Fig. \ref{figLenaRec8y57} (II)), and we used it as input in Algorithm \ref{alg1}. We obtained four reconstructed images using the LS, GM and the BMM estimators. Next, the Algorithm \ref{alg2} was performed. The studies were carry out considering $8\times8,$ and $57\times 57$ window sizes. In the first two columns in Figure \ref{figLenaRec8y57}, we can observe the results obtained considering $8\times 8$ windows. Visually, there are not great differences between the different images reconstructed. When analyzing Table \ref{tabSimilLenavsRecVent8_cont}, we verify this because the measured indexes are comparable to each other. On the other hand, the third and fourth columns of Figure \ref{figLenaRec8y57} show the results obtained by adjusting $57\times 57$ windows. It is observed that the image (l), corresponding to the difference between the image restored with BMM (l) and the one contaminated with additive noise, highlights the edges slightly more.\\

%%%%%%%%%%%%%%%%%%%%%%%%%%%%%%%%
 
%%%%%%%%%%%%%%%%%%%%%%%%%%%%%%%%

\begin{table}[h!]
\begin{center}
\caption{Similarity between the original image and the reconstructions of Lena with additive contamination using $8\times 8$ window size (Figure \ref{figLenaRec8y57}-(I) vs. figures \ref{figLenaRec8y57}-(a),(b),(c)).}
\begin{tabular}[c]{c c c c}
\hline
Estimate & SSIM & CQ(1,1) & CQ$_{\max}$ \\
\hline
LS   & 0.9836079 & 0.8416351 & 0.9588826 \\ 
GM   & 0.9390820 & 0.7821954 & 0.9103257 \\ 
BMM  & 0.9846007 & 0.8328356 & 0.9577393 \\ \hline
\end{tabular}
\label{tabSimilLenavsRecVent8_cont}
\end{center}
\end{table}

%%%%%%%%%%%%%%%%%%%%%%%%%%%%%%%%%%%%%%%%%%%%%%%%%%%%%%%%%%%%%%%%%%%%%%%%%%%%%%%%%%%%%%%%%%%%%%%
\section{Conclusions and discussions}
\label{sec:concl}
A new estimator called BMM was proposed to estimate the parameters in first-order two-dimensional autoregressive models with three parameters. The new estimator is a two-dimensional extension of the BMM estimator proposed by \cite{Mull}, for autoregressive models of time series. We also extended the definition of replacement contamination, given for one-dimensional models (\cite{Maro}), to the case of AR-2D models; this type of contamination includes additive-type contamination. The performance of the proposed estimator for AR-2D models with replacement contamination and without contamination was analyzed. Besides, the new estimator was compared with the classical least square estimator (LS) and robust estimators M, GM and RA. The comparative analysis was performed from six experiments, each of which involved several Monte Carlo studies, considering different replacement contamination patterns and varying the level of contamination and window sizes of observation. The LS estimator produced estimates that are very sensitive to the presence of atypical values, while the other estimators had better results. Using Monte Carlo simulation, we observed that the GM, RA and BMM estimators are highly superior to the M and LS estimators. However, the new estimator showed the best behavior, in both accuracy and precision, followed by the RA estimator in accuracy and the GM in precision. An analysis of the computational cost showed that the RA estimator is the most expensive, followed by the BMM, GM, M and LS estimators, in that order. Finally, in an application to real data, we introduced a variant of the algorithm developed by \cite{Oje2}, to perform image segmentation, using an AR-2D model with three parameters, and BMM estimators. In the light of the examples shown in Section \ref{sec:apli}, we conclude that the adapted  algorithm is able to highlight the borders and contours in the images.\\ 
The following proposals outline some directions for future work. In \cite{Oje}, the author established the asymptotic normality and consistency of the robust RA estimator for the parameter ${\bm \phi}$ of a two-dimensional autoregressive process. Although the estimators M, GM and BMM are reasonable to estimate parameter ${\bm \phi}$, their asymptotic behavior is still an open problem. In the context of image processing, in this work, the difference between a real image and an BMM approximated image was computed. The resulting image could contribute to solve problems like that border detection,  classification and restoration of images. It would be interesting to explore the limitations of a segmentation method based on the difference image between a real and a fitted image. It is also important to analyze the behavior of the BMM estimator in combination with image restoration techniques. The same relevance has the study of properties of BMM estimator, in particular, and robust estimators, in general, as alternatives to the least square estimators under non causal and semi causal AR-2D models.
%%%%%%%%%%%%%%%%%%%%%%%%%%%%%%%%%%%%%%%%%

\section*{Acknowledgement}

We thank to  Ph. D. Oscar Bustos and Ph. D. Ronny Vallejos for helpful comments and suggestions. The authors were supported by Secyt-UNC grant (Res. Secyt 313/2016.), Argentina. The first author was partially supported by CIEM-CONICET, Argentina.
%%%%%%%%%%%%%%%%%%%%%%%%%%%%%%%%%%%%%%%%%%%%%%%

%%%%%%%%%%%%%%%%%%%%%%%%%%%%%%%%%%%%%%%%%%%%%%%%% Caso I y II
\newpage

%%%%%%%%%%%%%%%%%%%%%%%%%%%%%%%%%%%%%%%%%%%%%%%%% Caso I y II

%\begin{sidewaystable}
\begin{sidewaystable*}%[h!]
\section*{Appendix}
\begin{center}
%{\scriptsize
\caption{Case (I). Estimate of $\phi_{1}=0.15$, $\phi_{2}=0.17$ and $\phi_{3}=0.2$ in an AR-2D model without contamination.}
\label{tabSinCont}
\resizebox*{22cm}{6cm}{
\begin{tabular}[c]{l c | c c c c c | c c c c c | c c c c c}
\hline
  &  & \multicolumn{5}{c}{$\phi_1$} & \multicolumn{5}{c}{$\phi_2$} & \multicolumn{5}{c}{$\phi_3$}\\
\hline
N & & LS & M & GM & RA & BMM-2D & LS & M & GM & RA & BMM-2D & LS & M & GM & RA & BMM-2D\\
\hline
8 & $\hat{\phi}_{i}$ & 0.14642 & 0.14638 & 0.14758 & 0.14797 & 0.15260 & 0.16328 & 0.16474 & 0.16422 & 0.16959 & 0.16968 & 0.18854 & 0.18636 & 0.18907 & 0.19012 & 0.19083\\ 
    & $mse_{\phi_i}$ & 0.01832 & 0.01955 & 0.02082 & 0.02509 & 0.02789 & 0.01658 & 0.01871 & 0.01914 & 0.02366 & 0.02766& 0.02046 & 0.02252 & 0.02255 & 0.03273 & 0.03233 \\ 
    & $\hat{\sigma}^2_{\hat{\phi}_{i}}$ & 0.01835 & 0.01957 & 0.02085 & 0.02513 & 0.02794 & 0.01657 & 0.01872 & 0.01915 & 0.02370 & 0.02771 & 0.02037 & 0.02238 & 0.02247 & 0.03270 & 0.03231 \\ \hline
16 & $\hat{\phi}_{i}$ & 0.14998 & 0.15056 & 0.15081 & 0.15153 & 0.15390 & 0.16794 & 0.16598 & 0.16706 & 0.16771 & 0.16953 & 0.19751 & 0.19826 & 0.19811 & 0.19560 & 0.20218 \\ 
      & $mse_{\phi_i}$ & 0.00430 & 0.00502 & 0.00498 & 0.00525 & 0.00586 & 0.00408 & 0.00458 & 0.00489 & 0.00497 & 0.00512 & 0.00453 & 0.00485 & 0.00520 & 0.00589 & 0.00537\\ 
      & $\hat{\sigma}^2_{\hat{\phi}_{i}}$ & 0.00430 & 0.00503 & 0.00499 & 0.00526 & 0.00586 & 0.00408 & 0.00457 & 0.00490 & 0.00497 & 0.00513 & 0.00454 & 0.00485 & 0.00521 & 0.00589 & 0.00537\\   \hline
32 & $\hat{\phi}_{i}$ & 0.15014 & 0.15025 & 0.14993 & 0.15053 & 0.15127 & 0.16881 & 0.16884 & 0.16906 & 0.16930 & 0.17019 & 0.19897 & 0.19842 & 0.19851 & 0.19828 & 0.19945\\ 
      & $mse_{\phi_i}$ & 0.00099 & 0.00108 & 0.00110 & 0.00113 & 0.00115 & 0.00099 & 0.00106 & 0.00114 & 0.00111 & 0.00108 & 0.00110 & 0.00120 & 0.00123 & 0.00128 & 0.00124\\ 
      & $\hat{\sigma}^2_{\hat{\phi}_{i}}$ & 0.00099 & 0.00108 & 0.00110 & 0.00113 & 0.00115 & 0.00099 & 0.00106 & 0.00114 & 0.00111 & 0.00109 & 0.00110 & 0.00120 & 0.00123 & 0.00128 & 0.00124\\  \hline
57 & $\hat{\phi}_{i}$ & 0.14937 & 0.14937 & 0.14951 & 0.14942 & 0.14956 & 0.16864 & 0.16840 & 0.16845 & 0.16851 & 0.16858 & 0.19934 & 0.19971 & 0.19962 & 0.19944 & 0.19944\\ 
      & $mse_{\phi_i}$ & 0.00031 & 0.00032 & 0.00033 & 0.00033 & 0.00033 & 0.00030 & 0.00031 & 0.00032 & 0.00033 & 0.00031 & 0.00031 & 0.00033 & 0.00034 & 0.00035 & 0.00033\\ 
      & $\hat{\sigma}^2_{\hat{\phi}_{i}}$ & 0.00031 & 0.00032 & 0.00033 & 0.00033 & 0.00033 & 0.00030 & 0.00031 & 0.00032 & 0.00033 & 0.00031 & 0.00031 & 0.00033 & 0.00034 & 0.00035 & 0.00033\\ \hline
\end{tabular}}
%\end{sidewaystable}
%\end{table*}
\vspace{2cm}
%{\scriptsize
\caption{Case (II). Estimate of $\phi_{1}=0.15$, $\phi_{2}=0.17$ and $\phi_{3}=0.2$  in an AR-2D model with additive contamination of $\mu =0$ and $\sigma^2=50$.}
\label{tabContAdit}
\resizebox*{22cm}{6cm}{
\begin{tabular}[c]{l c| c c c c c |c c c c c |c c c c c}
\hline
  &  & \multicolumn{5}{c}{$\phi_1$} & \multicolumn{5}{c}{$\phi_2$} & \multicolumn{5}{c}{$\phi_3$}\\
\hline
N & & LS & M & GM & RA & BMM-2D & LS & M & GM & RA & BMM-2D & LS & M & GM & RA & BMM-2D\\
\hline
8 & $\hat{\phi}_{i}$ & 0.04048 & 0.05018 & 0.11614 & {\bf 0.12502} & 0.10983 & 0.04915 & 0.04851 & 0.12014 & {\bf 0.12216} & 0.11082 & 0.06266 & 0.06973 & {\bf 0.15827} & 0.12456 & 0.11873\\ 
    & $mse_{\phi_i}$ & 0.03170 & {\bf 0.02101} & 0.02208 & 0.06115 & 0.02689 & 0.03560 & {\bf 0.02513} & 0.02582 & 0.06229 & 0.02848 & 0.04177 & {\bf 0.02690} & {\bf 0.02690} & 0.05767 & 0.02989\\ 
    & $\hat{\sigma}^2_{\hat{\phi}_{i}}$ & 0.01975 & {\bf 0.01107} & 0.02098 & 0.06065 & 0.02533 & 0.02104 & {\bf 0.01039} & 0.02338 & 0.06012 & 0.02503 & 0.02296 & {\bf 0.00995} & 0.02521 & 0.05208 & 0.02333\\ \hline
16 & $\hat{\phi}_{i}$ & 0.03640 & 0.03670 & 0.10622 & {\bf 0.17498} & 0.12390 & 0.03880 & 0.03902 & 0.11819 & {\bf 0.19017} & 0.13301 & 0.05228 & 0.05043 & 0.15777 & {\bf 0.16202} & 0.13817\\ 
      & $mse_{\phi_i}$ & 0.01750 & 0.01434 & {\bf 0.00669} & 0.01389 & 0.00729 & 0.02094 & 0.01872 & {\bf 0.00731} & 0.01249 & 0.00884 & 0.02581 & 0.02413 & {\bf 0.00677} & 0.01153 & 0.01208\\ 
      & $\hat{\sigma}^2_{\hat{\phi}_{i}}$ & 0.00460 & {\bf 0.00151} & 0.00478 & 0.01329 & 0.00662 & 0.00374 & {\bf 0.00157} & 0.00463 & 0.01211 & 0.00749 & 0.00399 & {\bf 0.00176} & 0.00499 & 0.01010 & 0.00827\\   \hline
32 & $\hat{\phi}_{i}$ & 0.03426 & 0.03537 & 0.11038 & 0.19341 & {\bf 0.13797} & 0.03831 & 0.04006 & 0.12820 & 0.21266 & {\bf 0.16113} & 0.04690 & 0.04852 & 0.15217 & 0.15033 & {\bf 0.16094}\\ 
      & $mse_{\phi_i}$ & 0.01446 & 0.01344 & 0.00259 & 0.00522 & {\bf 0.00180} & 0.01843 & 0.01721 & 0.00291 & 0.00588 & {\bf 0.00180} & 0.02460 & 0.02329 & {\bf 0.00352} & 0.00475 & 0.00354\\ 
      & $\hat{\sigma}^2_{\hat{\phi}_{i}}$ & 0.00107 & {\bf 0.00030} & 0.00103 & 0.00335 & 0.00166 & 0.00110 & {\bf 0.00033} & 0.00116 & 0.00407 & 0.00173 & 0.00117 & {\bf 0.00034} & 0.00124 & 0.00229 & 0.00202\\  \hline
57 & $\hat{\phi}_{i}$ & 0.03397 & 0.03567 & 0.11288 & 0.20015 & {\bf 0.14520} & 0.03860 & 0.03888 & 0.12653 & 0.21397 & {\bf 0.16261} & 0.04609 & 0.04812 & 0.15209 & 0.14793 & {\bf 0.17444}\\ 
      & $mse_{\phi_i}$ & 0.01378 & 0.01316 & 0.00172 & 0.00386 & {\bf 0.00049} & 0.01759 & 0.01729 & 0.00222 & 0.00332 & {\bf 0.00057} & 0.02401 & 0.02317 & 0.00266 & 0.00357 & {\bf 0.00125}\\ 
      & $\hat{\sigma}^2_{\hat{\phi}_{i}}$ & 0.00032 & {\bf 0.00009} & 0.00034 & 0.00135 & 0.00047 & 0.00032 & {\bf 0.00010} & 0.00033 & 0.00139 & 0.00052 & 0.00032 & {\bf 0.00010} & 0.00037 & 0.00086 & 0.00060\\ \hline
\end{tabular}}
\end{center}
\end{sidewaystable*}
\clearpage
\clearpage
\newpage
%\newcolumn
%%%%%%%%%%%%%%%%%%%%%%%%%%%%%%%%%%%%%%%%%%%%%%%%%% Caso III y IV
\begin{sidewaystable*}
\begin{center}
%\scriptsize{
\caption{Case (II): Estimate of $\phi_{1}=0.15$, $\phi_{2}=0.17$ and $\phi_{3}=0.2$ in an AR-2D model with additive contamination, window size $32\times 32$.}
\label{tabContAditPorc}
\resizebox*{22cm}{6cm}{
\begin{tabular}[c]{l c| c c c c c |c c c c c |c c c c c}
\hline
  &  & \multicolumn{5}{c}{$\phi_1$} & \multicolumn{5}{c}{$\phi_2$} & \multicolumn{5}{c}{$\phi_3$}\\
\hline
\% & & LS & M & GM & RA & BMM-2D & LS & M & GM & RA & BMM-2D & LS & M & GM & RA & BMM-2D\\
\hline
5\% & $\hat{\phi}_{i}$ & 0.05354 & 0.05835 & 0.12624 & 0.17486 & {\bf 0.14533} & 0.06474 & 0.06533 & 0.14652 & 0.19550 & {\bf 0.16565} & 0.07732 & 0.08454 & 0.17818 & 0.17824 & {\bf 0.18322}\\ 
    & $mse_{\phi_i}$ & 0.01035 & 0.00893 & 0.00164 & 0.00235 & {\bf 0.00139} & 0.012162 & 0.01153 & 0.00173 & 0.00245 & {\bf 0.00155} & 0.01613 & 0.01408 & {\bf 0.00164} & 0.00209 & 0.00188\\ 
    & $\hat{\sigma}^2_{\hat{\phi}_{i}}$ & 0.00104 & {\bf 0.00053} & 0.00108 & 0.00174 & 0.00137 & 0.00108 & {\bf 0.00058} & 0.00119 & 0.00181 & 0.00153 & 0.00108 & {\bf 0.00075} & 0.00116 & 0.00162 & 0.00160\\ \hline
10\% & $\hat{\phi}_{i}$ & 0.03426 & 0.03537 & 0.11038 & 0.19341 & {\bf 0.13797} & 0.03831 & 0.04006 & 0.12820 & 0.21266 & {\bf 0.16113} & 0.04690 & 0.04852 & 0.15217 &  0.15033 & {\bf 0.16094}\\ 
      & $mse_{\phi_i}$ & 0.01446 & 0.01344 & 0.00259 & 0.00522 & {\bf 0.00180} & 0.01843 & 0.01721 & 0.00291 & 0.00588 & {\bf 0.00180} & 0.02460 & 0.02329 & {\bf 0.00352} & 0.00475 & 0.00354\\ 
      & $\hat{\sigma}^2_{\hat{\phi}_{i}}$ & 0.00107 & {\bf 0.00030} & 0.00103 & 0.00335 & 0.00166 & 0.00110 & {\bf 0.00033} & 0.00116 & 0.00407 & 0.00173 & 0.00117 & {\bf 0.00034} & 0.00124 & 0.00229 & 0.00202\\   \hline
15\% & $\hat{\phi}_{i}$ & 0.02257 & 0.02493 &  0.09578 & 0.19227 & {\bf 0.13160} & 0.02793 & 0.02815 & 0.10832 & 0.20706 & {\bf 0.14783} & 0.03318 & 0.03389 & 0.13131 & 0.12771 & {\bf 0.14057}\\ 
      & $mse_{\phi_i}$ & 0.01729 & 0.01585 & 0.00397 & 0.00606 & {\bf 0.00232} & 0.02115 & 0.02035 & 0.00484 & 0.00628 & {\bf 0.00249} & 0.02881 & 0.02780 & 0.00582 & 0.00782 & {\bf 0.00553}\\ 
      & $\hat{\sigma}^2_{\hat{\phi}_{i}}$ & 0.00106 & {\bf 0.00021} & 0.00103 & 0.00428 & 0.00198 & 0.00096 & {\bf 0.00023} & 0.00103 & 0.00492 & 0.00200 & 0.00099 & {\bf 0.00021} & 0.00110 & 0.00259 & 0.00200\\  \hline
20\% & $\hat{\phi}_{i}$ & 0.01766 & 0.01886 & 0.08255 & {\bf 0.16534} & 0.12508 & 0.02144 & 0.02021 & 0.09349 & {\bf 0.17371} & 0.13534 & 0.02511 & 0.02647 & 0.11064 & 0.11122 & {\bf 0.12291}\\ 
      & $mse_{\phi_i}$ & 0.01852 & 0.01737 & 0.00580 & 0.00612 & {\bf 0.00273} & 0.02306 & 0.02259 & 0.00704 & 0.00566 & {\bf 0.00336} & 0.03155 & 0.03031 & 0.00919 & 0.01082 & {\bf 0.00820}\\ 
      & $\hat{\sigma}^2_{\hat{\phi}_{i}}$ & 0.00101 & {\bf 0.00017} & 0.00125 & 0.00589 & 0.00211 & 0.00100 & {\bf 0.00016} & 0.00118 & 0.00566 & 0.00216 & 0.00096 & {\bf 0.00020} & 0.00121 & 0.00295 & 0.00226\\ \hline
\end{tabular}}
\vspace{2cm}
%{\scriptsize
\caption{Case (III): Estimate of $\phi_{1}=0.15$, $\phi_{2}=0.17$ and $\phi_{3}=0.2$ in an AR-2D model with t-student contamination with 2.3 d.f., window size 57.}
\label{tabContStudent}
\resizebox*{22cm}{6cm}{
\begin{tabular}[c]{l c| c c c c c |c c c c c |c c c c c}
\hline
  &  & \multicolumn{5}{c}{$\phi_1$} & \multicolumn{5}{c}{$\phi_2$} & \multicolumn{5}{c}{$\phi_3$}\\
\hline
\% & & LS & M & GM & RA & BMM-2D & LS & M & GM & RA & BMM-2D & LS & M & GM & RA & BMM-2D\\
\hline
5\% & $\hat{\phi}_{i}$ & 0.12843 & 0.13359 & 0.14418 & 0.14803 & {\bf 0.14869} & 0.14606 & 0.15290 & 0.16332 & 0.16762 & {\bf 0.16841} & 0.17005 & 0.18190 & 0.19197 & 0.19234 & {\bf 0.19414}\\ 
    & $mse_{\phi_i}$ & 0.00099 & 0.00063 & 0.00038 & {\bf 0.00035} & 0.00044 & 0.00113 & 0.00064 & 0.00036 & {\bf 0.00033} & 0.00044 & 0.00157 & 0.00070 & {\bf 0.00042} & 0.00044 & 0.00047\\ 
    & $\hat{\sigma}^2_{\hat{\phi}_{i}}$ & 0.00053 & 0.00036 & {\bf 0.00035} & {\bf 0.00035} & 0.00043 & 0.00055 & 0.00035 & {\bf 0.00032} & 0.00033 & 0.00043 & 0.00067 & 0.00037 & {\bf 0.00036} & 0.00038 & 0.00044\\ \hline
10\% & $\hat{\phi}_{i}$ & 0.11243 & 0.11936 & 0.13663 & 0.14450 & {\bf 0.14653} & 0.12495 & 0.13453 & 0.15399 & 0.16093 & {\bf 0.16367} & 0.15040 & 0.16673 & 0.18506 & 0.18625 & {\bf 0.19089}\\ 
      & $mse_{\phi_i}$ & 0.00193 & 0.00133 & 0.00056 & {\bf 0.00042} & 0.00048 & 0.00271 & 0.00169 & 0.00060 & {\bf 0.00042} & 0.00049 & 0.00322 & 0.00159 & 0.00059 & 0.00057 & {\bf 0.00056}\\ 
      & $\hat{\sigma}^2_{\hat{\phi}_{i}}$ & 0.00052 & 0.00039 & {\bf 0.00038} & 0.00039 & 0.00047 & 0.00068 & 0.00043 & 0.00035 & {\bf 0.00034} & 0.00045 & 0.00076 & 0.00048 & {\bf 0.00037} & 0.00038 & 0.00047\\   \hline
15\% & $\hat{\phi}_{i}$ & 0.09909 & 0.10950 & 0.13231 & 0.14231 & {\bf 0.14422} & 0.11271 & 0.12404 & 0.14860 & 0.15876 & {\bf 0.16274} & 0.13272 & 0.14937 & 0.17524 & 0.17647 & {\bf 0.18357}\\ 
      & $mse_{\phi_i}$ & 0.00318 & 0.00207 & 0.00064 & {\bf 0.00041} & 0.00050 & 0.00396 & 0.00257 & 0.00081 & {\bf 0.00049} & 0.00058 & 0.00530 & 0.00300 & 0.00096 & 0.00092 & {\bf 0.00075}\\ 
      & $\hat{\sigma}^2_{\hat{\phi}_{i}}$ & 0.00059 & 0.00043 & {\bf 0.00033} & 0.00035 & 0.00047 & 0.00068 & 0.00046 & {\bf 0.00035} & 0.00037 & 0.00053 & 0.00078 & 0.00044 & {\bf 0.00035} & 0.00037 & 0.00049\\  \hline
20\% & $\hat{\phi}_{i}$ & 0.08895 & 0.09830 & 0.12530 & 0.13771 & {\bf 0.14167} & 0.09962 & 0.11246 & 0.14363 & 0.15533 & {\bf 0.16191} & 0.11802 & 0.13740 & 0.16826 & 0.16956 & {\bf 0.18116}\\ 
      & $mse_{\phi_i}$ & 0.00426 & 0.00308 & 0.00095 & {\bf 0.00052} & {\bf 0.00052} & 0.00556 & 0.00370 & 0.00099 & 0.00053 & {\bf 0.00042} & 0.00753 & 0.00434 & 0.00137 & 0.00132 & {\bf 0.00087}\\ 
      & $\hat{\sigma}^2_{\hat{\phi}_{i}}$ & 0.00053 & 0.00040 & {\bf 0.00034} & 0.00037 & 0.00045 & 0.00061 & 0.00039 & {\bf 0.00029} & 0.00031 & 0.00036 & 0.00081 & 0.00042 & {\bf 0.00037} & 0.00040 & 0.00052\\ \hline
\end{tabular}}
\end{center}
\end{sidewaystable*}
\newpage
\clearpage
\newpage
%%%%%%%%%%%%%%%%%%%%%%%%%%%%%%%%%%%%%%%%%%%%%%%%%% Caso V y VI
\begin{sidewaystable*}
\begin{center}
%{\scriptsize
\caption{Case (IV). Estimate of $\phi_{1}=0.15$, $\phi_{2}=0.17$ and $\phi_{3}=0.2$ in an AR-2D model with replacement contamination by another AR process of parameters $\tilde{\phi}_1=0.1$, $\tilde{\phi}_2=0.2$ and $\tilde{\phi}_3=0.3$, window size 32.}
\label{tabContReemARPorc}
\resizebox*{22cm}{6cm}{
\begin{tabular}[c]{l c| c c c c c |c c c c c |c c c c c}
\hline
  &  & \multicolumn{5}{c}{$\phi_1$} & \multicolumn{5}{c}{$\phi_2$} & \multicolumn{5}{c}{$\phi_3$}\\
\hline
\% & & LS & M & GM & RA & BMM-2D & LS & M & GM & RA & BMM-2D & LS & M & GM & RA & BMM-2D\\
\hline
5\% & $\hat{\phi}_{i}$ & 0.13738 & 0.13788 & 0.13763 & 0.13795 & 0.13951 & 0.15559 & 0.15667 & 0.15660 & 0.15707 & 0.15873 & 0.18439 & 0.18481 & 0.18530 & 0.18601 & 0.18667\\ 
    & $mse_{\phi_i}$ & 0.00115 & 0.00119 & 0.00120 & 0.00125 & 0.00116 & 0.00123 & 0.00125 & 0.00133 & 0.00128 & 0.00127 & 0.00126 & 0.00131 & 0.00140 & 0.00138 & 0.00128\\ 
    & $\hat{\sigma}^2_{\hat{\phi}_{i}}$ & 0.00099 & 0.00104 & 0.00105 & 0.00111 & 0.00105 & 0.00102 & 0.00108 & 0.00115 & 0.00112 & 0.00114 & 0.00102 & 0.00108 & 0.00119 & 0.00118 & 0.00111\\ \hline
10\% & $\hat{\phi}_{i}$ & 0.12977 & 0.13134 & 0.13044 & 0.13186 & 0.13395 & 0.14436 & 0.14571 & 0.14477 & 0.14603 & 0.14756 & 0.17055 & 0.17244 & 0.17132 & 0.17310 & 0.17487\\ 
      & $mse_{\phi_i}$ & 0.00136 & 0.00140 & 0.00145 & 0.00139 & 0.00136 & 0.00166 & 0.00165 & 0.00173 & 0.00170 & 0.00160 & 0.00192 & 0.00191 & 0.00199 & 0.00192 & 0.00187\\ 
      & $\hat{\sigma}^2_{\hat{\phi}_{i}}$ & 0.00096 & 0.00105 & 0.00107 & 0.00106 & 0.00111 & 0.00100 & 0.00107 & 0.00109 & 0.00113 & 0.00110 & 0.00106 & 0.00116 & 0.00117 & 0.00120 & 0.00124\\   \hline
15\% & $\hat{\phi}_{i}$ & 0.11879 & 0.12068 & 0.11975 & 0.12008 & 0.12333 & 0.13053 & 0.13242 & 0.13282 & 0.13331 & 0.13515 & 0.15953 & 0.16200 & 0.16142 & 0.16245 & 0.16436\\ 
      & $mse_{\phi_i}$ & 0.00197 & 0.00194 & 0.00203 & 0.00200 & 0.00188 & 0.00256 & 0.00249 & 0.00251 & 0.00247 & 0.00234 & 0.00255 & 0.00245 & 0.00249 & 0.00250 & 0.00236\\ 
      & $\hat{\sigma}^2_{\hat{\phi}_{i}}$ & 0.00100 & 0.00108 & 0.00112 & 0.00111 & 0.00117 & 0.00100 & 0.00108 & 0.00113 & 0.00113 & 0.00113 & 0.00091 & 0.00101 & 0.00100 & 0.00109 & 0.00110\\  \hline
20\% & $\hat{\phi}_{i}$ & 0.11002 & 0.11141 & 0.11104 & 0.11190 & 0.11384 & 0.11989 & 0.12046 & 0.11980 & 0.12060 & 0.12338 & 0.15032 & 0.15283 & 0.15146 & 0.15289 & 0.15556\\ 
      & $mse_{\phi_i}$ & 0.00257 & 0.00251 & 0.00253 & 0.00250 & 0.00238 & 0.00356 & 0.00362 & 0.00369 & 0.00364 & 0.00344 & 0.00344 & 0.00337 & 0.00348 & 0.00344 & 0.00317\\ 
      & $\hat{\sigma}^2_{\hat{\phi}_{i}}$ & 0.00097 & 0.00102 & 0.00101 & 0.00105 & 0.00107 & 0.00105 & 0.00117 & 0.00117 & 0.00120 & 0.00127 & 0.00097 & 0.00115 & 0.00113 & 0.00123 & 0.00120\\ \hline
\end{tabular}}
\vspace{2cm}
%{\scriptsize
\caption{Case (V). Estimate of $\phi_{1}=0.15$, $\phi_{2}=0.17$ and $\phi_{3}=0.2$ in an AR-2D model with replacement contamination by a white noise process of variance 50, window size 32.}
\label{tabContReemRBPorc}
\resizebox*{22cm}{6cm}{
\begin{tabular}[c]{l c| c c c c c |c c c c c |c c c c c}
\hline
  &  & \multicolumn{5}{c}{$\phi_1$} & \multicolumn{5}{c}{$\phi_2$} & \multicolumn{5}{c}{$\phi_3$}\\
\hline
\% & & LS & M & GM & RA & BMM-2D & LS & M & GM & RA & BMM-2D & LS & M & GM & RA & BMM-2D\\
\hline
5\% & $\hat{\phi}_{i}$ & 0.05141 & 0.05608 & 0.12961 & 0.18022 & 0.14770 & 0.05960 & 0.06437 & 0.14453 & 0.19548 & 0.16614 & 0.06973 & 0.07972 & 0.17067 & 0.17307 & 0.17888\\ 
    & $mse_{\phi_i}$ & 0.01085 & 0.00936 & 0.00152 & 0.00282 & 0.00141 & 0.01315 & 0.01174 & 0.00167 & 0.00230 & 0.00131 & 0.01828 & 0.01528 & 0.00203 & 0.00249 & 0.00205\\ 
    & $\hat{\sigma}^2_{\hat{\phi}_{i}}$ & 0.00113 & 0.00054 & 0.00110 & 0.00191 & 0.00141 & 0.00097 & 0.00058 & 0.00103 & 0.00166 & 0.00130 & 0.00132 & 0.00082 & 0.00117 & 0.00177 & 0.00161\\ \hline
10\% & $\hat{\phi}_{i}$ & 0.02990 & 0.03274 & 0.10976 & 0.19717 & 0.13843 & 0.03144 & 0.03431 & 0.12180 & 0.20953 & 0.15686 & 0.03895 & 0.04424 & 0.14775 & 0.14994 & 0.16107\\ 
      & $mse_{\phi_i}$ & 0.01550 & 0.01409 & 0.00274 & 0.00560 & 0.00178 & 0.02037 & 0.01871 & 0.00349 & 0.00467 & 0.00182 & 0.02701 & 0.02462 & 0.00392 & 0.00485 & 0.00347\\ 
      & $\hat{\sigma}^2_{\hat{\phi}_{i}}$ & 0.00108 & 0.00034 & 0.00112 & 0.00338 & 0.00165 & 0.00117 & 0.00030 & 0.00116 & 0.00311 & 0.00165 & 0.00107 & 0.00036 & 0.00119 & 0.00235 & 0.00196\\   \hline
15\% & $\hat{\phi}_{i}$ & 0.01994 & 0.02190 & 0.09124 & 0.19055 & 0.12807 & 0.01905 & 0.02397 & 0.10452 & 0.20163 & 0.14848 & 0.02439 & 0.02998 & 0.12630 & 0.12875 & 0.14057\\ 
      & $mse_{\phi_i}$ & 0.01796 & 0.01666 & 0.00453 & 0.00617 & 0.00237 & 0.02382 & 0.02154 & 0.00540 & 0.00536 & 0.00248 & 0.03185 & 0.02916 & 0.00673 & 0.00769 & 0.00576\\ 
      & $\hat{\sigma}^2_{\hat{\phi}_{i}}$ & 0.00104 & 0.00025 & 0.00108 & 0.00454 & 0.00190 & 0.00103 & 0.00022 & 0.00112 & 0.00436 & 0.00202 & 0.00101 & 0.00025 & 0.00130 & 0.00262 & 0.00224\\  \hline
20\% & $\hat{\phi}_{i}$ & 0.01369 & 0.01590 & 0.07760 & 0.16684 & 0.12467 & 0.01536 & 0.01802 & 0.08600 & 0.17541 & 0.13500 & 0.01779 & 0.02225 & 0.10617 & 0.10972 & 0.12265\\ 
      & $mse_{\phi_i}$ & 0.01970 & 0.01816 & 0.00627 & 0.00464 & 0.00308 & 0.02498 & 0.02326 & 0.00824 & 0.00463 & 0.00345 & 0.03430 & 0.03180 & 0.00991 & 0.01091 & 0.00807\\ 
      & $\hat{\sigma}^2_{\hat{\phi}_{i}}$ & 0.00112 & 0.00018 & 0.00103 & 0.00436 & 0.00245 & 0.00106 & 0.00016 & 0.00119 & 0.00461 & 0.00223 & 0.00110 & 0.00021 & 0.00111 & 0.00277 & 0.00209\\ \hline
\end{tabular}}
\end{center}
\end{sidewaystable*}
\clearpage
\newpage

%\begin{acknowledgements}
%\end{acknowledgements}

%\bibliography{ref}   % name your BibTeX data base

\begin{thebibliography}{99}


\bibitem{achim2006sar} Achim A, Kuruoglu EE, Zerubia J., SAR image filtering based on the heavy-tailed rayleigh model. IEEE Transactions on Image Processing 15(9):2686-2693 (2006).
\bibitem{All1} Allende H, Galbiati J. A non-parametric filter for digital image restoration, using cluster analysis. Pattern Recognition Letters 25(8):841-847 (2004).
\bibitem{All4} Allende H, Galbiati J, Vallejos R. Digital image restoration using autoregressive time series type models. Bulletin European Spatial Agency 434:53-59 (1998).
\bibitem{All2} Allende H, Galbiati J, Vallejos R. Robust image modeling on image processing. Pattern Recognition Letters 22(11):1219-1231 (2001).
\bibitem{Ays} Aysal TC, Barner KE. Quadratic weighted median filters for edge enhancement of noisy images.
IEEE Transactions on Image Processing 15(11):3294-3310 (2006).
\bibitem{Bar} Baran S, Pap G, van Zuijlen MC. Asymptotic inference for a nearly unstable sequence of stationary spatial ar models. Statistics and probability letters 69(1):53-61 (2004).
\bibitem{Bas} Basu S, Reinsel GC. Properties of the spatial unilateral first-order arma model. Advances in applied Probability 25(3):631-648 (1993).
\bibitem{bhandari2015improved} Bhandari A, Kumar A, Singh G. Improved knee transfer function and gamma correction based method for contrast and brightness enhancement of satellite image. AEU-International Journal of Electronics and Communications 69(2):579-589 (2015).
\bibitem{Bus1} Bustos O. Robust statistics in sar image processing. In: Image Processing Techniques, First Latino-American Seminar on Radar Remote Sensing, vol 407, p 81 (1997).
\bibitem{Bus4} Bustos, Oscar and Fraiman, Ricardo and Yohai, Victor J. (1984). Asymptotic behaviour of the estimates based on residual autocovariances for ARMA models. Robust and Nonlinear Time Series Analysis, Springer, pp. 26-49.
\bibitem{Bus6} Bustos O, Ojeda S, Vallejos R. Spatial arma models and its applications to image filtering. Brazilian Journal of Probability and Statistics pp 141-165 (2009).
\bibitem{Bus3} Bustos OH, Yohai VJ. Robust estimates for arma models. Journal of the American Statistical Association 81(393):155-168 (1986).
\bibitem{Bus5} Bustos OH, Ruiz M, Ojeda S, Vallejos R, Frery AC. Asymptotic behavior of ra-estimates in
autoregressive 2d processes. Journal of Statistical Planning and Inference 139(10):3649-3664 (2009).

\bibitem{Com} Comport AI, Marchand E, Chaumette F. Statistically robust 2-d visual servoing. IEEE Transactions on Robotics 22(2):415-420 (2006).
\bibitem{Got} Gottardo R, Raftery AE, Yee Yeung K, Bumgarner RE. Bayesian robust inference for differential gene expression in microarrays with multiple samples. Biometrics 62(1):10-18 (2006).
\bibitem{Ben} Hamza AB, Krim H. Image denoising: A nonlinear robust statistical approach. IEEE transactions on signal processing 49(12):3045-3054 (2001).
\bibitem{Hua} Huang HC, Lee TC. Data adaptive median filters for signal and image denoising using a generalized sure criterion. IEEE Signal Processing Letters 13(9):561-564 (2006).
\bibitem{Hub} Huber PJ, et al. Robust estimation of a location parameter. The Annals of Mathematical Statistics 35(1):73-101 (1964).
\bibitem{Ji} Ji Z, Huang Y, Xia Y, Zheng Y. A robust modified gaussian mixture model with rough set for image segmentation. Neurocomputing (2017).
\bibitem{Kas} Kashyap RL, Eom KB. Robust image modeling techniques with an image restoration application. IEEE Transactions on Acoustics, Speech, and Signal Processing 36(8):1313-1325 (1988).
\bibitem{Kim} Kim JH, Han JH Outlier correction from uncalibrated image sequence using the triangulation method. Pattern recognition 39(3):394-404 (2006).
\bibitem{Tzu} Lin TC. A new adaptive center weighted median filter for suppressing impulsive noise in images. Information Sciences 177(4):1073-1087 (2007).
\bibitem{Maro} Maronna R, Martin RD, Yohai V. Robust statistics. John Wiley and Sons, Chichester. ISBN  (2006)
\bibitem{Mull} Muler N, Pena D, Yohai VJ, et al. Robust estimation for arma models. The Annals of Statistics 37(2):816-840 (2009).
\bibitem{Oje} Ojeda S. Robust RA estimators for bidimensional autoregressive models. PhD thesis, Universidad Nacional de Córdoba. Facultad de Matemática, Astronomı́a y Fı́sica (1999).
\bibitem{Oje2} Ojeda S, Vallejos R, Bustos O. A new image segmentation algorithm with applications to image inpainting. Computational Statistics and Data Analysis 54(9):2082-2093 (2010).
\bibitem{Oje4} Ojeda SM, Vallejos RO, Lucini MM. Performance of robust ra estimator for bidimensional autoregressive models. Journal of Statistical Computation and Simulation 72(1):47-62 (2002).
\bibitem{Oje3} Ojeda SM, Vallejos RO, Lamberti PW. Measure of similarity between images based on the codispersion coefficient. Journal of Electronic Imaging 21(2):023019 (2012).
\bibitem{Oje5} Ojeda SM, Britos GM, Vallejos R. An image quality index based on coefficients of spatial association with an application to image fusion. Spatial Statistics 23:1-16 (2018).
\bibitem{Pis} Pistonesi S, Martinez J, Ojeda SM, Vallejos R. A novel quality image fusion assessment based on maximum codispersion. In: Iberoamerican Congress on Pattern Recognition, Springer, pp 383-390 (2015).
\bibitem{ponomarenko2009tid2008} Ponomarenko N, Lukin V, Zelensky A, Egiazarian K, Carli M, Battisti F. Tid2008-a database for evaluation of full-reference visual quality assessment metrics. Advances of Modern Radioelectronics 10(4):30-45 (2009).
\bibitem{ponomarenko2015image} Ponomarenko N, Jin L, Ieremeiev O, Lukin V, Egiazarian K, Astola J, Vozel B, Chehdi K, Carli M, Battisti F, et al. Image database tid2013: Peculiarities, results and perspectives. Signal Processing: Image Communication 30:57-77 (2015).
\bibitem{Pra} Prastawa M, Bullitt E, Ho S, Gerig G. A brain tumor segmentation framework based on outlier detection. Medical image analysis 8(3):275-283 (2004).
\bibitem{Qui} Quintana C, Ojeda S, Tirao G, Valente M. Mammography image detection processing for automatic micro-calcification recognition. Chilean Journal of Statistics (ChJS) 2(2) (2011).
\bibitem{Sad} Sadabadi MS, Shafiee M, Karrari M. Two-dimensional arma model order determination. ISA
transactions 48(3):247-253 (2009).
\bibitem{schabenberger2005statistical} Schabenberger O, Gotway CA. Statistical methods for spatial data analysis. Text in Statistical Sciences. CHAPMAN and HALL/CRC press (2005).
\bibitem{Sin} Singh M, Arora H, Ahuja N. A robust probabilistic estimation framework for parametric image models. In: European Conference on Computer Vision, Springer, pp 508-522 (2004).
\bibitem{Tar} Tarel JP, Ieng SS, Charbonnier P. Using robust estimation algorithms for tracking explicit curves. Computer VisionECCV 2002 pp 492-507 (2002).
\bibitem{Theo} Theodoridis S, Koutroumbas K. Pattern recognition. Elsevier (2003).
\bibitem{Tjo} Tj{\o}stheim D. Statistical spatial series modelling. Advances in Applied Probability 10(1):130-154 (1978).
\bibitem{Vall3} Vallejos RO, Garcı́a-Donato G. Bayesian analysis of contaminated quarter plane moving average models. Journal of Statistical Computation and Simulation 76(2):131-147  (2006).
\bibitem{Vall2} Vallejos RO, Mardesic TJ. A recursive algorithm to restore images based on robust estimation of nshp autoregressive models. Journal of Computational and Graphical Statistics 13(3):674-682 (2004).
\bibitem{Wan} Wang Z, Bovik AC. Image and multidimensional signal processing-a universal image quality index. IEEE Signal Processing Letters 9(3):81-84 (2002).
\bibitem{Whi} Whittle P. On stationary processes in the plane. Biometrika pp 434-449 (1954).
\bibitem{Yao} Yao Q, Brockwell PJ. Gaussian maximum likelihood estimation for arma models ii: spatial processes. Bernoulli 12(3):403-429 (2006)
\bibitem{Zie} Zielinski J, Bouaynaya N, Schonfeld D. Two-dimensional arma modeling for breast cancer detection and classification. In: Signal Processing and Communications (SPCOM), 2010 International Conference on, IEEE, pp 1-4 (2010).

\end{thebibliography}


\end{document}